\documentclass[twocolumn,prx,floatfix,longbibiliography,aps,10pt,superscriptaddress]{revtex4-2}
\usepackage[colorlinks=true,bookmarksnumbered,bookmarksopen=true,bookmarksopenlevel=-1]{hyperref}
\usepackage[english]{babel}
\usepackage{amsmath}
\usepackage{mathtools}
\usepackage{amssymb}
\usepackage[bbgreekl]{mathbbol}
\usepackage{graphicx}
\usepackage{bm}
\usepackage{xcolor}
\usepackage{wasysym}
\usepackage{hyperref}
\usepackage{upgreek}
\usepackage{scalerel}
\usepackage{multirow}
\usepackage{pgffor}
\usepackage{pdfpages}
\usepackage{float}
\usepackage{lipsum}
\usepackage{tabularx}
\usepackage[caption=false,position=top]{subfig}
\usepackage{booktabs}
\usepackage{textcomp}
\usepackage{silence}
\makeatletter
\AtBeginDocument{\let\LS@rot\@undefined}
\makeatother
\AtBeginDocument{%
	\newwrite\bibnotes
	\def\bibnotesext{Notes.bib}
	\immediate\openout\bibnotes=\jobname\bibnotesext
	\immediate\write\bibnotes{@CONTROL{REVTEX42Control}}
	\immediate\write\bibnotes{@CONTROL{%
			apsrev42Control,author="08",editor="1",pages="1",title="0",year="0"}}
	\if@filesw
	\immediate\write\@auxout{\string\citation{apsrev41Control}}%
	\fi
}%

\makeindex

\begin{document}

\newcommand{\UPPSALA}{\affiliation{Department of Physics and Astronomy, Uppsala University, Box 516, S-751 20 Uppsala, Sweden}}

\newcommand{\LUND}{\affiliation{Solid State Physics and NanoLund, Lund University, Box 118, S-221 00 Lund, Sweden}}

\newcommand{\COFIRST}{\thanks{TL and OAA contributed equally to this work.}}

\title{Superconductivity and magnetism in the surface states of ABC-stacked multilayer graphene}

\author{Oladunjoye A. Awoga}
    \email[e-mail:]{oladunjoye.awoga@ftf.lth.se}
	\COFIRST
	\UPPSALA
	\LUND
	
\author{Tomas L\"othman} 
    \email[e-mail:]{tomas.lothman@physics.uu.se}
	\COFIRST
	\UPPSALA
	
\author{Annica M. Black-Schaffer}	
	\UPPSALA
\date{\today}
\begin{abstract}
ABC-stacked multilayer graphene (ABC-MLG) exhibits topological surface flat bands with a divergent density of states, leading to many-body instabilities at charge neutrality. Here, we explore electronic ordering within a mean-field approach with full generic treatment of all spin-isotropic, two-site charge density and spin interactions up to next-nearest neighbor (NNN) sites. We find that surface superconductivity and magnetism are significantly enhanced over bulk values. We find spin-singlet $s$ wave and unconventional NNN bond spin-triplet $f$ wave to be the dominant superconducting pairing symmetries, both with a full energy gap. By establishing the existence of ferromagnetic intra-sublattice interaction, $(J_2<0)$ we conclude that the $f$-wave state is favored in ABC-MLG, in sharp contrast to bulk ABC-graphite where chiral $d$- or $p$-wave states, together with s-wave states, display stronger ordering tendencies albeit not achievable at charge neutrality. We trace this distinctive surface behavior to the strong sublattice polarization of the surface flat bands. We also find competing ferrimagnetic order, fully consistent with density functional theory (DFT) calculations. The magnetic order interpolates between sublattice ferromagnetism and antiferromagnetism, but only with the ratio of the sublattice magnetic moments ($R$) being insensitive to the DFT exchange correlation functional. We finally establish the full phase diagram by constraining the interactions to the $R$-value identified by DFT. We find $f$-wave superconductivity being favored for all weak to moderately strong couplings $J_2$ and as long as $J_2$ is a sufficiently large part of the full interaction mix. Gating ABC-MLG away from charge neutrality further enhances the $f$-wave state over the ferrimagnetic state, establishing ABC-MLG as a strong candidate for $f$-wave superconductivity.
\end{abstract}

\maketitle
\section{Introduction}

Graphene, the single layer version of graphite, has ever since its isolation \cite{Novoselov2004Electric} continued to provide surprises. Recently, electron interaction effects have been at the center of attention, as electronically ordered states, including  superconductivity as well as various magnetic or charge ordered or correlated insulating states, have been found in multiple few-layer graphene systems. This includes twisted bilayer and trilayer graphene~\cite{Cao2018Correlated,Cao2018Unconventional,Hao2021Electric,Park2021Tunable,Kim2022Evidence}, where a small, magic-angle, twist between layers generates a moir\'e pattern hosting low-energy flat bands \cite{Bistritzer2011Moire}, which dramatically boosts the low-energy density of states (DOS). With the kinetic energy quenched within these flat bands, interactions effects are bound to become prominent, resulting in the already observed plethora of ordered states. Likewise,  bilayer and trilayer graphene strongly biased such that a low-energy partially flat band emerges have recently also been found to host both superconductivity and magnetic or charge ordered states~\cite{Marchenko2018,Zhou2022Isospin,Zhang2023Enhanced,Zhou2021Superconductivity,Zhou2021Half}. While most properties of the ordered states in twisted or biased few-layer graphene systems are still intensively investigated~\cite{Po2018Origin,Guinea2018Electrostatic,wu2018theory,Peltonen2018,kennes2018strong,Lu2019Superconductors,Yankowitz2019Tuning,Chen2019Signatures,Alidoust2019Symmetry,Balents2020Superconductivity,Andrei2020Graphene,Christos2020Superconductivity,Chichinadze2020Nematic,Wu2020Harmonic,Fischer2021,Oh2021Evidence,Cao2021Nematicity,Yu2021Nematicity,Khalaf2021Charged,cea2021coulomb,chou2021correlation,lothman2022nematic,Fischer2022Unconventional,chatterjee2022intervalley,chou2021Acoustic,you2021kohnluttinger,ghazaryan2021unconventional,Szabo2022Metals,Cea2022Superconductivity,Lu2022Correlated,Ghazaryan2023Multilayer,Xu2018Topological,Durajski2019Superconductivity,Park2021Magic,Chou2022B,Chou2022C,Su2022Superconductivity,Cea2023Superconductivity,pangburn2022superconductivity,crepieux2022superconductivity,pangburn2022superconductivityIII}, these results attest to the importance of electron interactions in graphene systems, whenever the low-energy DOS is dramatically enhanced.

Just beyond the few-layer graphene systems, ABC-stacked multilayer graphene (ABC-MLG) \cite{Lipson1942Structure}, a finite stacking sequence of rhombohedral graphite, has notably been known for some time to host topological flat band surface states at zero energy (or charge neutrality)~\cite{Heikkila2011Dimensional,Heikkila2011Flat}, also more recently observed in experiments~\cite{Pierucci2015Evidence,Henck2018Flat}. The bulk topology generates the flat bands on surfaces perpendicular to the stacking direction, given a sufficient number of layers and thereby clearly appearing in thicker ABC-stack, beyond five to six layers. These surface flat bands generate a large density of states (DOS) at charge neutrality, thus opening the possibility for interaction effects and electronic ordering without any need for twists or bias. While ABC-stacking is not the most common or stable stacking, recent synthesis and characterization have identified both tall and wide area samples of ABC-MLG, even exceeding a dozen layers \cite{Norimatsu2010,Pierucci2015Evidence, Henni2016,Henck2018Flat,Yang2019, Yanmeng2020Electronic, Bouhafs2021, hagymasi2022signature}. This makes ABC-MLG a natural contender for considering interaction effects in an unperturbed and structurally clean graphene system, beyond the need for externally applied bias or moir\'e structures. 
 
A straightforward testimony to the power of flat bands for ordering comes from the Stoner criterion for ferromagnetism, stating that a magnetic state forms as soon as the interaction strength exceeds the inverse Fermi-energy DOS \cite{Blundell1967Mag}. Magnetic ordering is therefore always a possibility in flat band systems with large zero-energy DOS, including ABC-MLG, as it only requires weak interactions. In fact, ABC-MLG surfaces have already been found to ferrimagnetically order in \emph{ab initio} density functional theory (DFT) calculations, with works focusing on systems with $3$ to $12$ layers~\cite{Otani2010Intrinsic,Xiao2011Density,Cuong2012Magnetic,Pamuk2017}. Experiments have also found evidence of magnetism in trilayer~\cite{Lee2014competition,lee2019gate}, tetralayer~\cite{Myhro2018Large}, and many-layer ABC-MLG~\cite{Yanmeng2020Electronic,hagymasi2022signature}. While this magnetic ordering clearly demonstrates the enhanced ordering tendency, there exist many other possible many-body instabilities, including superconductivity, as also clearly demonstrated by recent results on twisted and biased few-layer graphene \cite{Cao2018Unconventional,Zhou2022Isospin,Zhou2021Superconductivity}. In fact, a flat band generically enhances all  ordering channels \cite{Lothman2017Universal}, including superconductivity, implying a wealth of putative states also in ABC-MLG, at least at low temperatures.

In terms of superconductivity, a flat band dispersion has been found to give a linear dependence for the superconducting critical temperature $T_{\rm c}$ on the pairing interaction strength~\cite{Kopnin2011High,Kopnin2013High,Munoz2013Tight,Lothman2017Universal}, in contrast to the exponentially suppressed BCS result, which thus enables higher-temperature superconductivity. However, boosting $T_{\rm c}$ via flat bands has historically been dismissed, since the conventional contribution to the superfluid weight also vanishes when flattening the band dispersion \cite{Scalapino1992Superfluid}. This barrier imposed by theory was recently removed by uncovering an additional geometric contribution to the superfluid weight, induced by non-trivial band topology \cite{Peotta2015Superfluidity, Liang2017Band,Xiang2019Geometrical, Verma2021}, also present in ABC-MLG \cite{Kopnin2011}.

When it comes to the actual possibilities of superconductivity in ABC-MLG, the discoveries of superconductivity in twisted and biased few-layer graphene are very promising. Not only do they show that superconductivity is compatible with a flat band dispersion, but, most importantly, show that all carbon-based material, and graphene systems in particular, have at least one mechanism for producing superconducting pairing~\cite{Cao2018Unconventional, Lu2019Superconductors, Yankowitz2019Tuning, Chen2019Signatures,BlackSchafferandHonerkamp2014,Dai2021Mott}. In fact, superconductivity found in biased trilayer graphene is in an ABC stack, where two distinct superconducting regions are reached by strong gating and displacement field bias~\cite{Zhou2021Superconductivity}. While there is as of yet no consensus on the superconducting mechanism, or even the superconducting order parameter symmetry, these results have already attracted considerable theoretical attention~\cite{chatterjee2022intervalley,chou2021Acoustic,you2021kohnluttinger,ghazaryan2021unconventional,Szabo2022Metals,Cea2022Superconductivity,Lu2022Correlated}. However, with three layers being insufficient to develop the topologically protected surface flat bands of ABC-MLG, trilayer graphene requires a strong displacement field bias to produce its partial band flattening leading to superconductivity. 

Considerations of superconductivity in ABC-MLG not requiring any twist or bias, but instead utilizing the natural topological flat band regime in thicker stacks, have so far been limited to modeling of conventional (on-site) $s$-wave superconductivity~\cite{Kopnin2011High,Kopnin2013High,Munoz2013Tight,Lothman2017Universal,Ojajaervi2018}, from effective phonon-mediated pairing. While a phonon-mediated pairing mechanism remains plausible, electronic interactions are also known to be important, with estimates from \emph{ab initio} calculations suggesting strong local and non-local repulsive Coulomb interaction in both single and multilayer graphene systems~\cite{Wehling2011strength}. 
Additionally, the rich phase diagrams of twisted and biased few-layer graphene, clearly show the importance of electron interactions, likely also for superconductivity. For example, intriguing similarities exists between twisted magic-angle bilayer graphene and the high-$T_{\rm c}$ cuprate superconductors, including characteristic domes of superconductivity next to correlated insulating states, pseudogap physics, and strange metal behavior \cite{Cao2018Correlated,Cao2018Unconventional}, all hinting at an electronic mechanism for superconductivity. With the large surface zero-energy DOS in ABC-MLG, we may expect similar prospects for electronic-driven superconductivity in the surface states of ABC-MLG.
	
In this work we address the possibility of electronic ordering in the surface flat bands of ABC-MLG systems arising from repulsive electronic interactions. We use a combination of approaches, including fully self-consistent tight-binding mean-field calculations and group theory analysis, as well as complementary DFT calculations. For a fully unbiased treatment, we start from completely general two-site spin-isotropic charge density and spin interactions. We then analyze within mean-field theory all superconducting, charge, and magnetic ordering channels, retaining all channels out to the next-nearest-neighbor (NNN) range.

We generally find strongly enhanced surface superconducting and magnetic orders due to the topological flat bands in thick ABC-MLG systems. For superconductivity we  find finite-ordering temperatures also for weak interactions for most superconducting symmetries. This is in sharp contrast to both few-layer graphene and bulk ABC-graphite where, due to their semimetallic dispersion, pairing is only possible beyond a critical interaction strength at charge neutrality. In particular, we establish isotropic spin-singlet $s$-wave pairing and unconventional NNN bond spin-triplet $f_{x\left(x^2-y^2\right)}$ wave to be the dominating pairing symmetries, both having a full energy gap. We find that the spin-triplet $f$-wave state is favored by ferromagnetic NNN bond Heisenberg interactions ($J_2$), while the $s$ wave requires NNN antiferromagnetic interactions or on-site attraction from electron-phonon interactions. With intra-sublattice interactions expected to be ferromagnetic in graphene, i.e.~$J_2<0$, and the $f$-wave state also surviving significant on-site repulsion $U_0$ expected to be present in graphene, while on-site repulsion instead hurts the $s$-wave state, we conclude that the spin-triplet $f$-wave pairing is the most likely superconducting symmetry in ABC-MLG. This is very different from bulk ABC-graphite and also few layer graphene, where instead the chiral $(d_{x^2-y^2}+id_{xy})$- or $(p_x+ip_y)$-wave states together with an $s$-wave state display the best ordering tendencies. We trace this different bulk versus surface behavior to the strong sublattice polarization of the surface states.	
	
For putative charge and magnetic ordering, we find a dominating ferrimagnetic state on the surface, also strongly enhanced by the surface flat band and not present in the bulk or few-layer graphene. This is in agreement with earlier DFT calculations, which captures all mean-field charge and magnetic orders (but not superconductivity) \cite{Otani2010Intrinsic, Xiao2011Density, Cuong2012Magnetic, Pamuk2017}. We further find that the ferrimagnetic ordering interpolates between pure sublattice ferromagnetism and complete antiferromagnetism, as any of the interaction strengths is tuned from weak to strong coupling. The ferrimagnetic state can therefore be quantified by its sublattice magnetic moment ratio $R = | m_{B1}/m_{A1} |$, with $m_{A1(B1)}$ being the magnetic moments on the surface A (B) sublattice. We also find that our complementary DFT calculations gives highly varying $m_{A1(B1)}$ depending on the choice of exchange-correlation functional, but their ratio notably stays approximately constant at $R \approx 0.42$. 
	
We finally treat the competition between ferrimagnetism and $f$-wave superconductivity in ABC-MLG. A ferromagnetic $J_2$ drives both states, but for a large range of weak to moderate $J_2$, we find the superconducting $f$-wave state to have a higher transition temperature. Thus, we find $f$-wave superconductivity as long as the proportion of $J_2$ in the full mix of interaction parameters is sufficiently large. We find that superconductivity is further favored over magnetic ordering by gating the system, such that the Fermi level moves away from the flat band. We finally establish the full phase diagram, including gating effects, as a function of relevant interaction parameters by utilizing the constant $R$-value extracted from the DFT calculations, and find that it contains a substantial portion of $f$-wave superconductivity. Taken together, our results establish that spin-triplet $f$-wave superconductivity is not only feasible from purely repulsive interactions on the surface of ABC-MLG but even the dominating ordered state in a large portion of the phase diagram. Gating the system away from charge neutrality further enhance the $f$-wave state compared to a competing ferrimagnetic state. Due to the distinctive properties of the surface flat band in ABC-MLG these results are notably different from ordered states in biased or twisted few layer graphene. Beyond direct predictions for ABC-MLG, our results also clearly illustrate the strong dependence on the normal state behavior for electronic ordering in graphene systems, leading directly to a surprising range of variations of electronic ordering in graphene-based systems.

We organize the remaining of this work as follows. We start in Section \ref{sec:NonInt} by recapitulating the properties of ABC-MLG, focusing especially on the surface flat bands. Then in Section \ref{sec:Int} we introduce a fully general treatment of all electron-electron interaction parameters up to the NNN range and perform a complete mean-field theory decomposition, taking all possible orders into account. In Section \ref{sec:SC} we investigate all possible superconducting orders, starting with bulk ABC-graphite as a reference, then focusing on ABC-MLG and its surface superconducting state, elucidating the layer dependence, superconducting pairing symmetries, and superconducting phase diagram containing both spin-singlet $s$- and spin-triplet $f$-wave pairing. In Section \ref{sec:Interactions} we provide estimation for the various interaction parameters and are thereby able to single out the spin-triplet $f$-wave state as the leading superconducting order. We then move onto charge and magnetic ordering in Section \ref{sec:Magnetism} where we establish a surface ferrimagnetic state based on our mean-field framework, and show how it can be quantified by the ratio $R$. We study the resulting competition between the $f$-wave superconducting and ferrimagnetic state in Section \ref{sec:SCPH}. With the help of both mean-field and DFT results we establish the overall phase diagram, also taking gating effects into account, which overall contains a large part with dominating $f$-wave superconductivity. Finally, in Section \ref{sec:Conc} we summarize our results.

\section{Normal State ABC-MLG} \label{sec:NonInt}
\begin{figure*}[!ht]
		\centering
		\includegraphics[width=1\textwidth]{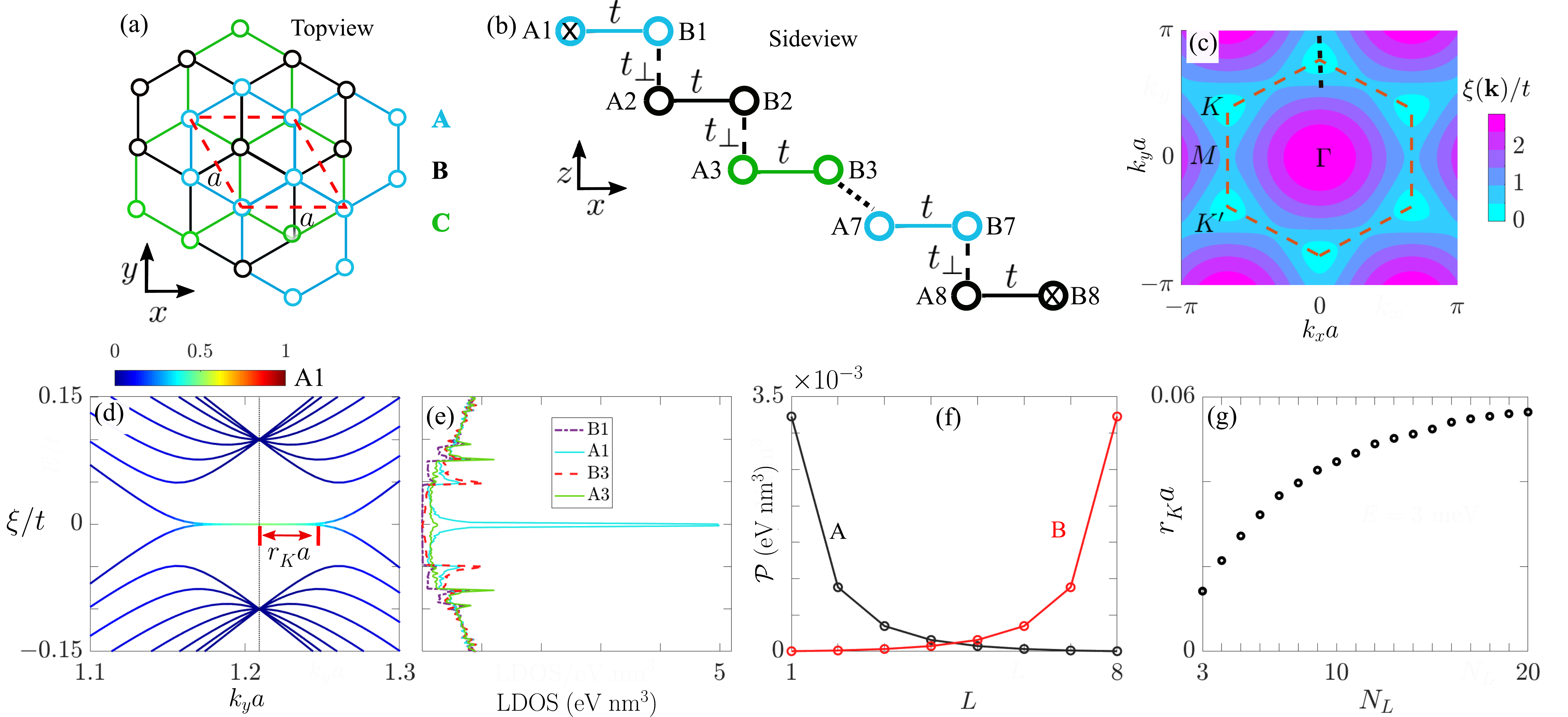}
		\caption{%
			{(a)} Top view of the atomic positions for three layers of graphene (blue,black, green) with  ABC stacking order. Red dashed region shows the unit cell of length $a$.
			{(b)} Side view of atomic positions for eight layers with bonds labeled by their tight-binding hopping amplitudes of Eq.~\eqref{eq:Normal}.
			{(c-f) }Normal state electronic structure of eight layer ABC-MLG: 
			{(c)} Contour plot of lowest positive energy band $\xi(\vec{k})$. Dashed brown hexagon marks the first Brillouin zone with high symmetry points $\Gamma$, $K$, $K'$, and $M$. 
			{(d)} Energy band plot $ \xi = \xi\left( k_x=0, k_ya \right)$ along the dashed vertical line in (c), passing through the $K'$-point. Band line color shows probability density on surface site A1 (or equivalently B8), marked by X in (b).
			{(e)} Local density of states (LDOS) on surface sites A1 and B1 and third layer sites A3 and B3. 
			{(f)} Integrated LDOS, $\mathcal{P} =\sum_\xi {\rm LDOS}\left(\xi\right)$ resolved by sublattice and layer for narrow low-energy window $\xi \in \left[E_{\rm F}-3\,{\rm meV},E_{\rm F}+3\,{\rm meV}\right]$ around $E_{\rm F}=0$. 
			{(g)} Surface flat band radius, $r_K$ as a function of number of layers $N_L$ in a ABC-MLG stack. The radius $r_K$, illustrated by a red line in (d), is estimated as the momentum where the flat band exits the narrow energy window used in (f).
			} 
		\label{Fig1_LatticeNormal}
	\end{figure*}

	In ABC-MLG, the graphene layers alternate cyclically between three positions: A, B, and C, see Fig.~\ref{Fig1_LatticeNormal}(a). Each graphene layer is in turn bipartite and when viewed from the side, ABC-MLG can be visualized as a staircase arrangement of dimers, see Fig.~\ref{Fig1_LatticeNormal}(b), in which the sublattice A atom is located above (or below) the B site atom of the adjacent layer. A consequence of this repeated staircase structure is that infinite ABC-MLG, i.e.~rhombohedral-graphite, has a simple unit cell with only two carbon atoms, with an in-plane unit cell length $a$.
	To model the normal state electronic structure of ABC-MLG, we use a tight-binding Hamiltonian. The energy scale is foremost set by the intralayer nearest-neighbor (NN) hopping amplitude $t=3.0$~eV within each graphene layer. Thereafter follows the interlayer hopping between different sublattices of neighboring layers $t_\perp=0.1t=0.3$~eV, since each carbon atom has another atom directly above or below it. A Hamiltonian consisting only of these two hopping amplitudes is already a good approximation for the non-interacting normal state electronic structure of ABC-MLG, as it captures the bulk topology and the associated topological surface flat bands~\cite{Kopnin2011High,Heikkila2011Flat}. For $N_L$ layers, this Hamiltonian reads \cite{Heikkila2011Flat,Xu2012,Lothman2017Universal}:
		\begin{equation}\label{eq:Normal}
			\begin{split}
				H_0 &=  -\mu \sum_{ iL\alpha} a_{iL\alpha}^\dagger a_{iL\alpha} - t \sum_{ \langle i,j\rangle L \alpha}  a_{iL\alpha}^\dagger a_{jL\alpha}  \\
				&+ t_\perp  \sum_{ \langle L,L'\rangle } \sum_{ \langle i,j\rangle \alpha} a_{iL\alpha}^\dagger a_{jL'\alpha },
			\end{split}
		\end{equation}
	where $a_{iL\alpha}^\dagger$ creates an electron with spin $\alpha$ at site $i$ in layer $L=1,\dots,N_L$ and neighboring sites (layers) are denoted by $\langle \cdots \rangle$. The number of electrons is regulated by the chemical potential $\mu$.

The low-energy electronic structure of normal state ABC-MLG originates from the single-layer graphene Dirac cones found at the $K$- and $K^\prime$ corner points of the graphene Brillouin zone. In ABC-MLG, the interlayer hybridization $t_\perp$  transforms the individual graphene Dirac points into Fermi spirals, twisting around the crystal momentum axis of the stacking direction $k_z$, connecting the $H-K-H$ and $H^\prime-K^\prime-H^\prime$ points \cite{Heikkila2011Dimensional}. Associated with these Fermi spirals is a topological winding number with the consequence that a topological surface flat band is formed within the projection of the Fermi spirals on each stack surface \cite{MCCLURE1969Electron, Heikkila2011Flat, Heikkila2011Dimensional}. This mimics the physics of the one-dimensional Su-Schrieffer-Heeger (SSH)-model, giving ABC-MLG very different properties compared to other graphene stackings~\cite{Su1979,Heikkila2011Dimensional,Heikkila2011Flat,Xiao2011Density}. 
A finite stack of ABC-MLG thus hosts two topological flat bands, one on each perpendicular surface, which are located around the $K$- and $K^\prime$-points and tied to the charge neutral zero energy point (the graphene Dirac point). This is illustrated in Fig.~\ref{Fig1_LatticeNormal}(c), where we plot the lowest positive energy electronic band of eight layer ABC-MLG. The vanishing dispersion of the flat bands is further seen in Fig.~\ref{Fig1_LatticeNormal}(d), where we plot the band structure along a line cut passing through one of the $K$-points, $\xi\left(0,k_y\right)$, as indicated by the dashed line in Fig.~\ref{Fig1_LatticeNormal}(c). 	
	
The flat dispersion results in a significantly enhanced DOS around zero energy, seen by plotting the local density of states (LDOS) in Fig.~\ref{Fig1_LatticeNormal}(e) for the same eight layer ABC-MLG. The large difference between the surface (cyan and purple) and third (red and green) layer LDOS clearly display the surface localization, as well as a strong sublattice dependence. Resolving the states further reveals that the flat bands exactly at the two $K$ points are almost completely localized on of the outer surface sites A1 and B8, marked by X in Fig.\ref{Fig1_LatticeNormal}(b). Moving slightly away from the $K$ points, other sites also contribute, shown by using the band color in Fig.~\ref{Fig1_LatticeNormal}(d) to indicate the  A1 site probability density.

We further quantify the sublattice layer dependence by integrating the site-dependent LDOS in a narrow energy window $\xi \in \left[E_{\rm F}-3\,{\rm meV},E_{\rm F}+3\,{\rm meV}\right]$ around zero energy, where $E_{\rm F}$ denotes the Fermi energy. We plot the resulting integrated density, $\mathcal{P}$, as a function of layer in Fig.~\ref{Fig1_LatticeNormal}(f) for the two sublattices A and B. Together Figs.~\ref{Fig1_LatticeNormal}(d-f) demonstrate that the topological surface states are strongly localized to the two surfaces and that the two surfaces have a near complete sublattice polarization that is reversed between them. We note that this sublattice polarization clearly sets many layer ABC-MLG apart from ABC-graphite, twisted bilayer graphene, and also to a large extent from few layer ABC-MLG where the sublattice polarization is non-existent or much less pronounced.

Due to their topological origin, the surface flat bands in ABC-MLG depends on having a sufficient number of layers in the stack to establish the bulk topology. To track the development of the surface flat band, we plot the flat band radius against the number of layers in Fig.~\ref{Fig1_LatticeNormal}(g), where we define the radius to be the length of the wave-vector around the $K$-point for which the band energy is below a fixed threshold, using the same energy window as in Figs.~\ref{Fig1_LatticeNormal}(e) and~\ref{Fig1_LatticeNormal}(f). As seen, the flat band radius grows monotonically with $N_L$, slowly saturating to the projected Fermi spiral area. Thus, there is a rapid increase in the DOS, which is proportional to the flat band area, with each added layer of ABC-MLG when going from only a few layers to thick samples.
		
Finally, we comment that in the NN hopping model of Eq.~\eqref{eq:Normal}, the bulk topology of ABC-MLG depends on  particle-hole symmetry, while longer-ranged intra-sublattice hopping terms break this symmetry. However, such longer range hopping is relatively small and adds only a small curvature to the flat bands. In fact, the surface flat bands have been shown to clearly develop also in DFT calculations at both surfaces as well as interfaces of ABC-MLG \cite{Xiao2011Density}, showing that the essential features are not dependent on the assumptions underpinning Eq.~\eqref{eq:Normal}.

\section{Electron Interactions} \label{sec:Int}
The large surface DOS makes ABC-MLG very susceptible to develop a variety of ordered states from interactions, a select few of which have already been explored \cite{Heikkila2011Dimensional, Heikkila2011Flat, Lothman2017Universal}. Here we proceed within a more systematic approach. We first note that graphene and graphite systems~\cite{Wehling2011strength,TancogneDejean2020}, including ABC-MLG~\cite{Zhang2010Band,Wang2013Flat}, have been found to have strong local and non-local Coulomb interactions. We therefore analyze all possible symmetry-breaking orders, including superconductivity, arising from local and non-local electronic interactions in ABC-MLG, by adding these interactions to the non-interacting normal state model of Eq.~\eqref{eq:Normal}. This allows us to determine the leading superconducting and magnetic instabilities, as well as their competition.

We keep our analysis generic and therefore include all spin-isotropic two-site effective density-density and spin-spin interactions within each graphene layer. We are able to  limit our analysis to these intralayer interactions, since the individual graphene layers of ABC-MLG are only coupled with weak van der Waals forces. 	
Since the in-plane interaction potential decays with distance \cite{Wehling2011strength}, we are also able to restrict ourselves to relatively short-range interactions. Still, we note that non-local interactions are important for ordering in ABC-MLG. In particular, we find that the strong surface sublattice polarization of the ABC-MLG surface states means that the next-nearest-neighbor (NNN) range is in fact the shortest interaction range beyond on-site interactions that connects carbon sites with a significant low-energy DOS, thereby producing  a strong ordering susceptibility. For this reason, we  consider interaction ranges up to NNN to capture these important effects. Thus, we extend the non-interacting normal state model of Eq.~\eqref{eq:Normal} by adding an interaction Hamiltonian $H_{\rm I}$ of the form:
		\begin{equation} \label{eq:Int}
			\begin{split}
			H_{\rm I} &=\frac{1}{2} \sum_{ijL\alpha\beta \delta \gamma} \Gamma_{i j}^{\alpha\beta \delta \gamma} a_{iL \alpha}^{\dagger} a_{jL \beta}^{\dagger} a_{jL \delta}a_{iL\gamma},\, \\
			\Gamma_{ij}^{\alpha\beta \delta \gamma} &=\frac{1}{2}\sum_\nu \left[U_{i j } \sigma_{\alpha \gamma}^{0} \sigma_{\beta \delta}^{0}+J_{i j } \sigma_{\alpha \gamma}^{\nu} \sigma_{\beta \delta}^{\nu}\right].
			\end{split}
		\end{equation}
		
Here the matrices $\sigma^0$ and $\sigma^\nu,\, \nu = x,y,z$, are the $2 \times 2$ identity matrix and Pauli matrices in the spin degrees of freedom (indexed by $\alpha, \beta, \gamma, \delta$), respectively, which we also collect in the four-vector $\sigma^\eta =\{\sigma^0,\, \sigma^\nu\}$.
The total interaction potentials $\Gamma_{ij}$ between sites $i$ and $j$ can always be divided into two distinct parts: $U_{ij}$ being the effective density-density Coulomb interactions and $J_{ij}$ being the effective isotropic Heisenberg spin-spin (or exchange) interactions. This captures all two-site electron interactions and we will consider all such terms up to NNN distance.

	\subsection{Mean-field decomposition}
To proceed we perform a complete Hartree-Fock-Bogoliubov mean-field decomposition of the total Hamiltonian  $H = H_0 + H_{\rm I}$. This allows us to access all possible orders, superconducting as well as charge and magnetic orders, in the ABC-MLG. Doing so we arrive at the mean-field interacting Hamiltonian: 
			\begin{align} \label{eq:MFInt} 
				H_{\rm MF} =  H_0 + 
					& 
					\sum_{i jL\eta} \big( \Delta_{j iL}^{\eta} \big(\hat{g}_{i jL}^{\eta}\big)^\dagger+ \text{H.c} \big) 
					+ 
					\big( m_{iiL}^{\eta}\hat{{n}}_{j jL}^{\eta}+ i\leftrightarrow j \big)\nonumber \\
					&  
					+ \big(\tilde{m}_{i jL}^{\eta} \hat{n}_{j iL}^{\eta}  +   i\leftrightarrow j \big) +{\rm constant~terms}
			\end{align}
		where
			\begin{subequations} \label{eq:MFSelfCon} 
				\begin{align}
					\Delta_{j iL}^{\eta} 		&= 
					\Gamma_{i j}^{\eta,+} \langle \hat{g}_{j iL}^{\eta}\rangle, 
					\;\; 
					\hat{g}_{j iL}^{\eta} = 
					\frac{1}{2} \sum_{\alpha \beta} 
					a_{jL\alpha} \left[\bar{\sigma}^{\eta}\right]_{\alpha \beta} 
					a_{iL \beta} 
					\label{eq:MFSelfConSC}
					\\
					m_{i jL}^{\eta} 				&= 
					\Gamma_{i j}^{\eta,-} \langle \hat{n}_{i i L}^{\eta} \rangle, 
					\;\; 
					\hat{n}_{ii L}^{\eta}=
					\frac{1}{2} \sum_{\alpha}  
					a_{iL\alpha}^\dagger \sigma^{\eta}_{\alpha \alpha} 
					a_{iL \alpha} 
					\label{eq:MFSelfConPHDir}
					\\
				 \tilde{m}_{i jL}^{\eta} 	&= 
					\tilde{\Gamma}_{ij}^{\eta,-}  \langle \hat{n}_{i jL}^{\eta} \rangle, 
					\;\; 
					\hat{n}_{i jL}^{\eta} = 
					\frac{1}{2} \sum_{\alpha \delta} 
					a_{iL\alpha}^\dagger \sigma^{\eta}_{\alpha \delta} 
					a_{jL \delta}. 
					\label{eq:MFSelfConPHEx}
				\end{align}
			\end{subequations}
	In order of appearance, the terms in the mean-field Hamiltonian~\eqref{eq:MFInt} are: the particle-particle (superconducting or Bogoliubov) channel with the order parameters $\Delta_{j iL}^{\eta}$ (here $\bar{\sigma}^{\eta}=i\sigma^y\sigma^\eta$), the direct particle-hole (Hartree) channel with order parameters $m_{i jL}^{\eta}$, and the exchange particle-hole (Fock) channel with the order parameters $ \tilde{m}_{i jL}^{\eta}$. The order parameters in each channel are defined self-consistently through Eqs.~\eqref{eq:MFSelfCon}. For each channel, the total coupling strengths, $\Gamma_{i j}^{\eta,\pm}$ are linear combinations of the effective interaction strengths $U_{ij}$ and $J_{ij}$, as tabulated in Table~\ref{Tab:MFInteraction}. In the table we divide the channels further into spin-singlet and spin-triplet configurations, based on the sigma matrices. Note that in Eq.~\eqref{eq:MFSelfCon}, non-local interactions connect order parameters at different sites. Additionally, the Bogoliubov and Fock channels, but not the Hartree channel, contain bond order parameters that also connect two different sites directly in the operator expectation value. The former is illustrated in Fig.~\ref{Fig2_SC}(a). 
		\begin{table}[!t] 
			\centering
			\renewcommand{\arraystretch}{1.5}
			\begin{tabular}{ |p{1.0cm}|p{2.7cm}|p{1.7cm}|p{2.6cm}|}
				\hline
				Spin	& Superconducting & PH Direct & PH Exchange	\\
				\hline
				Singlet 
					& $ \Gamma_{i j}^{0,+}						=	U_{i j}-3J_{i j} $ 
					& $ \Gamma_{i j}^{0,-}						=	U_{i j} $ 
					& $ \tilde{{\Gamma}}_{i j}^{0,-} 	= -\frac{U_{i j}+3J_{i j}}{2} $ 
				\\
				Triplet 
					& $ \Gamma_{i j}^{\nu,+}					= U_{i j}+J_{i j}$
					& $ \Gamma_{i j}^{\nu,-}					= J_{i j}$
					& $ \tilde{\Gamma}_{i j}^{\nu,-}	= -\frac{U_{i j}-J_{i j}}{2}$ 
				\\
				\hline
			\end{tabular}
			\caption{%
				Total mean-field coupling strengths in Eq.~\eqref{eq:MFSelfCon} for both superconducting and particle-hole (PH) channels in terms of the effective density-density Coulomb interactions $U_{ij}$ and isotropic Heisenberg spin-spin interactions $J_{ij}$ in Eq.~\eqref{eq:Int}.
			}
			\label{Tab:MFInteraction}
		\end{table}

	\subsection{Linearized gap equation}
		The complete mean-field decoupling of Eq.~\eqref{eq:MFInt} reduces the interacting system to one of non-interacting quasiparticles moving through the collective mean-field potentials defined by the non-linear self-consistency equations of Eq.~\eqref{eq:MFSelfCon}. These mean fields will often only renormalize the quasiparticle properties. There is, however, also the possibility that the interactions produce a qualitative different state through spontaneous symmetry breaking. Since we seek to analyze the possibility of such symmetry-breaking orders, we hereafter assume that all renormalizations have already been included in the normal state Hamiltonian~\eqref{eq:Normal} and its parameters, which therefore corresponds to what would be measured as the normal state above the ordering temperatures, as opposed to the bare parameters.
		
		The full nonlinear self-consistency equations of Eq.~\eqref{eq:MFSelfCon} are valid at any temperature as they completely define each order parameter. However, the order parameter of any symmetry breaking order only become finite below a certain critical transition temperature, $T_{\rm c}$. Right at $T_{\rm c}$ the order parameter is therefore vanishingly small. This allows the self-consistency of Eq.~\eqref{eq:MFSelfCon} to be linearized by a perturbation expansion in $H_{\rm MF}$ (see Appendix~\ref{app:StabMat} for details). After such linearization we arrive at the generic linearized gap equation (LGE)~\cite{Lothman2017Universal}:
		\begin{equation} \label{eq:StabMat}
			\begin{split}
			 \boldsymbol{D}^{\eta,\pm} & = \mathbb{M}^{\eta,\pm} \boldsymbol{D}^{\eta,\pm},\\
			\text{where}\quad \mathbb{M}^{\eta,\pm} & = - \mathbb{\Gamma}^{\eta,\pm}\sum_{\bar{i}\bar{j}\boldsymbol{k}}\mathbb{S}^{\bar{i}\bar{j},\eta,\pm}\left(\boldsymbol{k}\right)\chi_{\bar{i}\bar{j}}^\pm\left(\boldsymbol{k}\right),
			\end{split}
		\end{equation}
		with the order parameters $\boldsymbol{\Delta}^{\eta}$ collected in a column vector $\boldsymbol{D}^{\eta,+}$, and similarly $\boldsymbol{m}^{\eta}$ and $\tilde{\boldsymbol{m}}^{\eta}$  collected in $\boldsymbol{D}^{\eta,-}$. Here, $\bar{i},\,\bar{j}$ label the bands of the normal state, $H_0$. In particular, because only the superconducting orders breaks the U(1) gauge symmetry, there is no coupling between the superconducting particle-particle ($+$ superscript) and particle-hole, i.e., the density and magnetic, ($-$ superscript) orders and thus Eq.~\eqref{eq:StabMat} is in fact two equations for these two separate channels. 		
		 
		Central to the LGE is $\mathbb{M}^{\eta,\pm}$, the stability matrix, which is the temperature dependent response matrix. Its matrix elements are a combination of the interaction coupling strengths $\Gamma_{ij}^{\eta,\pm}$, stored in $\mathbb{\Gamma}^{\eta,\pm}$, the structure factors expressed by the symmetry matrix $\mathbb{S}^{\eta,\pm}$ (derived in Appendix~\ref{app:StabMat}), and the temperature-dependent particle-particle and particle-hole susceptibilities:
			\begin{equation} \label{eq:Susp}
				\chi_{\bar{i}\bar{j}}^\pm\left(\beta\xi_{ \bar{i}^{}},\beta\xi_{ \bar{j}^{}},\boldsymbol{k}\right)
				= 
				\beta
				\frac{ 
					\tanh \left(\frac{\beta_{\rm } \xi_{\bar{i}}\left(\boldsymbol{k}\right)}{2}\right) \pm \tanh \left(\frac{\beta_{\rm } \xi_{\bar{j}}\left(\boldsymbol{-k}\right)}{2}\right)
				}{
					\beta \xi_{\bar{i}} \left(\boldsymbol{k}\right) \pm \beta \xi_{\bar{j}}\left(\boldsymbol{-k}\right)
				},
			\end{equation}
		where $\beta = (k_{\rm B}T)^{-1}$ is the inverse temperature, $k_{\rm B}$ is the Boltzmann constant, and $\xi_{\bar{i}}^{}$ are the band energies of the normal state $H_0$. 				

		For a given set of interactions and at an arbitrary temperature, the LGE Eq.~\eqref{eq:StabMat} does not generically have a solution. Rather, Eq.~\eqref{eq:StabMat} can only be satisfied when the stability matrix $\mathbb{M}$ has at least one eigenvalue that is unity. At high temperatures, all the eigenvalues are generally suppressed by the temperature factor ($\beta$) in the susceptibility, Eq.~\eqref{eq:Susp}. With decreasing temperatures, the eigenvalues tend to grow. The temperature where an eigenvalue eventually reaches unity marks the critical temperature $T_{\rm c}$ where Eq.~\eqref{eq:StabMat} then has a solution and an ordering transition takes place. The type of order and its spatial pattern is completely specified by that solution's corresponding eigenvector $\boldsymbol{D}^{\eta,\pm}$. The LGE Eq.~\eqref{eq:StabMat} is therefore an eigenvalue problem of the temperature dependent stability matrix, $\mathbb{M}^{\eta,\pm}$, and by looking at the  eigenvalues of $\mathbb{M}$ it is possible to determine the order competition. The order with the largest eigenvalue becomes the dominant order, but can also be followed by sub-dominant orders, arising at lower temperatures. Two orders with similar eigenvalues can likewise be viewed as directly competing.
		
\subsection{Order parameter symmetries} 
The LGE Eq.~\eqref{eq:StabMat} has the same symmetry as the normal state Hamiltonian Eq.~\eqref{eq:Normal} and therefore the stability matrix $\mathbb{M}$ is block diagonal in a symmetry basis of the irreducible representations (irreps) of the normal state Hamiltonian. Only below $T_{\rm c}$ is the symmetry reduced, when a symmetry-breaking order parameter becomes finite. 
		
Since the normal state Hamiltonian Eq.\eqref{eq:Normal} is spin-isotropic, the mean-field channels first split into two spin-symmetry sectors, spin-singlet ($\eta=0$) and spin-triplet ($\eta=\nu$). The spin-triplet sector is further completely degenerate, and we can thus here focus only on the $\hat{z}$ channel, with no loss of generality. With only in-plane interactions, we can resolve the order parameters as functions of the in-plane bond ranges by writing any parameter as $D_{L,i j}=D_{L,i,i+r}=D_{L,r}$, with $r=0,1,2$ for on-site (ON), nearest-neighbor (NN), and next nearest neighbor (NNN) orders, respectively. Because each of these bond ranges is a close set under the symmetry transformations of the normal state Hamiltonian $H_0$, we can construct a real space symmetry basis for all orders labeled by spin sector $\eta$, bond range $r$, and irreps of the lattice symmetry of $H_0$. Consequently, any order parameter $\boldsymbol{D}^{\eta,\pm}$ can be written as a linear combination in this basis:
			\begin{equation}
			\label{eq:symbasis}			
				\boldsymbol{D}^{\eta,\pm} = \sum_{L,r,\chi} {A}_{L,r}^{\eta,\chi} \hat{\boldsymbol{D}}_{L,r}^{\eta,\chi,\pm},
			\end{equation}
		where $\chi$ labels all the in-plane irreps in layer $L$. Since any solution to Eq.~\eqref{eq:StabMat} has a definite symmetry, the non-zero amplitudes ${A}_{L}^{\eta}$ can only belong to one irrep, which is then the same in all layers of a ABC-MLG stack. We can therefore drop the $L$ subscript. The real-space symmetry basis states in Eq.~\eqref{eq:symbasis} also translate to a corresponding expression in reciprocal space:
			\begin{equation}
			\label{eq:symk}
			\boldsymbol{D}_{L,r}^{\eta, \chi,\pm} \left( \boldsymbol{k} \right) \equiv	\boldsymbol{D}_{r}^{\eta, \chi,\pm} \left( \boldsymbol{k} \right) 
				= 
				\sum_{n}
				\hat{\boldsymbol{D}}_{ \boldsymbol{g}_{n}^{\left( r \right)}}^{\eta,\chi,\pm}
				e^{ i \boldsymbol{k} \cdot \boldsymbol{g}_n^{\left(r\right)} },
			\end{equation} 			
		where $\boldsymbol{k}$ is the in-plane crystal momentum in the first Brillouin zone and $\hat{\boldsymbol{D}}_{ \boldsymbol{g}_{n}^{\left( r \right)}}^{\eta,\chi,\pm}$ is the on-site or bond amplitude on the in-plane on-site or bond vector $\boldsymbol{g}_n^{\left(r\right)}$, as illustrated in Fig.~\ref{Fig2_SC}(a) for sublattice A. 
			
We concretize this general derivation by tabulating in Table~\ref{Tab:Symmetry} all possible basis states in Eq.~\eqref{eq:symbasis} for the superconducting order possibilities, using the notation $\vec{\Delta}_r^{\eta}$ for $\hat{\boldsymbol{D}}_{r}^{\eta,\chi,+}$, where the rows represent the possible irreps $\chi$ for each range $r$ and divided into the two different spin sectors. Note that in the spin-singlet sector only even irreps (gerade) are allowed, while spin-triplet only allows for odd representations (ungerade), due to the Fermi-Dirac statistics of the superconducting Cooper pairs. For each irrep and bond range, we extract the corresponding real-space structure for $\vec{\Delta}_r^{\eta}$ (second column). In reciprocal space we tabulate the lowest order expansion around the Brillouin zone center $\Gamma$ (third column). This choice of high-symmetry Brillouin zone point makes it possible to directly match up the resulting basis function to the irreps of the $D_{6h}$ point group of the honeycomb lattice (fourth column)~\cite{Black-Schaffer07, BlackSchafferandHonerkamp2014, Lothman14,Awoga2017Domain,Awoga2018Probing,SigristUeda1991}. Other reciprocal space choices, such as the $K$ point, do not  change the overall decomposition but can mix up the gerade (ungerade)  and spin singlet (spin triplet) as they are not centered at zero momentum.

Our goal next is to analyze which of the ordering channels have the highest $T_{\rm c}$ and are thus dominant, starting with superconductivity. Before proceeding we note that below $T_{\rm c}$ all multidimensional irreps generally split~\cite{SigristUeda1991}, with the dominating channel being the one with the lowest energy determined by higher order (non-linear) terms in the self-consistent equations, Eq.~\eqref{eq:MFSelfCon}. Moreover, below $T_{\rm c}$ it is in principle possible to also produce co-existing orders from different spin channels and irreps, which can only be accessed by solving the full non-linear self-consistency equations~\eqref{eq:MFSelfCon}. In practice this is, however, seldom seen and then only at notably lower temperatures, and we will therefore for now ignore that possibility and focus on the order with the highest critical temperature.
		 
\section{Superconductivity} \label{sec:SC}
Having performed a full symmetry analysis of all possible order parameter symmetries, we next determine which superconducting orders can actually be realized in ABC-stacked graphene. We consider here both finite and infinite ABC-stacked graphene. The infinite-layer case, corresponding to bulk ABC-graphite, serves as a baseline to separate bulk from surface contributions. Since the spin-singlet and -triplet sectors never mix at $T_{\rm c}$, we consider each spin sector separately. To clearly identify each contribution and possible order, we at first also turn off the coupling between the different pairing range $r$ (ON, NN, NNN), treating them also separately before a joint analysis in Sec.~\ref{sec:SurfaceSCInt}. Note however that spin-triplet ON ordering is forbidden by Fermi-Dirac statistics, resulting in total five possible ordering channels, all tabulated in Table~\ref{Tab:Symmetry}. For each of these channels, we solve Eq.~\eqref{eq:StabMat}, using $320\times320$ in-plane $\boldsymbol{k}$-points to find both the complete spatial structure of the leading superconducting order parameter $\boldsymbol{\Delta}^\eta= \boldsymbol{D}^{\eta,+}$, i.e.~both layer distribution and in-plane irrep, and its $T_{\rm c}$ as a function of the coupling strength $\Gamma_{r}^{\eta,+}$ in that particular channel, given by Table~\ref{Tab:MFInteraction}.

\subsection{Bulk ABC-stacked graphite}\label{sec:BulkSC}
\begin{figure*}[!t]
		\centering 
		\includegraphics[width=1\textwidth]{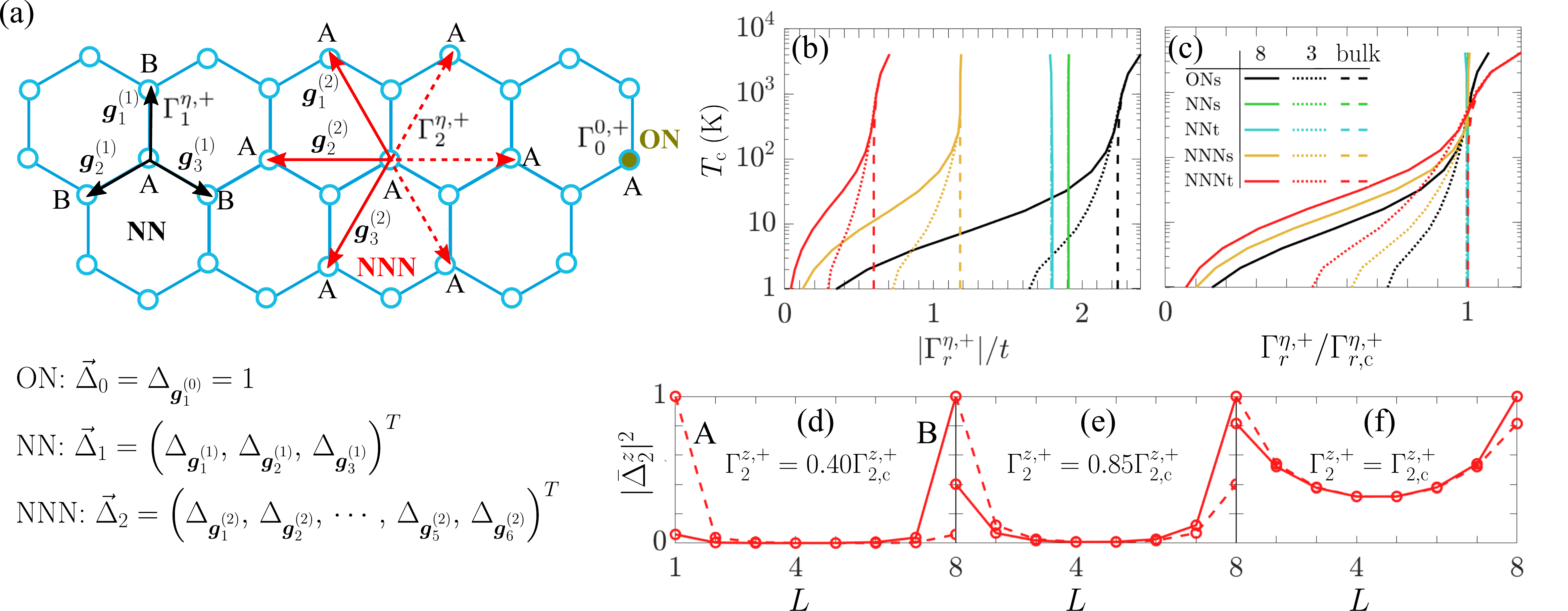}
		\caption{%
			(a) Illustration of the in-plane graphene bond vectors, $\boldsymbol{g}_{n}^{\left(r\right)}$, labeled by the bond number $n$, range $r=0,1,2$ for ON, NN, and NNN, respectively, as well as total coupling strengths, $\Gamma_r^{\eta,+}$. For NNN, dashed vectors represent $\boldsymbol{g}_{4,5,6}^{\left(2\right)}$. The superconducting order parameter vectors $\vec{\Delta}_r$ are  also written out in terms of their bond components for each $r$. 
			(b) Superconducting critical temperature, $T_{\rm c}$, as function of channel coupling strength $|\Gamma_r^{\eta,+}|$ for eight- (solid) and three-layer (dotted) ABC-MLG, as well as bulk ABC-stacked graphite (dashed), all at $\mu=0$. 
			(c) Same as (b) but with normalized coupling strengths, $\overline{\Gamma}_r^{+,\eta}=\Gamma_r^{+,\eta}/\Gamma_{r,\rm c}^{+,\eta}$. 
			(d) Normalized spin-triplet NNN superconducting order parameter, $|\bar{{\Delta}}_2^{z}|^2=|{\Delta}_2^{z}|^2/\max \left(|{\Delta}_2^{z}|^2\right)$, as a function of layer in the weak coupling regime on sublattices A (dashed) and B (solid). (e), (f) Same as (d) but for intermediate (e) and strong (f) coupling regimes, with coupling strengths given in figures.
		}
		\label{Fig2_SC}
	\end{figure*}

	We start by considering bulk ABC-stacked graphite. For the pristine situation with $\mu = 0$ we plot $T_{\rm c}$ for each channel in Fig.~\ref{Fig2_SC}(b) (dashed lines) as a function of the absolute value of that channel's coupling strength $\Gamma_{r}^{\eta,+}$. As seen, in each pairing channel there is no superconductivity below a channel dependent critical coupling strength, $\Gamma_{r,c}^{\eta,+}$. Estimates of the critical coupling strengths can be extracted from Fig.~\ref{Fig2_SC}(b) as:
		 $\{| \Gamma_{0, \rm c}^{0,+} |,\,
		  | \Gamma_{1, \rm c}^{0,+},\, |,\,
		  | \Gamma_{2, \rm c}^{0,+} |\} = 
		  \{2.3t,
		  1.9t,
		  1.2t\}$
		  for the spin-singlet channel and
		$  \{| \Gamma_{1, \rm c}^{z,+},\, |,\,
		  | \Gamma_{2, \rm c}^{z,+} |\} = 
		  \{1.8t,
		  0.6t\}$
		  for the spin-triplet channel. Assuming bulk ABC-stacked graphite to not be superconducting, these critical coupling strengths set a natural upper limit for the values we later investigate for the coupling strengths.
	It is the nodal electronic structure of ABC-graphite with its vanishing DOS at $\mu = 0$ that produces these critical coupling strengths, thus with no superconducting states possible for any interactions lower than these critical couplings. This is in contrast to the standard BCS relationship, which produces a finite (but possibly very small) $T_{\rm c}$ for any infinitesimal attraction. The situation in bulk ABC-stacked graphite is analogous to that of single layer graphene where the Dirac cones similarly form a nodal electronic structure. To emphasize this similarity, we find that the spin-singlet NN channel critical coupling strength $| \Gamma_{1, \rm c}^{0,+} | \approx 1.9t$ is almost identical to that of  single layer graphene reported before~\cite{Black-Schaffer07,BlackSchafferandHonerkamp2014}. 
	
	Since we find that ABC-stacked graphite has large critical coupling strengths in all superconducting channels, it means that pristine ABC-stacked graphite is by itself an unlikely host of superconductivity. To later contrast with finite ABC-MLG, we however still report the superconducting symmetries above the critical coupling strengths in ABC-stacked graphite and summarize the result in the bulk prevalence column of Table~\ref{Tab:Symmetry}. First, the spin-singlet ON channel has necessary $s$-wave spatial symmetry. Next, we find that the leading order in the spin-singlet NN channel is the two-fold degenerate $d$-wave solution belonging to the $E_{2g}$ representation. At $\mu=0$ this $d$-wave solution is fully degenerate with the rotationally symmetric $s_{\rm ext}$-wave solution ($A_{1g}$), whereas the $d$-wave solution becomes dominant for any finite $\mu$, just as in single-layer graphene~\cite{Black-Schaffer07,BlackSchafferandHonerkamp2014}. In contrast, in the spin-singlet NNN channel, the $d$-wave order  is always the dominating solution over the symmetric $s_{\rm ext}$-wave solution, with further increasing dominance at finite $\mu$. Both the ON and NNN pairing have equal order parameter magnitudes on both sublattices. For spin-triplet superconductivity we find that the two-fold degenerate $p$-wave solution of the $E_{1u}$ representation to be dominant in both the NN and NNN channels.
	
	\begin{table*}[th]
		\centering
		\renewcommand{\arraystretch}{2}
		\begin{tabular}{|p{0.8cm}| p{2.4cm} p{1.2cm} p{0.8cm}  p{0.95cm}  | p{0.6cm}  p{0.4cm}| p{2.4cm} p{1.9cm} p{1.2cm}  p{1.45cm}| p{0.6cm}  p{0.4cm}|}
			\hline
			\hline
			\multirow{3}{*}{$\quad r$} & \multicolumn{6}{c|}{spin-singlet, $\eta=0$} & \multicolumn{6}{c|}{spin-triplet, $\eta=\nu$} \\
			\cline{2-13}
			&  \multirow{2}{*}{$\qquad \vec{\Delta}_r$} & \multirow{2}{*}{func.} & \multirow{2}{*}{irrep.} & \multirow{2}{*}{char.} & \multicolumn{2}{c|}{Prevalence} &  \multirow{2}{*}{$\qquad \vec{\Delta}_r$} & \multirow{2}{*}{func.} & \multirow{2}{*}{irrep.} & \multirow{2}{*}{char.} & \multicolumn{2}{c|}{Prevalence} \\
			\cline{6-7}
			\cline{12-13}
			&  &  & & & bulk  & surf. &  & & & & bulk  & surf. \\
			\hline
			ON & $1$  &  $1$ & $A_{1g}$ & $s_{\rm on}$ & $\checkmark$  & $\checkmark$ & $-$  & $-$ & $-$ & $-$  & $-$ & $-$\\
			\hline
			\rule{0pt}{5ex}

			\multirow{2}{*}{NN} &  $\frac{\left(1,\,1,\, 1\right)^T}{\sqrt{3}}$  & $k_x^2+k_y^2$ & $A_{1g}$ & $s_{\rm ext}$   & $\circ$ & $-$ & $-$ &  $-$ &$-$ &  $-$ & $-$ & $-$\\ 
			\rule{0pt}{5ex}
			&$\frac{\left(2,\,{1}^\prime,\, {1}^\prime\right)}{\sqrt{6}}^T$ & $k_x^2-k_y^2$ &\multirow{2}{*}{$E_{2g}$}  &  $d_{x^2-y^2}$ &\multirow{2}{*}{$\checkmark$}  & \multirow{2}{*}{$-$}&  $\frac{\left(2,\,{1}^\prime,\, {1}^\prime\right)}{\sqrt{6}}^T$ &
			$k_y$ & \multirow{2}{*}{$E_{1u}$} &  $p_y$ &\multirow{2}{*}{$\checkmark$} & \multirow{2}{*}{$-$}\\
			& $\frac{\left(0,\,1,\, {1}^\prime\right)}{\sqrt{2}}^T$& $k_xk_y$ &  & $d_{xy}$  & & & $\frac{\left(0,\,1,\, {1}^\prime\right)}{\sqrt{2}}^T$ &
			$k_x$ & &  $p_{x}$  & &\\
			\hline
			\rule{0pt}{5ex}
			\multirow{2}{*}{NNN} & $\frac{\left(1,\,1,\, 1,\,1,\, 1,\, 1\right)}{\sqrt{6}}^T$ &
			$k_x^2+k_y^2$ &$A_{1g}$ &  $s_{\rm ext}$  & $\star$  & $\checkmark$ & $\frac{\left(1,\,{1}^\prime\, 1,\,{1}^\prime,\, 1,\, {1}^\prime\right)}{\sqrt{6}}^T$ 
			& $k_x\left(k_x^2-3k_y^2\right)$ &$B_{2u}$ &  $f_{x\left(x^2-3y^2\right)}$ &\multirow{2}{*}{$-$} & \multirow{2}{*}{$\checkmark$}\\
			\rule{0pt}{5ex}
			\multirow{2}{*}{} & $\frac{\left({1}^\prime,\,2,\, {1}^\prime,\, {1}^\prime,\,2,\, {1}^\prime\right)}{\sqrt{12}}^T$ &
			$k_x^2-k_y^2$ & \multirow{2}{*}{$E_{2g}$} &  $d_{x^2-y^2}$  & \multirow{2}{*}{$\checkmark$} & \multirow{2}{*}{$-$} &  $\frac{\left(1,\,2,\, 1,\, {1}^\prime,\,{2}^\prime,\, {1}^\prime\right)}{\sqrt{12}}^T$ &
			$k_x$ & \multirow{2}{*}{$E_{1u}$} &  $p_x$ &\multirow{2}{*}{$\checkmark$} & \multirow{2}{*}{$-$}\\
			& $\frac{\left({1}^\prime,\,0,\, 1,\,{1}^\prime,\,0,\, 1\right)}{2}^T$ &
			$k_xk_y$ & &  $d_{xy}$  & & & $\frac{\left({1}^\prime,\,0,\, 1,\,{1}^\prime,\,0,\, 1\right)}{2}^T$ &
			$k_y$ & &  $p_{y}$ &  &\\
			\hline
			\hline
		\end{tabular}\\
\caption{%
	Symmetry allowed superconducting states for pairing with range $r=0,1,2$ encoding on-site (ON), nearest-neighbor (NN), and next-nearest-neighbor (NNN) pairing, respectively, divided into spin-singlet $\eta=0$ and spin-triplet $\eta = \nu$, with $\nu = 1,2,3$. Results are based on $D_{6h}$ point group for the normal state Hamiltonian~\eqref{eq:Normal}. Columns represent range $r$, order parameter vector in real space $\vec{\Delta}_r$ [see Fig.~\ref{Fig2_SC}(a) and with notation ${h}^\prime=-h$], func.~meaning the expansion in reciprocal space of the basis function around the $\Gamma$ point up to cubic order, irrep.~meaning the irreducible representation, char.~meaning the nature of the orbital part of the superconducting order parameter, and their prevalence in bulk (graphite) and surf.~(ABC-MLG). For the prevalence, $\checkmark$ implies dominant state at $T_{\rm c}$, $\circ$ implies degenerate with dominant state at $\mu=0$ but  sub-dominant at $\mu\neq 0$ at $T_{\rm c}$, $\star$ implies sub-dominant at $T_{\rm c}$, and $-$ implies the state does not occur or its eigenvalue at $T_{\rm c}$ is less than $10 \%$ of that of the dominant state. Note that $E_{1u}$ and $E_{2g}$ irreps are two-fold degenerate. 
		}
		\label{Tab:Symmetry}
	\end{table*}

\subsection{ABC-MLG}\label{sec:SurfaceSC}
In contrast to bulk ABC-graphite, the topological flat bands on the surfaces of sufficiently thick stacks of ABC-MLG drastically increase the possibilities for superconductivity. In Fig.~\ref{Fig2_SC}(b) we also plot the critical temperatures $T_{\rm c}$ of the superconducting channels for three (dotted) and eight (solid) layer ABC-MLG as a function of the coupling strengths $\Gamma_{r}^{\eta,+}$. It is immediately evident that the onset of superconductivity, at any specified temperature, occurs at much lower coupling strengths compared to the bulk ABC-stacked graphite results (dashed) for almost all symmetries. In particular, for thick slabs, we find that the surface states initially generate a linear increase in $T_{\rm c}$ for weak coupling strengths, which is consistent with the $T_{\rm c}$ dependence earlier derived for flat band systems~\cite{Kopnin2013High,Lothman2017Universal}. For increasing coupling strengths, $T_{\rm c}$ instead starts grows exponentially with the channel coupling strength, leading to a more traditional BCS-like increase in $T_{\rm c}$. This is expected, as for larger strengths, also higher energy states besides the flat band states begin to contribute to the susceptibility and hence the superconducting $T_{\rm c}$. 
The significant enhancement of the superconducting susceptibilities channels becomes particularly clear when we in Fig.~\ref{Fig2_SC}(c) plot $T_{\rm c}$ as a function of the normalized coupling strengths $\bar{\Gamma}_{r}^{\eta,+} = \Gamma_{r}^{\eta,+} / \Gamma_{r, \rm c}^{\eta,+}$ for each channel. By doing this normalization with respect to the bulk ABC-stacked graphite critical couplings, we directly quantify the large effect of the surface states in ABC-MLG, including clearly illustrating how eight layers are much more prone to superconductivity than three layers. This strong surface enhancement of superconductivity is the first important result of this work.

The notable exception to the strong enhancement of superconductivity in ABC-MLG is for NN range pairing, for which there is instead no change compared to the bulk, independent on spin channel (blue and green curves). This is in sharp contrast to both doped graphene and magic-angle twisted bilayer graphene, which both have been shown to host enhanced NN pairing compared to the pristine graphene baseline, resulting in either chiral $\left(d+id\right)$ or nematic $d$-wave superconducting pairing \cite{BlackSchafferandHonerkamp2014,lothman2022nematic,Fischer2021}. We attribute the lack of any possibility for $d$-wave pairing in ABC-MLG to its strong, near complete, sublattice polarization of the surface flat bands (see Fig.~\ref{Fig1_LatticeNormal}). This sublattice polarization makes it essentially impossible to have NN, or any inter-sublattice, superconducting pairing.

\subsubsection{Layer dependence} \label{sec:Layerdep}	
We next quantify the layer and sublattice dependence of the superconducting states. In Figs.~\ref{Fig2_SC}(d) - \ref{Fig2_SC}(f) we plot the layer and sublattice resolved magnitude of the normalized superconducting order parameter going from weak [Fig.~\ref{Fig2_SC}(d)] and intermediate [Fig.~\ref{Fig2_SC}(e)] to strong coupling [Fig.~\ref{Fig2_SC}(f)] strengths, where we use spatially normalized magnitudes as $|\bar{\boldsymbol{\Delta}}^{z}|^2=|\boldsymbol{\Delta}^{z}|^2 / \max \left(|\boldsymbol{\Delta}^{z}|^2\right)$.
We here illustrate the behavior by plotting the NNN spin-triplet state, but find similar behavior also for the spin-singlet ON and NNN states. 
In the weak coupling regime, which we define as  $0 < |\Gamma_{0,2}^{\eta,+}| \ll |\Gamma_{0,2, \rm c}^{\eta,+}|$, i.e.,~very weak coupling relative to the critical coupling in bulk ABC-stacked graphite, we find superconductivity only localized to the outer surface layers. In fact, we find that the order parameter magnitude generally follows the surface state LDOS [see Fig.~\ref{Fig1_LatticeNormal}(f)].  Additionally, as seen in Fig.~\ref{Fig2_SC}(d), the order parameter has a large sublattice staggering within each layer that reflects the strong normal state sublattice polarization.  This means that both the LDOS and superconducting state live essentially exclusively on one sublattice on each surface. 
Increasing the coupling strength and the system enters an intermediate regime, $0 \ll |\Gamma_{0,2}^{\eta,+}| \lesssim 0.85 |\Gamma_{0,2 \rm c}^{\eta,+}|$, where the superconducting order parameters on the two outer surfaces begin to couple, even for an eight layer stack [see Fig.~\ref{Fig2_SC}(e)]. Simultaneously, the order parameter also begins to spread to both sublattices, even if it maintains a large sublattice staggering. Finally, in the strong coupling regime, the superconducting order parameter becomes finite throughout the entire ABC-MLG stack [see Fig.~\ref{Fig2_SC}(f)]. This happens because the coupling strength approaches that of the critical coupling of the bulk. As seen, for eight layers this occurs around $|\Gamma_{0,2}^{\eta,+}| \approx 0.85 |\Gamma_{0,2 \rm c}^{\eta,+}|$, while thicker stacks naturally requires a coupling strength closer to the bulk critical coupling for the full stack stack to become superconducting. 

\begin{figure}[!t]
		\includegraphics[width=0.4\textwidth]{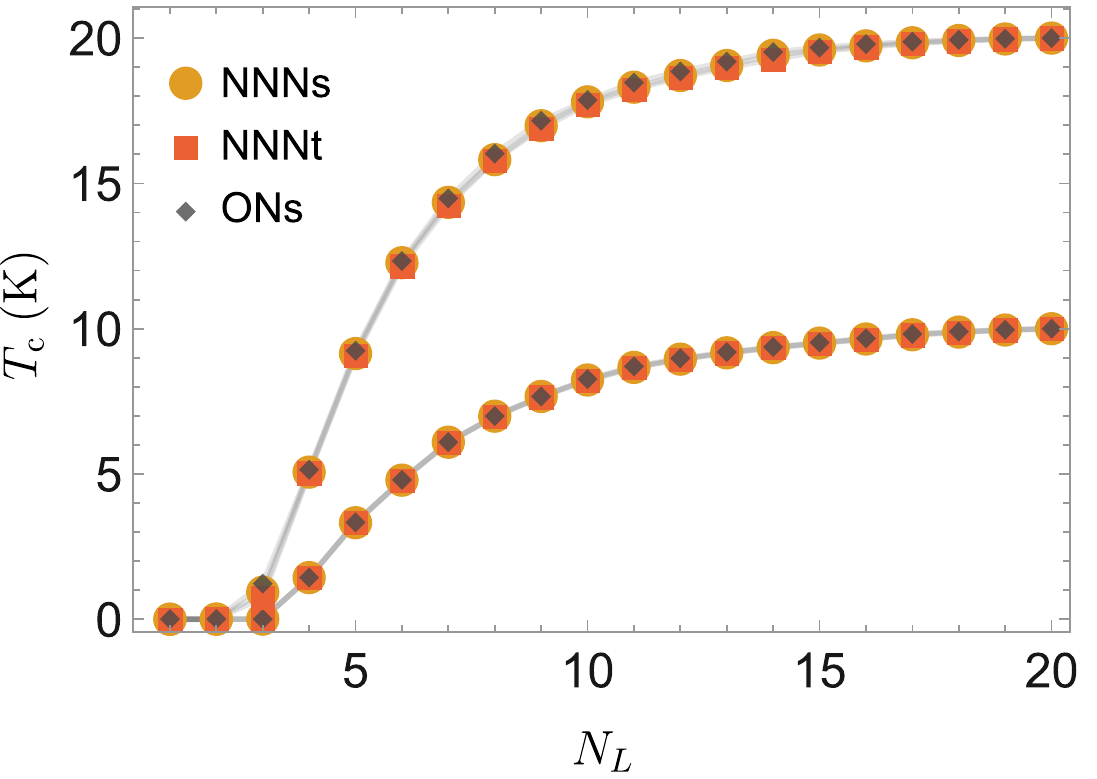}
		\caption{%
			Critical temperatures $T_{\rm c}$ of the three superconducting channels, ONs, NNNs, and NNNt, as a function of the number of layers in the ABC-MLG stack, where coupling strengths of each channel ${\Gamma}_r^{\eta,+}$ have been selected to produce $T_{\rm c}=20$K (top data set) and $T_{\rm c}=10$K (bottom data set) at $20$ layers: $\{| {\Gamma}_0^{0,+} |, |{\Gamma}_1^{0,+}|, |{\Gamma}_2^{z,+}| \} = \{1.17t,0.46t, 0.17t\}$ in the $20$K data set and $\{1.60t,0.28t,0.07t\}$ in the $10$K data set. 
		}
		\label{Fig3_TcVLayer}
	\end{figure}

The results in Figs.~\ref{Fig2_SC}(d) - \ref{Fig2_SC}(f) clearly illustrate the large effect of the surface states on superconductivity. ABC-MLG does however require a certain thickness before the topological surface flat band states are fully formed, see Fig.~\ref{Fig1_LatticeNormal}(g). Thus, $T_{\rm c}$ is also expected to grow with additional layers in ABC-MLG. To illustrate this stack thickness dependence, we plot the $T_{\rm c}$ of all three enhanced superconducting channels: spin-singlet ON (ONs), NNN (NNNs), and spin-triplet NNN (NNNt) pairing, all as a function of the number of layers in Fig.~\ref{Fig3_TcVLayer}. To isolate the layers dependence from other factors, we fix the coupling strengths ${\Gamma}_r^{\eta,+}$ of each channel to produce a fixed $T_{\rm c}$ of either $20$K or $10$K for a $20$ layer stack. 
As seen, all superconducting channels display the same $T_{\rm c}$-layer dependence. Thus, the topological flat band is equally important for enhancing all superconducting tendencies in ABC-MLG, independent on both the superconducting pairing range and spatial symmetry. This complements our first important result of this work, stating that the surface states enhance all superconducting states equally. Moreover, to achieve a higher $T_{\rm c}$, thicker slabs are generally required to reach saturation towards the maximum $T_{\rm c}$ for surface superconductivity. We also note in Fig.~\ref{Fig3_TcVLayer} that for ABC-MLG with three layers or less we find only a vanishingly small $T_{\rm c}$, consistent with essentially no flat band formed for such thin stacks. Thus, to find superconductivity in pristine (undoped and unbiased) ABC-MLG, we need to consider stacks substantially thicker than trilayer graphene.

\subsubsection{Superconducting symmetries} \label{sec:SCsym}		
\begin{figure}[!t]
			\centering
			\includegraphics[width=0.45\textwidth]{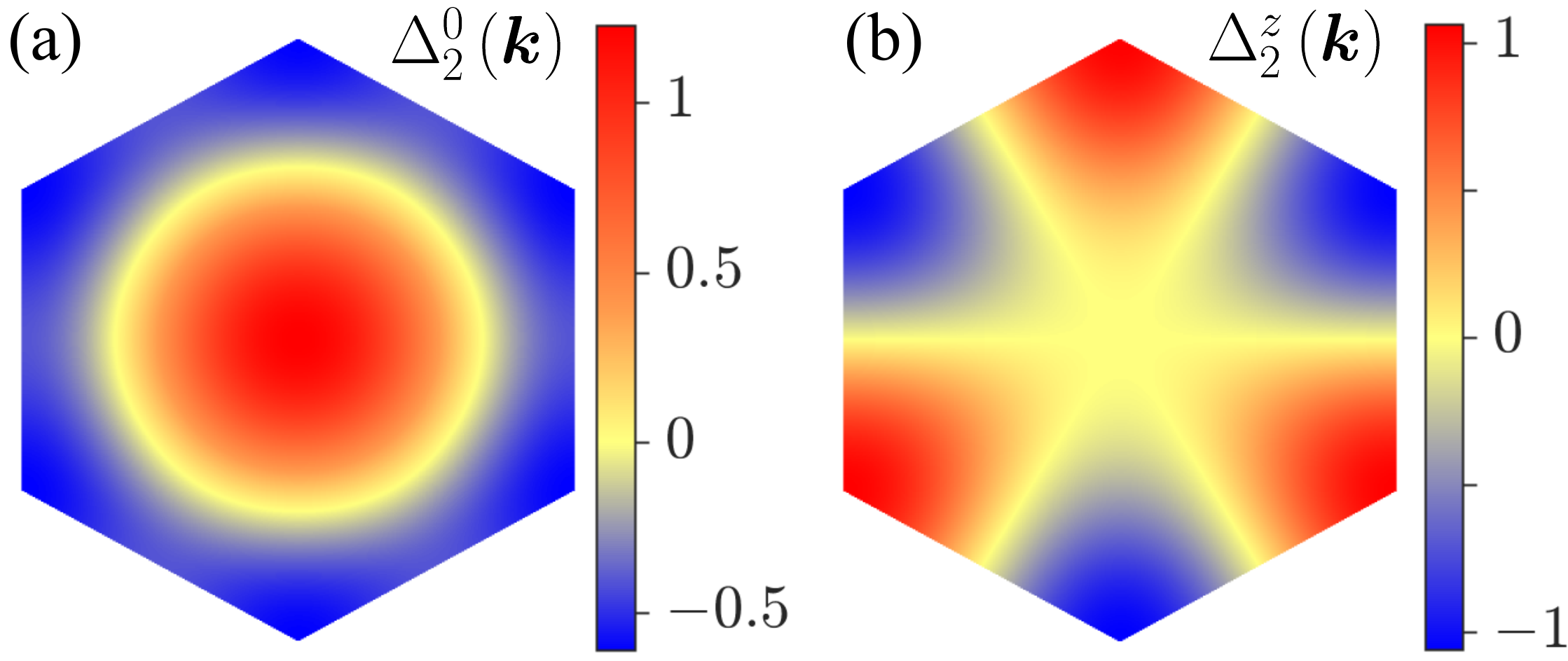}
			\caption{%
				Momentum dependence of the form factors, $\boldsymbol{\Delta}_2^{\eta}\left(\boldsymbol{k}\right) = \boldsymbol{D}^{\eta,+}\left(\boldsymbol{k}\right)$, in the first Brillouin zone for the NNN spin-singlet ($\eta=0$) $s_{\rm ext}$-wave state (a) and spin-triplet ($\eta=z$) $f_{x\left(x^2-3y^2\right)}$-wave state (b).
			}
			\label{Fig4_OPSym}
		\end{figure}

Having shown the strongly enhanced critical temperature in ABC-MLG, we next look at the order symmetry in each of the achievable surface superconducting channels, i.e.,~ON and NNN pairing. In the spin-singlet channel, we find that the surface flat bands support either on-site $s_{\rm on}$-wave or NNN extended $s_{\rm ext}$-wave symmetry superconductivity. Both of these states are isotropic, preserving the full symmetry of the normal state Hamiltonian, Eq.~\eqref{eq:Normal}, thus belonging to the $A_{1g}$ irrep of the $D_{6h}$ symmetry group. They also both have a fully gapped energy spectrum. For the $s_{\rm on}$-wave state the gap is constant. In contrast, for the $s_{\rm ext}$-wave state, we display the reciprocal space form factor $\boldsymbol{\Delta}_2^{0}\left(\boldsymbol{k}\right)$ in Fig.~\ref{Fig4_OPSym}(a), illustrating a nodal line encircling the $\Gamma$ point. Yet, close to charge neutrality, the Fermi surface is far away from this nodal line and the spectrum is thus fully gapped.  

		\begin{figure*}[!t]
			\centering
			\includegraphics[width=0.95\textwidth]{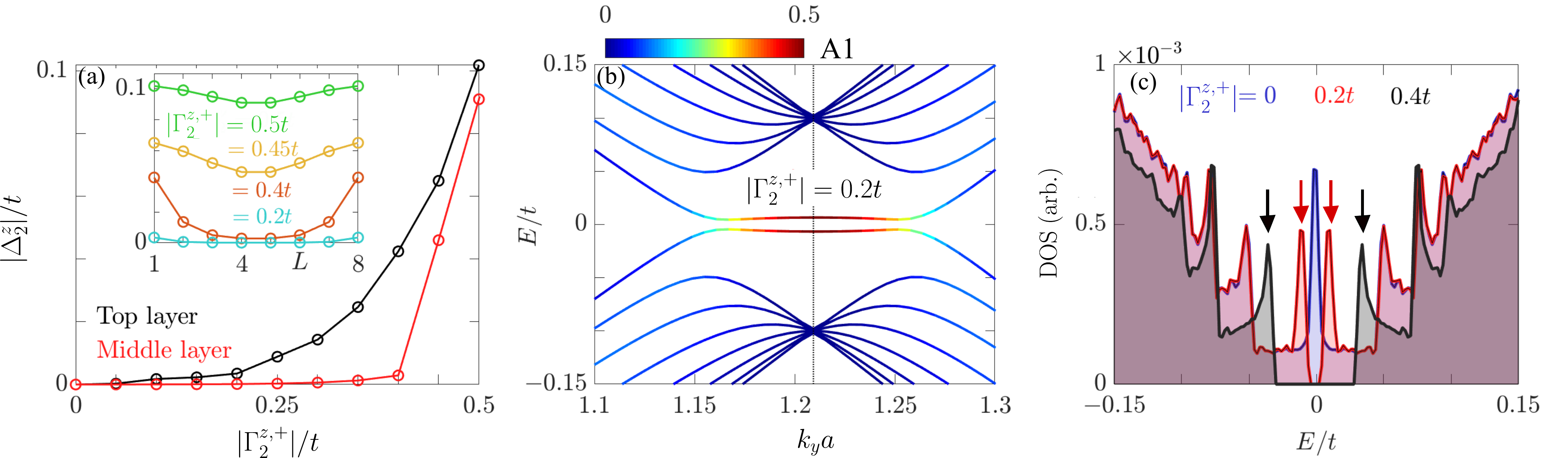}
			\caption{%
				Self-consistently calculated features of the spin-triplet NNN $f$-wave superconducting state at $T=0$ and $\mu=0$ for eight layer ABC-MLG using the BdG formalism.
					{(a)} Magnitude of order parameter $|\Delta_2^{z}|$ as a function of $|\Gamma_2^{z,+}|$. Inset: Order parameter layer profile for several $|\Gamma_2^{z,+}|$. 
					{(b)} BdG energy spectrum for $|\Gamma_2^{z,+}|=0.2t$. Color code shows probability amplitude on the A1-site (same result for B8-site).
					{(c)} Electronic DOS of the normal state (blue), and the $f$-wave superconducting state with $|\Gamma_2^{z,+}| = 0.2t, 0.4t$. Arrows mark the superconducting coherence peaks. 
								}
			\label{Fig5_fwave}
		\end{figure*}
		
In the spin-triplet channel, here only for the NNN pairing range, we find that the dominant order is an unconventional $f_{x\left(x^2-3y^2\right)}$-wave symmetry superconducting state belonging to the $B_{2u}$ irrep. When measured against the normalized coupling strength $\bar{\Gamma}_{r}^{\eta,+}$ in  Fig.~\ref{Fig2_SC}(c), we find that this $f$-wave state is the most enhanced superconducting channel, giving the highest $T_{\rm c}$ for all normalized coupling strengths.
 One explanation for this significant enhancement of the $f$-wave state is that it is also a fully gapped superconducting state, for all filling factors around the charge neutral level. The full gap of the $f$-wave state can be traced back to the fact that the three nodal lines of the pairing potential, shown in Fig.~\ref{Fig4_OPSym}(b) pass through the $M$-points of the Brillouin zone, which therefore do not intersect with the flat bands residing around the $K$ and $K'$-points. This full gap results in a large superconducting condensation energy, which in turn makes the $f$-wave energetically favorable. To summarize these results, we tabulate the symmetries of the leading superconducting states for ABC-MLG in each channel in Table~\ref{Tab:Symmetry} in the surface prevalence column. We note the dominant surface superconducting symmetries are essentially completely opposite to results in the bulk. While the bulk will host NN or NNN spin-singlet $d$-wave or spin-triplet $p$-wave states, the surface will only host NNN spin-singlet $s_{\rm ext}$ or spin-triplet $f$-wave states. In fact, the only dominant superconducting symmetries appearing for both bulk and surface is the on-site spin-singlet $s$-wave state, which is to be expected since it does not have any real- or reciprocal space structure. This distinct difference between surface and bulk superconductivity is surprising and a second major result of this work.

With the $f$-wave state being the most enhanced surface superconducting state, we next investigate it further by additionally solving self-consistently for spin-triplet NNN pairing at zero temperature $T=0$ for an eight layer stack. For this we use Eq.~\eqref{eq:MFSelfConSC} within the Bogoliubov-de-Gennes (BdG) formalism~\cite{DeGennes}, which then gives an excellent complement to the LGE results in Fig.~\ref{Fig2_SC} representing the solution at $T_{\rm c}$. In Fig.~\ref{Fig5_fwave}(a) we show the $T=0$ self-consistent result for the amplitude $|\boldsymbol{\Delta}_2^{z}|$ on the surface (black) and middle (red) layers as a function of coupling strength $\Gamma_2^{z,+}$ with the full-layer profile in the inset, which fully corroborates the results at $T_{\rm c}$
Moreover, tracing the order parameter as a function of $\Gamma_2^{z,+}$, we find that it increases monotonically everywhere, but for $0<|\Gamma_2^{z,+}\lesssim| 0.4t$ the order parameter is only localized to the surfaces, illustrating the strongly enhanced surface superconductivity. (The slight difference between self-consistent upper coupling bound $|\Gamma_2^{z,+}|\approx 0.42t$ and that of the LGE, $|\Gamma_2^{z,+}| \approx 0.85|\Gamma_{2,\rm c}^{z,+}| = 0.51t$ obtained in Section~\ref{sec:Layerdep}, stems from the inclusion of all higher order terms in self-consistency calculation while LGE in contrast contains only first order terms.)
	The $f$-wave superconducting state also opens a superconducting gap on the surfaces, splitting the surface flat bands, as illustrated in Fig.~\ref{Fig5_fwave}(b), where we plot the BdG quasi-particle spectrum $E \left(\boldsymbol{k}\right)$. The color code shows the probability density of the bands on site A1 (same result for B8, thus summing to 1), which show that the gapped flat band states fully reside on the outer layers with essentially full sublattice polarization. Finally, in Fig.~\ref{Fig5_fwave}(c), we show the DOS as a function of energy, $E$, for the normal state (blue), and the $f$-wave superconducting state at $|\Gamma_2^{z,+}| = 0.2t, 0.4t$ (red, black). While the normal state DOS is gapless with large surface DOS at $E=0$, the superconducting DOS has a finite spectral gap which increases with $\Gamma_2^{z,+}$.

	\subsubsection{Phase diagram} \label{sec:SurfaceSCInt}
		In the previous subsections, we showed that the topological flat band surface states of ABC-MLG generate an enhanced susceptibility in the spin-singlet on-site $s$-wave and NNN extended $s$-wave channels, as well as in the unconventional spin-triplet NNN $f$-wave superconducting channel. So far, we have however only considered each pairing-range $r$ and spin-channel separately and also regarded the channel coupling strengths $\Gamma_{r}^{\eta,+}$ as independent system parameters. While this is appropriate for a general analysis of the available possibilities, we note that the different ordering channels are generally not independent. Rather, the coupling strengths $\Gamma_{r}^{\eta,+}$ are linear combinations of the effective interactions, $U_{r}$ and $J_{r}$, see Table~\ref{Tab:MFInteraction}. Next, we therefore analyze all pairing channels jointly to show their possible couplings and dependence on the effective interactions $J_{r}$ and $U_r$ in order to extract the phase diagram. 
\begin{figure*}[!t]
	\centering
	\includegraphics[width=1.00\textwidth]{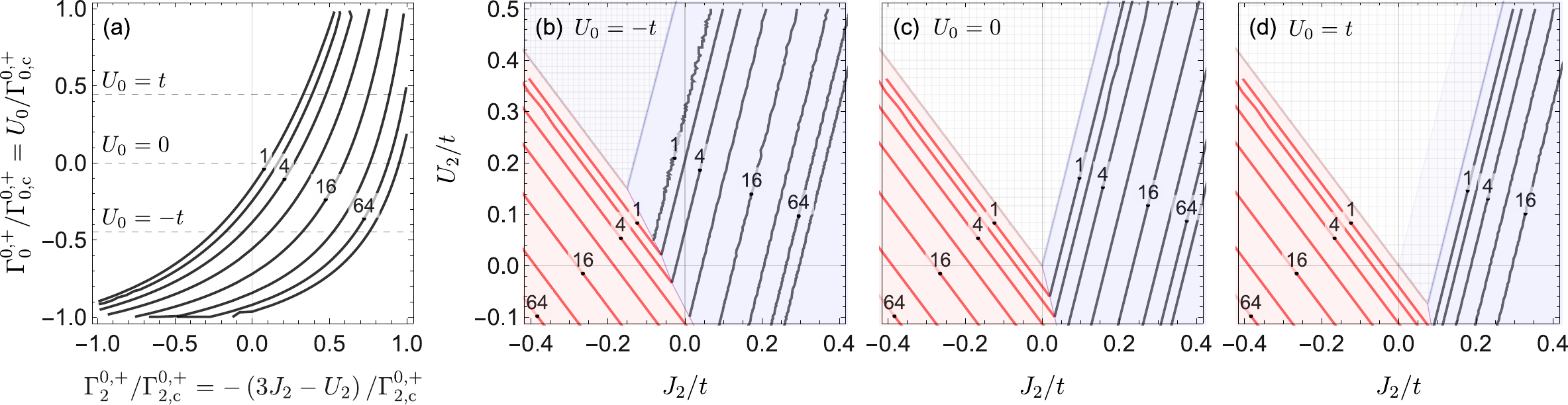}
	\caption{
		(a) Contour plot for $T_{\rm c} =1, 4, \ldots, 64$ K as function of the two singlet-pairing channel coupling strengths $\Gamma_{0}^{0,+}$ and $\Gamma_{2}^{0,+}$. Dashed lines marks cuts at specific $U_0$.
		(b) - (d) Superconducting phase diagram and contour plot of $T_{\rm c}$ for spin-singlet (blue) and spin-triplet (red) pairing as a function of $J_2/t$ and $U_2/t$ for (b) $U_0=-1t$, (c) $U_0=0$, (d) $U_0=1t$;  Hatched-marked regions indicates absence of superconductivity, either when forbidden as in (c) and parts of (d), or when $T_{\rm c} < 10^{-2}$ K as in (b) and (d) then shown as blue shaded hatched regions. All phase diagrams are for undoped, $\mu =0$, eight layer ABC-MLG.				}
	\label{fig:PD}
\end{figure*}

		When considering the pairing channels together, they are also allowed to couple in the LGE, especially mixing different ranges. We however find that the NN pairing ranges do not couple sufficiently to the other pairing ranges to affect the analysis for interaction weaker than their corresponding bulk critical coupling strength, which we use as an upper limit of parameters that we consider. We therefore exclude the NN pairing channels from the rest of our analysis. The NNN spin-triplet range is therefore the only spin-triplet channel left in our analysis, and with spin-channels not mixing with each other, we can treat the NNN spin-triplet channel as uncoupled. In contrast, the two spin-singlet $s$-wave pairing states, the ON $s_{\rm on}$-wave and NNN $s_{\rm ext}$-wave states, couple even to first order in the LGE in a joint analysis, since both states belong to the same $A_{1g}$ symmetry, see Table~\ref{Tab:Symmetry}. 
	
		In Fig.~\ref{fig:PD} we analyze the phase diagram for the relevant combinations of $U_{r}$ and $J_{r}$. We start in Fig.~\ref{fig:PD}(a) by plotting the joint $T_{\rm c}$ produced by the two spin-singlet $s$-wave states as a function of the two normalized coupling strengths $\bar{\Gamma}_{0}^{0,+}$ and $\bar{\Gamma}_{2}^{0,+}$. We here use the normalized strength to be able to cut off the phase diagram at $\bar{\Gamma}_{0,2}^{0,+}=1$ as that corresponds to the bulk critical coupling strength. We find that the spin-singlet $s$-wave state is favored by either an effective on-site attraction, $U_0 < 0$, a NNN attraction $U_2 < 0$, or an antiferromagnetic NNN interaction $J_2>0$. These values all produce the highest $T_{\rm c}$, appearing in the lower right corner of the phase diagram. Moreover, Fig.~\ref{fig:PD}(a) also shows that a net attraction for one of the pairing ranges, i.e.~$U_0$ or $U_2$, can overcome a net repulsion in the other range, i.e.~$U_2$ or $U_0$, to still produce a finite $T_{\rm c}$.
		
		To draw a complete phase diagram, the spin-singlet $T_{\rm c}$ of Fig.~\ref{fig:PD}(a) has to be compared with the $T_{\rm c}$ of the spin-triplet channel. In Figs.~\ref{fig:PD}(b-d), we overlay the $T_{\rm c}$ contours of both the spin-singlet $s$-wave and the spin-triplet $f$-wave states for three different fixed values of the on-site interaction $U_0$ and as a function of the remaining interactions $J_2$ and $U_2$. For $U_0=0$, shown in Fig.~\ref{fig:PD}(c), the dominant superconducting order is directly determined by the sign of $J_2$. A ferromagnetic NNN interaction, $J_2<0$, favors spin-triplet $f$-wave pairing, while an antiferromagnetic, $J_2>0$, favors the spin-singlet $s$-wave pairing. A repulsive $U_2>0$ reduces the total coupling constant of both pairing channels, which produces a wedge-shaped region with boundaries $\Gamma_{2}^{z,+}=0$ and $\Gamma_{2}^{0,+}=0$ (hatched-marked region) inside of which both the spin-singlet and -triplet channels have a net repulsion, leading to a phase space region with no pairing. 
		
		Adding a repulsive on-site term $U_0=t$ in Fig.~\ref{fig:PD}(d), we find that the spin-singlet $s$-wave state is suppressed. The $T_{\rm c}$ of the $f$-wave is, however, completely unaffected by $U_0$, since by symmetry the spin-triplet $f$-wave pair amplitude vanishes on-site and thus completely avoids any on-site repulsion. The result is that the spin-singlet state is suppressed and the non-superconducting wedge-shaped region grows at the expense of the spin-singlet state region. We note that, technically, this wedge shape region actually splits in two. Inside the original wedge, both channels have a net repulsion, devoid of all pairing. In contrast, between the old and new wedge-boundary (light blue hatched region), one channel is still technically attractive, which in principle gives a finite $T_{\rm c}$, but due to the strong on-site repulsion, this $T_{\rm c}$ is vanishingly small. To be able to draw a realistic phase diagram in Fig.~\ref{fig:PD}(d), we have included a cut-off of $T_{\rm c} \le 10^{-2}$ K to define this second boundary.
		
		Finally, if we instead add an attractive $U_0=-t$, as in Fig.~\ref{fig:PD}(b), then the $T_{\rm c}$ of the spin-singlet channel increases. Such attractive $U_0$ can be viewed as the effective interaction arising from putative electron-phonon coupling in ABC-MLG graphite and is, as such, still relevant to consider. Since the ON spin-singlet channel is now attractive everywhere in the phase diagram, the entire prior non-superconducting wedge is now transformed into a region with finite but vanishingly small $T_{\rm c}$, indicated by the light blue region in Fig.~\ref{fig:PD}(c). Moreover, the attractive on-site interaction further shifts the wedge shaped region which now also encroaches on the $f$-wave region which consequently shrinks. 
		Together, Figs.~\ref{fig:PD}(a) - \ref{fig:PD}(d) give a complete picture of the superconducting phase diagram for all interactions out to the NNN coupling range, also including the effect of electron-phonon attraction. We find that the spin-singlet $s$ wave (jointly ON and NNN pairing) and the spin-triplet $f$ wave  (NNN pairing) are both present even for repulsive Coulomb interactions $U_0$ and $U_2$ and that the sign of the Heisenberg interaction $J_2$ determines which of these states is materialized. This constitutes a third important result in this work.

	\section{Interaction Estimates} \label{sec:Interactions}	
		In the previous section, we established the generic superconducting phase diagram of ABC-MLG. We on purpose kept the discussion general, without assuming any specific interaction strengths in order to show the full range of possibilities, finding that electronic interactions in ABC-MLG can support both spin-singlet isotropic $s$-wave and unconventional spin-triplet $f$-wave superconductivity depending on the interactions. Here we aim to give an estimate of the actual interaction strengths in ABC-MLG, in order to finally discern the most likely superconducting state in ABC-MLG.
				
		We note first that the effective density-density Coulomb interactions have already been calculated from \emph{ab initio} using the constrained random phase approximation (cRPA) \cite{Wehling2011strength}, which find a strong on-site interaction in both single-layer free-standing graphene ($U_0=9.3$~eV) and in graphite ($U_0=8.1$~eV). Similar estimates were later also found in a DFT$+U+V$ framework, for both graphene and graphite ($U_0=7.6$~eV) \cite{TancogneDejean2020}. These estimates produce a dielectric constant in graphene close to the experimentally observed value \cite{Wehling2011strength,TancogneDejean2020} and are also comparable to values $\left(U_0\sim10\right)$ eV used in quantum chemistry models of hydrocarbon and the Pariser-Parr-Pople model \cite{Bursill1998,Verges2010}. 
			
		The very strong on-site repulsive Coulomb interactions found in graphene and graphite significantly impact the range of likely ordering also in ABC-MLG. Foremost, it suggests that the effective Coulomb interactions is repulsive and therefore significantly detrimental to the isotropic spin-singlet pairing in ABC-MLG, as found in Fig.~\ref{fig:PD}. Conventionally, the main pairing mechanism for the isotropic spin-singlet pairing channel would be an effective electron-phonon coupling, which here is effectively modeled as an attractive $U_0<0$. We note, however that in single-layer graphene the strong out-of-plane vibrations do not couple to the out-of-plane $p_z$ orbitals of the low-energy electronic structure to first order \cite{Pietronero1980,Thingstad2020}. There is likewise no coupling between the out-of-plane $p_z$ orbitals and in-plane vibrational modes due to symmetry \cite{Falkovsky2008}. Consequently, the effective dimensionless electron-phonon pairing is small in graphene and any Coulomb repulsion has further been shown to dramatically suppress both the pairing and the critical temperature of phonon-driven superconductivity in single layer graphene \cite{Thingstad2020}. Despite this lack of electron-phonon coupling in graphene, we nonetheless include the possibility of an effective on-site attraction $U_0 < 0$ in Fig.~\ref{fig:PD}(b), since there exist additional phonon modes in ABC-MLG due to the additional layers. For instance, in-plane acoustic longitudinal phonon modes have been suggested as a mechanism for superconductivity in both trilayer and tetralayer ABC graphene \cite{chou2021Acoustic,Chou2022C}. This might thus open for phonon-driven $s$-wave superconductivity also in ABC-MLG, although any such electron-phonon coupling would have to first overcome the strong repulsive Coulomb interaction before generating superconducting pairing. 
		
		Beyond suppressing any putative $s$-wave state, the strong on-site repulsive Coulomb interactions also have large implications for the exchange interactions in ABC-MLG. The direct Coulomb exchange interaction produces a ferromagnetic exchange between orbital sites \cite{Wang2013Flat,SongJin2021}. On the other hand, the strong coupling limit of the on-site Coulomb repulsion $U_0$ is captured by the $t$-$J$ model \cite{Lee2006}, featuring an antiferromagnetic super-exchange on NN bonds. Thus, while the NN direct exchange in graphene has been calculated to be ferromagnetic with $J_{\rm NN, FM} \approx 1.6$~eV, using \emph{ab initio} cRPA, this contribution is dwarfed by the antiferromagnetic NN super-exchanged $J_{\rm  NN, AF} = 4  t^2 / U \approx 3.5$~eV \cite{Hadipour2015}. We can thus here estimate the net effective Heisenberg interaction $J_1 = J_{\rm NN, AF} - J_{\rm NN, FM} \approx 1.9$~eV to be antiferromagnetic on NN bonds \cite{Hadipour2015}. 	

		In contrast, the NNN bonds super-exchange is suppressed by the smaller NNN hopping amplitude $t_2 \sim 0.1t$ \cite{Reich2002,Neto2009}, which can thus be expected to be 100 times smaller than the NN super-exchange. The NNN direct ferromagnetic Coulomb exchange remains finite but is also similarly small \cite{SongJin2021}. The question here remains as to their balance. From a basic point of view, the bipartite lattice structure of graphene has been shown to dictate the sign of the bare spin susceptibility with an alternating sign between the two sublattices \cite{Saremi2007}. Due to the role  the bare spin susceptibility plays in generating effective interactions, e.g.,~in the random phase and fluctuation-exchange approximation \cite{Scalapino1995,Bickers1989,Roemer2020}, the effective exchange is expected to inherit this alternating sign structure of the bare susceptibility. Consequently, we expect the NN effective exchange $J_1$ to be antiferromagnetic $J_1>0$, just as we concluded in the previous paragraph, while we expect the NNN term to be ferromagnetic, $J_2<0$. Corroborating this basic expectation, the effective interaction generated in the functional renormalization group (fRG) flow has been shown to have an alternating sign structure in single layer, bilayer, and in both ABC- and ABA-stacked trilayer graphene \cite{Honerkamp2008,Scherer2012,Scherer2012a}, when starting either from a single repulsive on-site Hubbard term or a full set of longer range \emph{ab initio} cRPA Coulomb repulsion terms~\cite{Wehling2011strength}. Additional results for trilayer ABC-MLG also point to the projected Coulomb interactions being overall ferromagnetic for NNN \cite{Wang2013Flat}. Additionally, a ferromagnetic $J_2$ would also help explain the ferrimagnetic state found in \emph{ab initio} spin-polarized DFT calculations of ABC-MLG~\cite{Otani2010Intrinsic,Xiao2011Density,Cuong2012Magnetic,Pamuk2017}, as we discuss further in Sec.~\ref{sec:Magnetism}. Taken together, we expect antiferromagnetic $J_1>0$, while a ferromagnetic $J_2<0$.
							
		Notably, an antiferromagnetic $J_1$ has been proposed as a mechanism for generating chiral $d+id$ superconductivity in doped graphene \cite{BlackSchafferandHonerkamp2014}, as well as chiral or nematic $d$-wave superconductivity in magic-angle twisted bilayer graphene \cite{lothman2022nematic,Fischer2021}. However, our results  in Fig.~\ref{Fig2_SC} show that a similar mechanism cannot exist for ABC-MLG, since no NN pairing is enhanced by the surface state due to its sublattice polarization. This effectively excludes NN or $J_1$ driven pairing in ABC-MLG. Left are then only ON and NNN pairing. With the expectation that $J_2<0$ and ferromagnetic, we conclude from the phase diagrams in Fig.~\ref{fig:PD}, that the most likely superconducting state in ABC-MLG is the spin-triplet $f$-wave state. The only other realistic possibility is a phonon-driven spin-singlet $s$-wave, but due to weak electron-phonon coupling and strong repulsive Coulomb interactions in graphene, we expect that to be a less likely option. This argument for spin-triplet $f$-wave pairing in ABC-MLG is the fourth important result of this work.

\section{Magnetism}~\label{sec:Magnetism}			
	In the previous two sections we have investigated the superconducting channels in ABC-MLG and given an estimate of the interaction strengths, respectively, together strongly pointing to a spin-triplet $f$-wave state. However, the divergent DOS of the topological flat bands also enhance orders in the particle-hole channels, capturing charge and magnetic orders \cite{Lothman2017Universal}. Intense order competitions are therefore likely.  In this section, we therefore start by addressing all possible ordering in the particle-hole channel in ABC-MLG. 
		
	Suggestively, \emph{ab initio} spin-polarized DFT calculations of ABC-MLG have already found a magnetically ordered electronic structure~\cite{Otani2010Intrinsic,Xiao2011Density,Cuong2012Magnetic,Pamuk2017,Henck2018Flat}, and a Hubbard model, producing an intrinsic magnetic gap, has been invoked to explain transport experiments of ABC-trilayer graphene \cite{Xu2012}. Further signatures of magnetic ordering in ABC-MLG have also been found in multiple experiments~\cite{Lee2014competition,Myhro2018Large,Yanmeng2020Electronic,hagymasi2022signature}. These results all indicate a strong tendency for magnetic ordering, although we at the same time note that spin-polarized DFT calculations are unable to capture superconducting orders and hence these results cannot resolve any competition between superconductivity and magnetism, which is instead our goal of the next section.

	To be able to compare superconductivity with magnetism, we need to treat them on an equal footing. We therefore here first need to solve our tight-binding self-consistent mean-field equations for the particle-hole channels. Ordering in the particle-hole channels originates from the Hartree, or direct, and Fock, or exchange, terms in the mean-field decomposition of Eq.~\eqref{eq:MFInt}. Accounting for all these direct and exchange terms out to the NNN range, tabulated for each channel in Table~\ref{Tab:MFInteraction}, we solve the self-consistency equations~\eqref{eq:MFSelfConPHDir}~and~\eqref{eq:MFSelfConPHEx}. We find that the magnetic orders (spin-triplet in Table~\ref{Tab:MFInteraction}) generally dominate over the charge orders (spin-singlet in Table~\ref{Tab:MFInteraction}), in agreement with previous results \cite{Otani2010Intrinsic, Xiao2011Density, Cuong2012Magnetic,Pamuk2017}. Additionally, we find that the Fock exchange bond magnetization orders $\tilde{\boldsymbol{m}}_{ij}^\nu$ are negligible, compared to the orbital magnetization from the direct terms, except the on-site term where we have $\tilde{\boldsymbol{m}}_{ii}^\nu \equiv \boldsymbol{m}_{ii}^\nu$. Based on these observations, we are able to greatly simplify our analysis and hereafter focus exclusively on the orbital magnetization with only the channel coupling strengths $\Gamma_{r=0}^{z,-} +\tilde{\Gamma}_{r=0}^{z,-} = U_0/2$, $\Gamma_{1}^{z,-} = J_1 $, and $\Gamma_{2}^{z,-} = -J_{2}$ being non-zero. Here we have used the fact that for $r=0$, the direct and exchange terms contribute additively, but for simplicity we will ignore the factor of $\tfrac{1}{2}$ henceforth. We have also already evoked the expected signs of the interactions in ABC-MLG, i.e.,~a negative sign has been included in $U_0$ and $J_2$ with regards to Table~\ref{Tab:MFInteraction}, as established in the previous section: ON repulsion $U_0 > 0$, antiferromagnetic NN Heisenberg term $J_1 > 0$, and ferromagnetic NNN Heisenberg term $J_2 < 0$.	
	
		\begin{figure}[t]
			\includegraphics[width=0.42\textwidth]{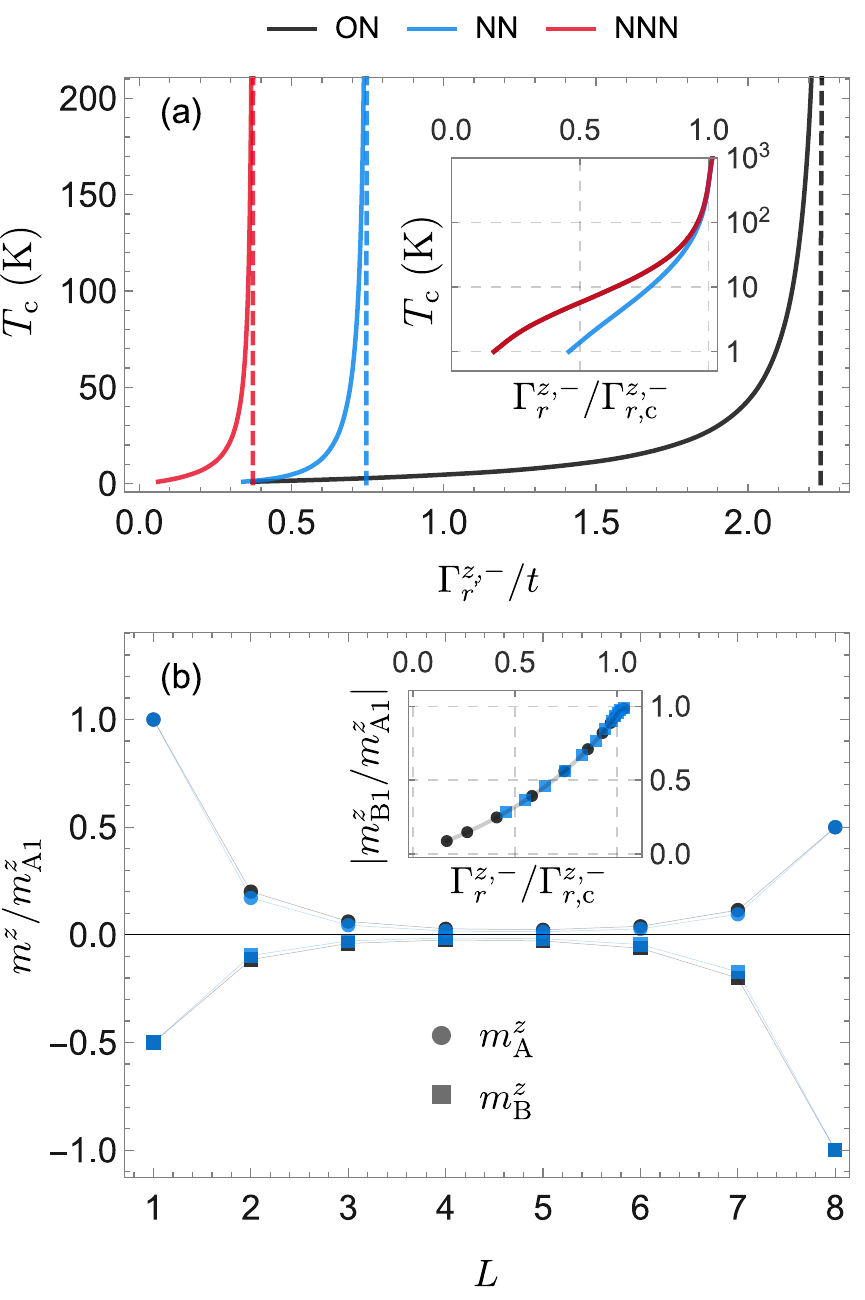}
			\caption{%
				Magnetic ordering in ABC-MLG.
				({a}) %
					Magnetic critical temperature $T_{\rm c,mag}$ from ON (black), NN (blue), and NNN (red) interactions as a function of corresponding coupling strengths: 
						$\Gamma_{r=0}^{z,-} = U_0 $,
						$\Gamma_{1}^{z,-} = J_1 $,
						and 
						$\Gamma_{2}^{z,-} = -J_2$, for eight layers (solid) and bulk ABC-stacked graphite (dashed)
					Inset: Same data but on a log-scale and for normalized couplings strengths $\bar{\Gamma}_{r}^{z,-} = \Gamma_{r}^{z,-} / \Gamma_{r,c}^{z,-}$, with $\Gamma_{r,c}^{z,-}$ defined as the critical coupling for ordering in bulk ABC-stacked graphite. 
				({b}) %
					Layer profile of the sublattice resolved magnetic moments for $\bar{\Gamma}_{r}^{z,-}=0.5$, normalized by the A-sublattice surface magnetic moment $m_{A1}$. Inset: Ratio of surface sublattice magnetic moments $|m_{B1}/m_{A1}|$ as a function of $\bar{\Gamma}_{r}^{z,-}$. Note, ON and NNN results are identical when normalized. 
			}
			\label{Fig:Magnetism}
		\end{figure}
				
To proceed, we first analyze the magnetic ordering that arises by considering each individual interaction range separately, just as we did earlier for superconductivity. In Fig.~\ref{Fig:Magnetism}(a) we show the magnetic critical temperature $T_{\rm c}$ in both bulk (dashed) and eight layer ABC-MLG (solid), produced by either ON, NN, or NNN interactions. In the case of bulk ABC-stacked graphite, there exist a critical coupling for each interactions below which there is no ordering, very similar to the case of superconductivity in Fig.~\ref{Fig2_SC}. Numerically, we find that the bulk critical coupling strengths are: $\Gamma_{0,c}^{z,-} = U^{-}_{0,c} = 2.23t$, $\Gamma_{1,c}^{z,-} = J^{-}_{1,c} = 0.75t$, and $\Gamma_{2,c}^{z,-} = -J^{-}_{2,c} = 0.37t$. Since bulk ABC-stacked graphite is not magnetic, we will henceforth assume that $\Gamma_r^{z,-}$ are all below these values.
In contrast, for eight layer ABC-MLG, we find that the flat band surface states enhance the susceptibilities of all interaction ranges, which now all display ordering far below their corresponding bulk critical coupling strengths. We note that this is different from the case of superconductivity in Fig.~\ref{Fig2_SC}, where NN pairing was not  at all affected by the surface states. However, as we scale the couplings strengths by their corresponding bulk critical coupling, i.e.~$\bar{\Gamma}_{r}^{z,-} = \Gamma_{r}^{z,-} / \Gamma_{r,c}^{z,-}$, and use this as the normalized interaction strength to plot $T_{\rm c}$ the inset of Fig.~\ref{Fig:Magnetism}(a), we find that NN range magnetism produces lower $T_{\rm c}$ compared to both the ON and NNN range. Just as in the superconducting case, we attribute this relative suppression of NN magnetism to the large sublattice polarization on the surfaces of ABC-MLG. Additionally, we note that ON and NNN magnetism both map on to equivalent mean-field models for periodic systems, despite them originating from very different interactions, but with distinct couplings strengths. 
	
We next consider the type of magnetic order generated by the ON, NN, and NNN interactions. In Fig.\ref{Fig:Magnetism}(b) we plot the sublattice resolved layer profile of the magnetic moments for  the different interactions. Note that we only show the moments for the ON and NN ranges, since the ON and NNN produce the same moments for the fixed normalized coupling strength $\bar{\Gamma}_{r}^{z,-}=0.5$. Overall, we find very similar magnetic state for the ON, NN, and NNN ranges: large magnetic moments are localized on the outermost surfaces and the magnetization profile generally follows the LDOS of the topological flat bands in Fig.~\ref{Fig1_LatticeNormal}(f) and therefore also the profile of the superconducting order displayed in Fig.~\ref{Fig2_SC}(d). Moreover, we find that the magnetic moments change signs between the two sublattices, but the sublattices do not have the same moment magnitudes, thus creating an in-plane ferrimagnetic ordering. In-between layers we find an antiferromagnetic ordering since an A site sits next to B-sites of the neighboring layer. This magnetic order, emerging from our unconstrained self-consistent tight-binding model matches the ground state found in spin-polarized DFT~\cite{Otani2010Intrinsic,Xiao2011Density,Cuong2012Magnetic,Pamuk2017}, both in terms of symmetry and in the qualitative profile of magnetic moments, both of which remain qualitatively unchanged for all interaction strengths below the bulk critical coupling strengths, where instead the whole stack becomes magnetic. 	

In particularly, the close agreement in spatial magnetic moment structure between all mean-field results can be quantified by plotting the ratio between the surface ferrimagnetic moments, $R = |m_{B1}/m_{A1}|$ as a function of the normalized coupling strengths $\bar{\Gamma}_{r}^{z,-}$  in the inset of Fig.\ref{Fig:Magnetism}(b). Specifically, we find the ratio $R$ to be an approximately linear function that is zero at no interaction $\bar{\Gamma}_{r}^{z,-}=0$ and unity at the bulk critical coupling $\bar{\Gamma}_{r}^{z,-}=1$. This means that there is a gradual, approximately linear, connection between a ferromagnetic state with no magnetic moment on the minority lattice to a completely balanced antiferromagnetic state close to the bulk critical coupling. Inbetween these two extremes, the ABC-MLG surface state is in a ferrimagnetic state. 
	
\begin{figure}[]
				\includegraphics[width=0.5\textwidth]{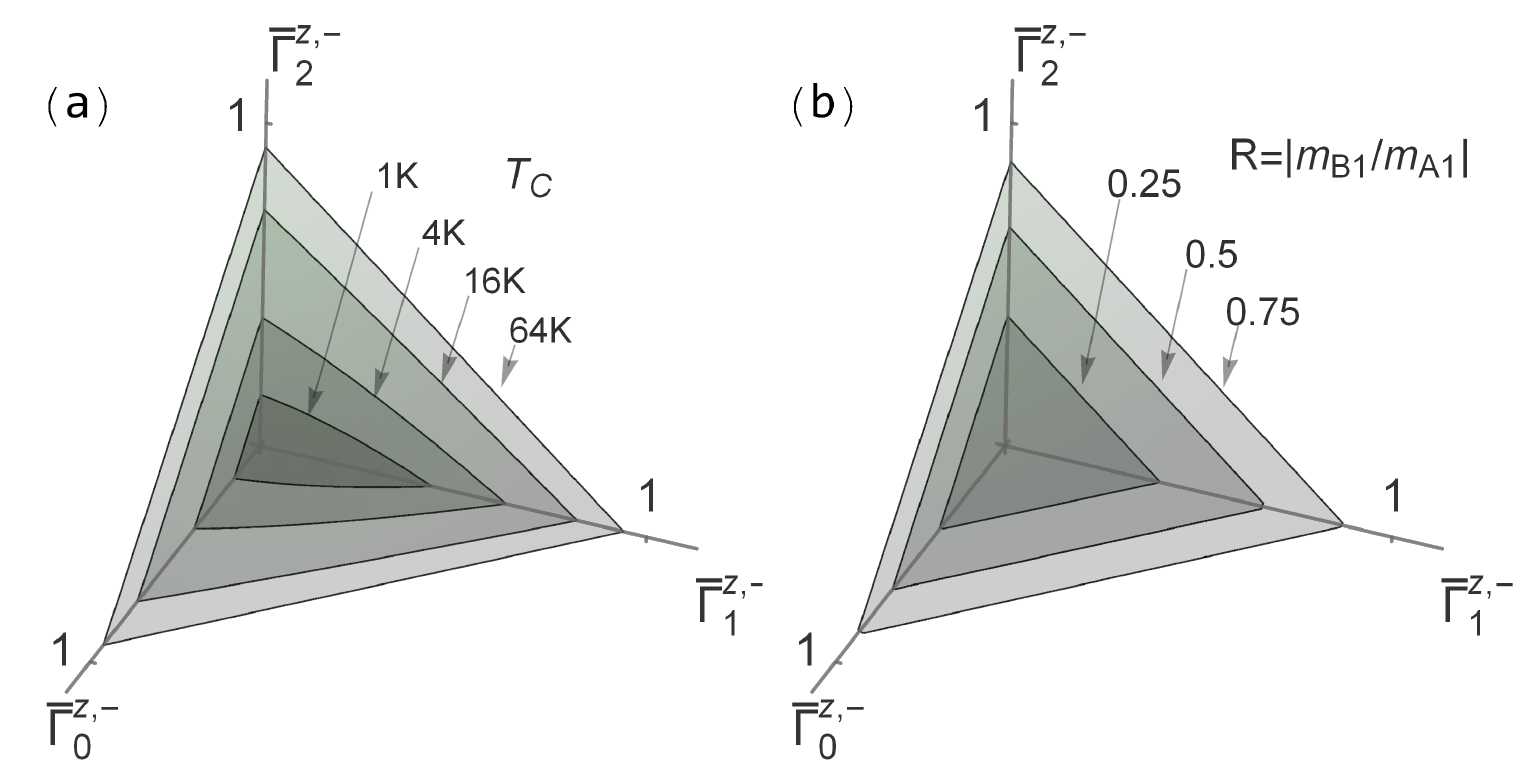}
				\caption{%
					({a}) Constant magnetic critical temperature $T_{\rm c}$ surfaces (a) and constant surface sublattice magnetic moment ratio $R$ (b) as functions the three normalized couplings strengths $\bar{\Gamma}_{r=0,1,2}^{z,-}$, forming simple simplexes.
				}
				\label{Fig:Simplex}
			\end{figure}

So far we have treated each interaction range separately and discussed its magnetic order. Finally, we here consider all interactions up to the NNN range jointly. We find that the magnetizations of all three different interaction ranges add linearly and additively, to an excellent approximation. Consequently, we find that the surfaces of constant $T_{\rm c}$ as a function of all three $\bar{\Gamma}_{r=0,1,2}^{z,-}$ are all simple simplexes when plotted in a three-dimensional space with the three different interactions ranges as orthogonal axes in Fig.~\ref{Fig:Simplex}(a). In Fig.~\ref{Fig:Simplex}(b) we also show that the same holds true for the surface ratio $R$. We thus conclude that the surface state of ABC-MLG easily harbors a magnetic surface state, with an in-plane ferrimagnetic structure. This state is driven by any of the interactions $U_0>0, J_1>0, J_2<0$ and if present together, the resulting magnetization combine in an additive fashion. Peculiarly, we find that the ratio $R$ of magnetic moments on the two sublattice sites is independent of the interaction range and increases approximately linearly with the total interaction strength. Taken together, this conclusion about the magnetic state constitute the fifth significant result of this work.
	
\section{Superconductivity versus Magnetism} 
\label{sec:SCPH}
Having established both superconductivity and magnetism in the previous sections, we in this section address the resulting order competition and how it changes with interaction strengths and also gating. The latter is particularly easy in stacked systems, where a dual gate setup enables simultaneous  control of both band filling and band tuning by field gating and displacement field bias \cite{Oostinga2008, Zhang2010Band, Yanmeng2020Electronic, Koshino2010}. Such  control has already been used to map the phase diagram of twisted bilayer graphene which includes superconducting domes ~\cite{Cao2018Unconventional, Lu2019Superconductors, Yankowitz2019Tuning, Chen2019Signatures,Dai2021Mott}, as well as used to uncover two distinct superconducting phase space regions in trilayer graphene~\cite{Zhou2021Superconductivity}, where a displacement field bias was used to produce a partial band flattening as the stack is not thick enough for zero-energy surface states. We here first continue within the same mean-field framework as in the previous sections, and then also complement with \emph{ab initio} DFT calculations. For the gating we primarily study a homogenous shift of the chemical potential, but note that a displacement field has similar effects \cite{Lothman2017Universal}.
	
\subsection{Mean-field analysis}
\begin{figure*}[t]
			\includegraphics[width=0.975\textwidth]{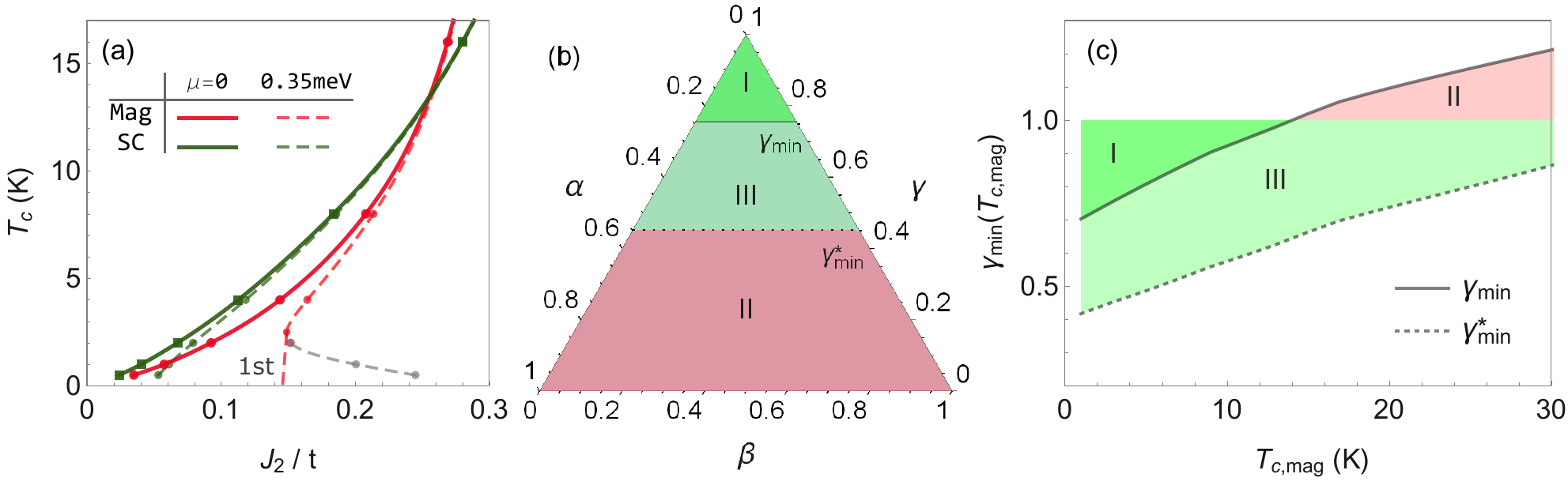}
			\caption{%
				Order competition between ferrimagnetism and $f$-wave superconductivity in eight layer ABC-MLG. 	
				({a}) %
					Critical temperatures $T_{\rm c,mag}$ of magnetism (Mag) and $T_{\rm c, sc}$ of triplet $f$-wave superconductivity (SC) as a function of the effective ferromagnetic NNN Heisenberg interaction $J_2$ at charge neutrality $\mu = 0$ (solid line) and  $\mu = 0.35$~meV (dashed line). Grey line are also solution to the LGE, Eq.~\eqref{eq:StabMat}, but to an unstable equilibrium where the magnetic state free energy is higher than the normal state free energy; instead magnetic state has a first order transition to the normal state (vertical dashed red line, marked by 1st). 
				({b}) %
					Ternary plot of the phase diagram on the simplex $\alpha + \beta + \gamma  = 1 $, ($\alpha, \beta, \gamma \ge 0$) corresponding to the interaction strengths 
					$
						\{
							\Gamma_{0}^{z,-},
							\Gamma_{1}^{z,-},
							\Gamma_{2}^{z,-}
						\}
						=
						\{
						  U_0,
							J_1,
							-J_2
						\}
						=
						\{
							\alpha \Gamma_{0}^{z,-} (T_{\rm c, mag}), 
							\beta \Gamma_{1}^{z,-} (T_{\rm c, mag}),  
							\gamma \Gamma_{2}^{z,-} (T_{\rm c, mag})
						\}
					$,
					where $\Gamma_{r}^{z,-} (T_{\rm c, mag})$ is the coupling strength that by itself gives $T_{\rm c, mag} = 3$~K.
					Note that $T_{\rm c,mag}$ is here constant over the entire simplex, as established in Fig.~\ref{Fig:Simplex}. 
					Region {I} (green) is superconducting, with $T_{\rm c, sc} > T_{\rm c, mag}$ both at $\mu=0$ and finite $\mu$. 
					Region {II} (red) is magnetic, with $T_{\rm c, sc} < T_{\rm c, mag}$ at all $\mu$. 
					Region {III} (light green) is magnetic or superconducting depending on the chemical potential $\mu$, with $T_{\rm c, sc} < T_{\rm c, mag}$ at $\mu=0$, but for a finite $\mu$, $T_{\rm c, sc} > T_{\rm c, mag}$.				({c}) %
					Region I boundary $\gamma_{\rm min}$ (solid) and region III boundary $\gamma_{\rm min}^{*}$ (dashed), defined in (b) as a function of $T_{\rm c, mag}$ with regions I-III indicated. Note that solutions with $\gamma_{\rm min} > 1$ have no region I in the phase diagrams, since the region I boundary $\gamma_{\rm min}$ lies outside the phase diagram; the system is therefore magnetic at $\mu=0$. 
			}
			\label{Fig:J2MagSc}
		\end{figure*}

With both superconductivity and magnetism both generated by the $J_2$ interaction, we start by limiting our analysis to when only $J_2$ is present, as that will inform the rest of our analysis. In Fig.~\ref{Fig:J2MagSc}(a) we show the critical temperature of both the spin-triplet $f$-wave superconducting state, now called $T_{\rm c, sc}$, and the ferrimagnetic order, now called $T_{\rm c, mag}$, as a function of $J_2 $ for eight layer ABC-MLG. At charge neutrality, i.e.~$\mu = 0$ (solid lines), $T_{\rm c, sc}$ is higher for all $|J_2| \lesssim 0.25t$. This is in spite of the magnetic state having a lower bulk critical coupling strength of $|J^{-}_{2,c}| \approx 0.4t$ compared to the critical coupling strength $|J^{+}_{2,c}| \approx 0.6t$ for the superconducting state. In fact, the magnetic state only overtakes the superconducting state for $J_2$ approaching the bulk critical strength $J^{-}_{2,c}$. Thus, we find that the $f$-wave superconducting state is the dominant order for a wide range of weak to moderate $J_2$ values.

Electronic gating further offers an experimentally accessible tool to tune ABC-MLG and the competition between magnetism and superconductivity. We here investigate the effects of applying a homogeneous chemical potential $\mu$ to the whole stack to illustrate the possibilities. In Fig.~\ref{Fig:J2MagSc}(a) we also display $T_{\rm c, sc}$ and $T_{\rm c, mag}$ for $\mu = 0.35$~meV (dashed lines). As seen, both $T_{\rm c, sc}$ and $T_{\rm c, mag}$ are reduced for all $J_2$. However, the reduction of $T_{\rm c, mag}$ is much  larger compared to the reduction in $T_{\rm c, sc}$. For example, even if finite doping requires a finite interactions strength for ordering, $|J_2| \sim 0.15t$ is needed for magnetic ordering at $\mu = 0.35$~meV, which is considerably larger than the $|J_2| \sim 0.05t$ needed for superconductivity to first appear at the finite doping. In fact, only at $|J_2| \sim 0.15t$ does the magnetic state appear, but now only via a first order transition line marked by 1st in Fig.~\ref{Fig:J2MagSc}(a) where the free energy of the magnetically ordered state crosses the free energy of the normal state \cite{Lothman2017Universal}. Consequently, a finite $\mu$ strongly increases the advantage of the superconducting state, although both $T_{\rm c}$s are reduced. This advantage of superconductivity over magnetism at finite $\mu$ can be attributed to a universal feature of ordering in flat band systems. Within mean-field theory it has been shown that all superconducting, or particle-particle, orders maintain a higher critical temperature following a shift of the Fermi energy away from the flat band or DOS peak, compared to any competing charge or magnetic, i.e.~particle-hole, orders \cite{Lothman2017Universal}. The result is that a flat band supports superconducting orders over a much wider range of filling factors and superconducting domes are thus likely to form on the flanks of flat bands. To summarize, the sixth important result of this work is that $f$-wave superconductivity is the dominant order for weak to moderate $J_2$ interaction in ABC-MLG and that it becomes further favored over the ferrimagnetic state with gating or doping away from charge neutrality.
	
So far in this section we have only investigated the individual effect of $J_2$. We next consider an arbitrary interaction mix of the expected interactions in ABC-MLG as established in Sec.~\ref{sec:Interactions}: ON repulsion $U_0 > 0$, antiferromagnetic NN Heisenberg term $J_1 > 0$, and ferromagnetic NNN Heisenberg term $J_2 < 0$. From Sec.~\ref{sec:Magnetism} we already know that each of the interactions additively increases $T_{\rm c, mag}$. In contrast, $J_2$ is the only interaction that generates the $f$-wave superconducting state. Because of this fact, we anticipate to find that a generic interaction mix needs a minimal proportion, here defined as $\gamma_{\rm min}$ of $J_2$ to maintain the $T_{\rm c, sc} > T_{\rm c, mag}$. To fully establish this general statement for an arbitrary interaction mix however requires a careful parametrization of the interactions. 
	
To find a suitable way to study any interaction mix beyond just $J_2$, we start by noting that the analysis is greatly aided by the observations that the constant $T_{\rm c,mag}$ surfaces in Fig.~\ref{Fig:Simplex}(a) are simple simplexes. To an excellent approximation, $T_{\rm c, mag}$ is therefore given by the additive contributions of $U_0$, $J_1$, and $J_2$. Thus, if we define ${\Gamma}_{r}^{z,-} (T_{\rm c, mag})$ to be the value of ${\Gamma}_{r}^{z,-}$ that generates the critical temperature $T_{\rm c, mag}$ at $\mu=0$ when ${\Gamma}_{r}^{z,-}$ is the only coupling present, then it is also true that the mix of interactions given by
		$
			\{
				\Gamma_{0}^{z,-},
				\Gamma_{1}^{z,-},
				\Gamma_{2}^{z,-}
			\} = 	\{
				\alpha \Gamma_{0}^{z,-} (T_{\rm c, mag}), 
				\beta \Gamma_{1}^{z,-} (T_{\rm c, mag}),  
				\gamma \Gamma_{2}^{z,-} (T_{\rm c, mag})
			\}
		$
	also generates $T_{\rm c, mag}$ over the entire simplex spanned $\alpha + \beta + \gamma  = 1$, here assuming $\alpha, \beta, \gamma \ge 0$. As a consequence, we first choose to parametrize any mix of interactions first by the relative proportion of $\Gamma_{0}^{z,-}=U_0 $, $\Gamma_{1}^{z,-}=J_1$ and $\Gamma_{2}^{z,-}=-J_2$, defined by the coefficients $\alpha, \beta,$ and $ \gamma$ respectively, and study its influence. Second, we also study the total net interaction strength given by the critical temperature $T_{\rm c, mag}$ that this mix generates at $\mu=0$. 
	
Having established a useful interaction parametrization, we first look at the effect of tuning the relative proportions of the individual interactions, keeping the total net strength constant such that $T_{\rm c, mag}=3$~K at $\mu=0$. In Fig.~\ref{Fig:J2MagSc}(b) we show the full phase diagram for eight layer ABC-MLG in the $\alpha, \beta, \gamma$ parameter space, or equivalently viewed as plotting the phase diagram on a constant $T_{\rm c, mag}$ simplex surfaces such as those in Fig.~\ref{Fig:Simplex}(a). At the top vertex the interaction mix consists solely of $J_2$ and the superconducting order has the highest critical temperature with $T_{\rm c, sc} > T_{\rm c, mag}$, as already known from Fig.~\ref{Fig:J2MagSc}(a). Away from the top vertex, the proportion $\gamma$ of $J_2$ in the interaction mix decreases and, since $J_2$ is the only interaction supporting the superconducting state, there exists a minimal proportion $\gamma_{\rm min}$ below which the magnetic state instead has the highest critical temperature. Region I spanning $\gamma > \gamma_{\rm min}$ therefore represents the parameter space where ABC-MLG is superconducting at charge neutrality. Conversely, for smaller proportions $J_2$, in regions II and III ABC-MLG is magnetically ordered at $\mu=0$. However, because a finite $\mu$ give a competitive advantage to superconductivity, as already established in Fig.~\ref{Fig:J2MagSc}(a), we find that in the intermediate region III, defined as $\gamma_{\rm min} > \gamma > \gamma_{\rm min}^{*}$), ABC-MLG is superconducting for at least at some finite $\mu$.

In order to easily establish the position of $\gamma_{\rm min}^{*}$ in parameter space, we use Ref.~\cite{Lothman2017Universal}, which derives that, assuming an isolated flat band at the Fermi level for $\mu=0$ and that the interactions are independent on $\mu$, superconducting domes exists at finite $\mu$ flanking a magnetic, or more generically a particle-hole, order whenever $T_{\rm c, sc} >  T_{\rm c, mag} /2 $ at $\mu=0$. Using this result together with the data in Fig.~\ref{Fig:J2MagSc}(a) we can position $\gamma_{\rm min}^{*}$ in Fig.~\ref{Fig:J2MagSc}(b). We note, however, that the assumptions behind this derivation start to breakdown for large enough $T_{\rm c, mag}$ due to the proximity to the bulk critical coupling strengths in combination with the flat band not being entirely isolated. Thus, $\gamma_{\rm min}^{*}$ is best viewed as a lower bound, which we estimate should be valid for  $T_{\rm c, mag} \lesssim 10$~K. Nonetheless, Fig.~\ref{Fig:J2MagSc}(b) clearly illustrates that there exists a large extended region III, where $J_2$ comprises only a modest proportion of the total interaction, yet ABC-MLG is still superconducting and not magnetic. 
	
Next, we show in Fig.~\ref{Fig:J2MagSc}(c) how the phase diagram in Fig.~\ref{Fig:J2MagSc}(b) evolves as we tune the total interaction strength, here parameterized by $T_{\rm c, mag}$ at $\mu=0$. In particular, we show how the phase boundaries $\gamma_{\rm min}$ and $\gamma_{\rm min}^{*}$ evolve with $T_{\rm c, mag}$. We find that $\gamma_{\rm min}$ and $\gamma_{\rm min}^{*}$ increases with $T_{\rm c,mag}$, a behavior that can be derived to the fact that the difference between $T_{\rm c, sc}$ and $T_{\rm c, mag}$ decreases as $J_2$ increases, as seen  in Fig.~\ref{Fig:J2MagSc}(a). In particular, the region {I} of Fig.~\ref{Fig:J2MagSc}(c) remains finite until the two curves cross, $T_{\rm c, sc} =  T_{\rm c, mag}$, in Fig.~\ref{Fig:J2MagSc}(a), beyond which the systems is magnetic at $\mu=0$. Still, even as $T_{\rm c, mag}$ increases further there is a considerable Region {III} for which ABC-MLG is superconducting for some finite doping. Thus, there exists a wide range of coupling strengths for which only a modest proportion of $J_2$ generates  superconducting phase space regions (green and light green). Finally, we note that we have in the above results completely treated the most relevant interactions of Eq.~\eqref{eq:Int} out to the NNN range. In terms of superconductivity, we however also know from Fig.~\ref{fig:PD} that $U_2$ is an important parameter, although it has little to no effect on the ferrimagnetic state. In Fig.~\ref{fig:PD} it is evident that the effect of $U_2$ on the NNN spin-triplet $f$-wave state is a simple shift of the coupling strength. Thus, the net effect of incorporating a finite $U_2 < |J_2|$ is to renormalize both $\gamma_{\rm min} \mapsto  \gamma_{\rm min} (1 - U_2/|J_2|)^{-1}$ and $\gamma_{\rm min}^{*} \mapsto \gamma_{\rm min}^{*} (1 - U_2/|J_2|)^{-1}$. This somewhat shrinks the superconducting regions of the phase diagram.
	
The final result of incorporating all relevant interaction strengths out to the NNN range is that the superconducting $f$-wave state surprisingly dominates over the ferrimagnetic state, in fact, as soon as the system exceeds a minimal proportion of $J_2$ in the total interaction strengths. The superconducting state is further favored by finite doping $\mu$, such that a substantial part of the interaction parameter space results in a superconducting state, as displayed in Figs.~\ref{Fig:J2MagSc}(b) and~\ref{Fig:J2MagSc}(c). This abundance and resilience of the superconducting state when incorporating the relevant interactions constitutes the seventh important result of this work. 

\subsection{Density functional theory analysis}
In the previous subsection we established the general phase diagram and  competition between the ferrimagnetic and the spin-triplet NNN $f$-wave states in ABC-MLG within unconstrained mean-field theory for all interaction ranges out to NNN. Here we complement this analysis with \emph{ab initio} spin-polarized DFT calculations. While DFT cannot capture the superconducting state, it can provide additional information on the magnetic state. Multiple DFT results already exist \cite{Otani2010Intrinsic,Xiao2011Density,Cuong2012Magnetic,Pamuk2017}, but we are here primarily concerned with being able to explicitly match with our mean-field theory results, as they offer the direct comparison with the superconducting state. To proceed, we model an eight layer ABC-MLG system with the intralayer (interlayer) carbon distance $a_{\rm CC}=1.42~\text{\AA}\left(c=3.347~\text{\AA}\right)$~\cite{Aoki2007} in a slab geometry using QUANTUM ESPRESSO~\cite{Giannozzi2009,Giannozzi2017}. We assume in-plane translational invariance and use a $204\times204$ in-plane $k$-point sampling. Additionally, we surround the ABC-MLG stack with 30~\AA~of padded vacuum. Due to highly varying results in the previous literature \cite{Otani2010Intrinsic,Xiao2011Density,Cuong2012Magnetic,Pamuk2017}, we use and compare the results of multiple different exchange-correlation (XC) functionals, including also van der Waals corrections to accurately capture the weak but important interlayer couplings~\cite{Grimme2006Semi}.
		
\begin{figure}[tb]
			\includegraphics[width=0.44\textwidth]{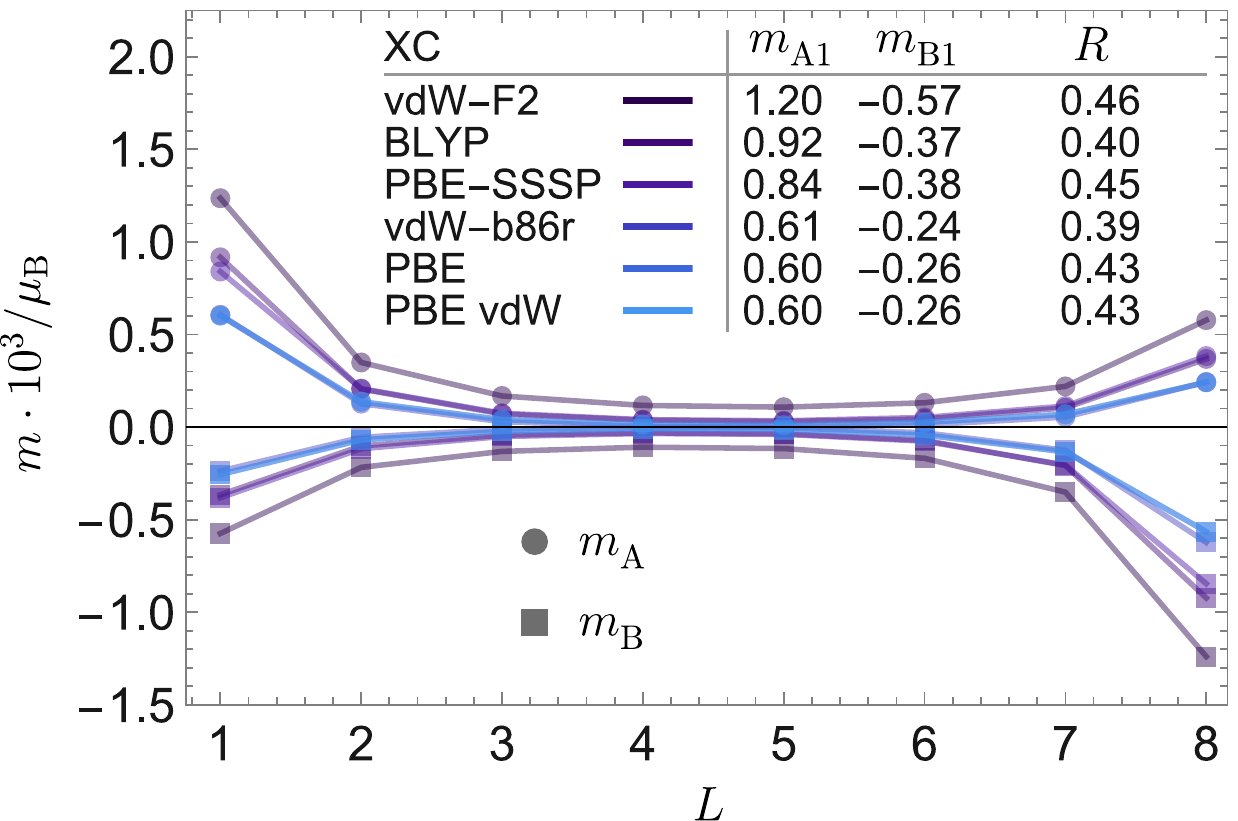}
			\caption{%
				Layer profile of sublattice resolved magnetic moments $m$ (units of Bohr magneton $\mu_B$) calculated within DFT for different XC functionals. Legend summarizes magnetic moments on A1 and B1 surface sites as well as the ratio $R=|m_{B1}/m_{A1}|$, sorted from largest to smallest $m_{\rm A1}$. 		
				 Here, PBE-SSSP is a PBE XC functional based on the Standard Solid State Pseudopotentials (SSSP)~\cite{Prandini2018Precision,Prandini2018A}, while all other XC functionals are built-in in QUANTUM ESPRESSO. Note that vdW-F2 and vdW-b86r are special XC functionals for van der Waals systems. We also included PBE vdW, which includes a non-local van der Waals interaction, in addition to the pure PBE. 
			}
			\label{Fig:DFTMoments}
		\end{figure}

In Fig.~\ref{Fig:DFTMoments} we show the sublattice and layer-resolved magnetic moments calculated for six different choices of XC functionals (see legend in Fig.~\ref{Fig:DFTMoments}). For all choices, the spin-polarized DFT calculations give a magnetically ordered state at $T=0$, with intralayer ferrimagnetism localized to the surfaces with a strong decay of the magnetic moments in to the center of the ABC-MLG slab. This is qualitatively the same magnetic state we find within the mean-field theory modeling in Sec.~\ref{sec:Magnetism} and also consistent with previous DFT results~\cite{Otani2010Intrinsic,Xiao2011Density,Cuong2012Magnetic,Pamuk2017}. Yet, we also find that the size of the magnetic moments depends strongly on the choice of XC functional. Overall, the sizes of the magnetic moments are small and of the order of $\sim 10^{-3}\,\mu_{B}$, $\mu_B$ being the Bohr magneton, but there is a factor of $2$ difference depending on the choice of XC functional. Notably, while van der Waals interactions are expected to be important in ABC-MLG, we find no notable improvement in the consistency of the results when including or not including van der Waals interactions. With this strong variations in magnetic moments, and the fact that all interaction ranges investigated within our mean-field treatment produces the ferrimagnetic state, we can unfortunately not use these DFT-calculated magnetic moments to get a better handle on the realistic magnitudes of the mean-field interaction parameters $U_0$, $J_1$, and $J_2$.
	
Despite the large variation in the calculated magnetic surface moments, we can still actually extract useful data. For this, we turn to our mean-field results in Fig.~\ref{Fig:Magnetism}(b), which established that the ratio of the surface magnetic moments $R=|m_{B1}/m_{A1}|$ to an excellent approximation are independent on the interaction range and instead only depend on normalized channel coupling strengths. We therefore also compute $R$ from our DFT results and we find that, despite the large magnitude variations, $R$ remains remarkably stable across all different XC functionals, ranging from just $R=0.39$ to $0.46$. This both establishes consistency between the mean-field theory and DFT results and that $R$ is an efficient tool to capture the magnetic structure of the ferrimagnetic state.

\begin{figure}[t]
			\includegraphics[width=0.3\textwidth]{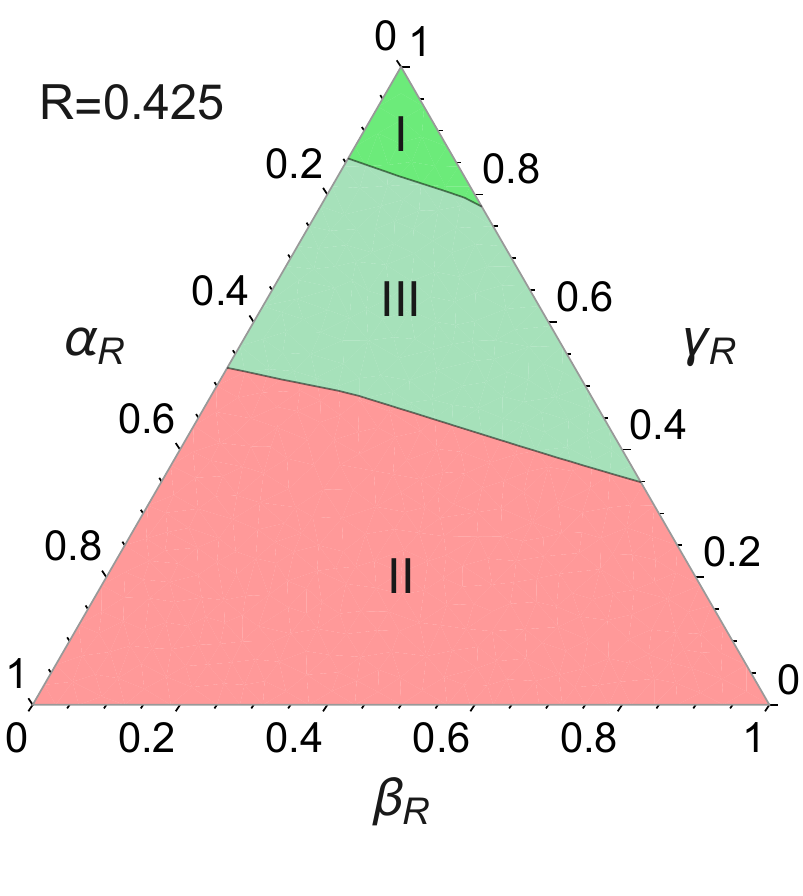}
			\caption{%
				Ternary plot of phase diagram on the simplex $\alpha_R + \beta_R + \gamma_R  = 1 $, ($\alpha_R, \beta_R \gamma_R \ge 0$) corresponding to the interaction strengths
					$
						\{
							\Gamma_{0}^{z,-},
							\Gamma_{1}^{z,-},
							\Gamma_{2}^{z,-}
						\}
						=
						\{
							U_0,
							J_1,
							-J_2
						\}
						=
						\{
							\alpha_R \Gamma_{0}^{z,-} (R), 
							\beta_R \Gamma_{1}^{z,-}  (R),  
							\gamma_R \Gamma_{2}^{z,-} (R)
						\}
					$,
				where $\Gamma_{r}^{z,-} (R)$ is the coupling strength that by itself gives $R=0.425$, established by the DFT results, i.e.~similar phase diagram to Fig.~\ref{Fig:J2MagSc}(b), but where instead the simplex $T_{\rm c,mag} =3$~K was chosen. Note that $R$ is here constant over the entire simplex, as established in Fig.~\ref{Fig:Simplex}. Region {I} (green) is superconducting, with $T_{\rm c, sc} > T_{\rm c, mag}$ both at $\mu=0$ and finite $\mu$. 
					Region {II} (red) is magnetic, with $T_{\rm c, sc} < T_{\rm c, mag}$ at all $\mu$. 
					Region {III} (light green) is magnetic or superconducting depending on the chemical potential $\mu$, with $T_{\rm c, sc} < T_{\rm c, mag}$ at $\mu=0$, but for a finite $\mu$, $T_{\rm c, sc} > T_{\rm c, mag}$.					
			}
			\label{Fig:MomentPD}
		\end{figure}

We next use the consistency of $R$ to extract the full ternary phase diagram of magnetism and superconductivity. In Fig.~\ref{Fig:Simplex}(b) we established that constant $R$ surfaces are to excellent approximation simple simplexes when plotted against all three ranges of interactions, just as the constant $T_{\rm c,mag}$ surfaces in Fig.~\ref{Fig:Simplex}(a). In  Fig.~\ref{Fig:J2MagSc}(b) we utilized the latter result and plotted the full ternary phase diagram with magnetism and superconductivity of ABC-MLG for a fixed $T_{\rm c,mag}$. Here we similarly extract the phase diagram using a constant $R$ simplex, choosing $R=0.425$ as an average extracted from the DFT calculations.  Note that we could not easily have performed this analysis earlier only within the mean-field theory results as we then did not know the size of $R$. In Fig.~\ref{Fig:MomentPD}, we plot the ternary phase diagram plot of ABC-MLG on the constant $R=0.425$ surface. Each point in Fig.~\ref{Fig:MomentPD} thus correspond to an interaction mix, 	
		$
			\{
				\Gamma_{0}^{z,-},
				\Gamma_{1}^{z,-},
				\Gamma_{2}^{z,-}
			\}
			=
				\{
				U_0,
				J_1,
				-J_2
		        \}
			=
			\{
				\alpha_R \Gamma_{0}^{z,-} (R), 
				\beta_R \Gamma_{1}^{z,-}  (R),  
				\gamma_R \Gamma_{2}^{z,-} (R)
			\}
		$,
	where $\alpha_R + \beta_R + \gamma_R = 1 $, $\alpha_R, \beta_R, \gamma_R \ge 0$, and the coupling strengths $\Gamma_{r}^{z,-} (R)$ are defined to produce a magnetic order with $R=0.425$ when it is the only coupling in the mix. 	
	We find that the constant $R$ ternary phase diagram of Fig.~\ref{Fig:MomentPD} shares its basic features with the constant $T_{\rm c, mag}$ ternary phase diagram of Fig.~\ref{Fig:J2MagSc}(b). Both phase diagrams contain three different regions. In region I, close to the top vertex, superconducting order dominates both at $\mu=0$ and for finite $\mu$, since then the interaction mix consist primarily of $J_2$ which favors the superconducting state. In contrast, close to the other two other vertices, $U_0$ or $J_1$ is the main ingredient of the interaction mix, and the magnetic state dominates in Region II, while in region III a superconducting state is still present at finite $\mu$.  
	The only real difference between the constant $T_{\rm c, mag}$ and constant $R$ phase diagram is that the phase boundaries in Fig.~\ref{Fig:MomentPD} are slanted instead of horizontal as in Fig.~\ref{Fig:J2MagSc}(b). The origin of this slant can be found in the normal directions of the constant $R$ and $T_{\rm c, mag}$ simplexes in Fig.~\ref{Fig:Simplex}, which in turn can be traced back to the lower $T_{\rm c,mag}$ produced by the NN ranged interaction compared to the ON and NNN interactions seen in Fig.~\ref{Fig:Magnetism}(a), while $R$ is less affected.
	The existence of a constant  $R$ value, largely independent on XC functional, and thus the possibility to extract the phase diagram in Fig.~\ref{Fig:MomentPD} is the eighth major result of this work. It thoroughly establishes that only a moderate portion of $J_2$ is needed in order to achieve superconductivity in ABC-MLG as guided by DFT calculations, possibly as domes flanking a ferrimagnetic state.
	
\section{Conclusions}~\label{sec:Conc}
In this work we analyze the electron interaction driven ordering possibilities in ABC-stacked multilayer graphene (ABC-MLG), focusing on thick stacks that host topological surface flat bands at charge neutrality. We consider all spin-symmetric short-ranged two-site interactions, out to the next-nearest-neighbor (NNN) range within a fully general mean-field treatment, complemented with general group theory analysis, as well as spin-polarized DFT calculations. For these interactions, the infinite layered ABC-MLG, i.e.,~bulk ABC-stacked graphite, remains unordered in all ordering channels, due to its vanishing DOS at charge neutrality demanding a too large finite critical interaction strength for ordering. In contrast, the topological flat band surface states of finite stacks of ABC-MLG strongly enhance most order possibilities.

For superconducting ordering, we find that all pairing channels hosts a finite critical temperature $T_{\rm c,sc}$ for surface superconductivity also for weak interactions, with the sole exception of pairing in the nearest-neighbor (NN) range, which we attribute to an almost complete sublattice polarization of the surface states preventing NN pairing. We further establish that the superconducting phase diagram is dominated by an isotropic spin-singlet $s$-wave pairing and unconventional NNN bond spin-triplet $f_{x\left(x^2-y^2\right)}$-wave pairing, which both have a full energy gap. We find that the spin-triplet $f$-wave state is favored by predominantly ferromagnetic NNN bond Heisenberg interactions ($J_2$), while the $s$ wave is favored by antiferromagnetic such interactions. With intra-sublattice interactions having been identified as ferromagnetic in graphene, i.e.~$J_2<0$, we conclude that the fully gapped unconventional spin-triplet $f$-wave state is the most likely superconducting state on the surface of ABC-MLG. We further note that this $f$-wave state survives significant on-site repulsion $U_0$ that hurts the spin-singlet $s$-wave state, which further promotes tenability of the $f$ wave. Finally, electron-phonon coupling could in principle still favor an $s$-wave state, but due to it being very weak in graphene and in face of strong on-site repulsion, we estimate that the $f$-wave superconductivity is the most favorable superconducting state in ABC-MLG. These results are in notable contrast with results in bulk ABC-stacked graphite as well as few layer graphene without the ABC-MLG surface state, where instead the chiral $\left(d_{x^2-y^2}+id_{xy}\right)$- or the $\left(p_x+ip_y\right)$-wave state displays the best ordering tendencies. We establish that this very different bulk versus surface behavior stems from the sublattice polarization of the surface states. 

The topological surface flat bands also generically promote particle-hole ordering in ABC-MLG. By treating all charge and magnetic ordering possibilities on an equal footing, we find that ABC-MLG has a ferrimagnetic order on the surface for all relevant spin-symmetric short-ranged interactions with mean-field theory, with results fully agreeing with spin-polarized DFT calculations. Thus, while bulk ABC-graphite is not magnetically ordered, ABC-MLG exhibits a finite $T_{\rm c,mag}$ also for weak interactions. We further find that the ferrimagnetic order interpolates between complete sublattice ferromagnetism for weak coupling to a complete antiferromagnetic state close to the critical bulk coupling strengths, quantified by the ratio between the surface sublattice magnetic moments $R = | m_{B1}/m_{A1} |$ ranging from zero and one.  Additionally, we find that differently ranged interactions contribute additively to the ferrimagnetic order and therefore the constant $T_{\rm c,mag}$ surface is a simple simplex in the $\{U_0,J_1,-J_2\}$ interaction space, which encodes the relevant interactions: repulsive on-site Coulomb $U_0>0$, antiferromagnetic $J_1>0$, and ferromagnetic $J_2<0$.

Having both  $f$-wave superconductivity and ferrimagnetic ordering as possibilities in ABC-MLG, we finally establish their competition and the resulting phase diagram as a function of interaction strengths. Both ferrimagnetism and $f$-wave superconductivity are driven by a ferromagnetic $J_2$, but we find that the susceptibility for the superconducting $f$-wave state is larger, resulting in $T_{\rm c,sc}>T_{\rm c,mag}$ over a large range of weak to moderate $J_2$ interaction strengths. Consequently, we find that the $f$-wave state is the dominant order in ABC-MLG, as long as the proportion of $J_2$ in the full mix of interaction parameters is sufficiently large. Moreover, we find that superconductivity is further favored by gating, resulting in a homogeneous doping or an electric field gradient perpendicular to the ABC-MLG stack, which suppresses the magnetic order much more than the superconducting order. Thus, $f$-wave superconductivity generally appears upon doping, even if the $J_2$ proportion is relatively small. By using spin-polarized DFT calculations we are able to further specify the phase diagram. We find that the sizes of the magnetic moments are highly sensitive to the choice of exchange-correlation functional used for the DFT calculations. Still, the surface sublattice moment ratio $R$ is remarkably stable, which we utilize to construct an improved phase diagram, with a similarly notable superconducting portion. Overall, our results establish that spin-triplet $f$-wave superconductivity is possible from purely repulsive interactions on the surface of ABC-MLG and feasible over a large region of the phase diagram in competition with a ferrimagnetic state. The $f$-wave state further gains a competitive advantage when doping slightly away from from charge neutrality. We finally note that due to the special properties of the surface flat band in ABC-MLG, these results are notably different from doped or twisted few layer graphene, which clearly illustrates the fascinating possibilities and variations of electronic ordering in graphene systems.

\section*{Acknowledgement}
We thank S.~Nahas for useful discussions on the DFT calculations and C.~Bena for general discussions. We acknowledge financial support from the Swedish Research Council (Vetenskapsr\aa det Grant No.~2018-03488) and the Knut and Alice Wallenberg Foundation through the Wallenberg Academy Fellows program, as 	well as the EU-COST Action CA-21144 Superqumap. Simulations were enabled by resources provided by the Swedish National Infrastructure for Computing (SNIC) at the Uppsala Multidisciplinary Center for Advanced Computational Science (UPPMAX), partially funded by the Swedish Research Council through Grant No.~2018-05973. In addition, O.A.A acknowledges funding from NanoLund.

\appendix
\onecolumngrid
\section{Derivation of linearized gap equation}\label{app:StabMat}
	To  arrive at the LGE, Eq.~\eqref{eq:StabMat} in the main text, we carry out a perturbation expansion of the mean-field interaction Hamiltonian, $H_{\rm MF}$. This is done by expanding the interacting density matrix $\hat{\rho}=e^{-\beta (H_{\rm MF}-H_0)}$ in terms of the noninteracting density matrix, $\hat{\rho}_0=e^{-\beta H_0}$ and keep up to linear terms in the interaction obtaining, 
	\begin{equation}\label{eq:DensityMat}
	\hat{\rho} \approx \hat{\rho}_0- \hat{\rho}_0\int_{0}^{\beta} {\rm d} \beta^\prime e^{\beta^\prime H_0} (H_{\rm MF}-H_0)^{\rm MF}e^{-\beta^\prime H_0}, 
	\end{equation}
	where here $\beta =1/\left(k_{\rm B}T\right)$ is the inverse temperature, for details see e.g.~ Refs.~\cite{Lothman2017Universal,Feynman1998Stat}. The averages of the bilinear operators in the self-consistency equations, Eq.~\eqref{eq:MFSelfCon}, can now be written as
	\begin{equation}\label{eq:LinearSelf}
	 \langle a_i a_j \rangle= {\rm Tr} \left[\hat{\rho}  a_i a_j\right].
	 \end{equation}
	But we also know that $ {\rm Tr} \left[\hat{\rho}_0  a_i a_j\right]=0$ since the order parameters vanish in the noninteracting phase. Thus, only the second term in the expansion in Eq.~\eqref{eq:DensityMat} contributes to the first order perturbation and thus the LGE.

	To proceed, we let $\hat{S}\left(\boldsymbol{k}\right)$ be the unitary matrix that diagonalizes the noninteracting Hamiltonian, given in real space by Eq.~\eqref{eq:Normal} in the main text, and $\xi_{\bar{i}}\left(\boldsymbol{k}\right)$ the corresponding energy bands, where $\bar{i}$ labels the bands. Then, the real space operators $a_{jL\alpha}^{}$ used in the main text can be written in terms of band operators $c_{\boldsymbol{k}\bar{j}\alpha}^{}$ as
	\begin{equation}\label{eq:BandOP}
	a_{jL\alpha}^{}=\sum_{\boldsymbol{k} \bar{j}}\hat{S}_{L,j}^{\bar{j}}\left(\boldsymbol{k}\right)e^{i\boldsymbol{k}\cdot \boldsymbol{r}_j}c_{\boldsymbol{k}\bar{j}\alpha}^{},
	\end{equation}
	 where $\boldsymbol{r}_j$ is position vector of the site $j$. Using Eq.~\eqref{eq:DensityMat} and Eq.~\eqref{eq:BandOP}  in Eq.~\eqref{eq:LinearSelf}, and putting the result into the self-consistency equations, Eq.~\eqref{eq:MFSelfCon} in the main text, we obtain the LGE as  
	\begin{equation}\label{eq:LGEapp}
		{D}_{L,i j}^{\eta,\pm} =\sum_{\boldsymbol{k}\bar{i},\bar{j}}
			\sum_{rs} \Gamma_{ij}^{\eta,\pm} \mathcal{S}_{L,i j;L, sr }^{\bar{i}\bar{j},\eta,\pm}\left(\boldsymbol{k}\right) \chi_{\bar{i},\bar{j}}^\pm \left(\boldsymbol{k}\right) {D}_{L,sr}^{\eta,\pm},
		\end{equation}
	where the order parameter ${D}_{ij}^{\eta,+}={\Delta}_{i j}^{\eta}\, \left({D}_{ij}^{\eta,-}={m}_{ij}^{\eta},\, \tilde{m}_{ij}^{\eta}\right)$ encodes superconducting (particle-hole) order.  The mean-field decomposed coupling strengths $\Gamma_{ij}^{\eta,\pm}$ are listed in Table~\ref{Tab:MFInteraction}, while the susceptibilities $\chi_{\bar{i},\bar{j}}^\pm$ are given in Eq.~\eqref{eq:Susp}, both in the main text. Finally, the factor $\mathcal{S}$ contains symmetry-related information and is given by 
	%
		\begin{equation}\label{eq:SymSC}
				\begin{split}
			\mathcal{S}_{L,i j;L, sr }^{\bar{i},\bar{j},\eta,+}\left(\boldsymbol{k}\right)   = \frac{1}{4}
			\sum_{ \alpha \beta} & 
			 \left(\left[\bar{\sigma}^{\eta}\right]_{\beta \alpha} \hat{S}_{L, i}^{\bar{i}}\left (\boldsymbol{k}\right) \hat{S}_{L,j}^{\bar{j} *}\left (\boldsymbol{k}\right) e^{i\boldsymbol{k}\cdot \left(\boldsymbol{r}_j-\boldsymbol{r}_i\right)} \right. 
			 - \left. \left[\bar{\sigma}^{\eta}\right]_{\alpha \beta}\hat{S}_{L,i}^{\bar{j}*}\left (\boldsymbol{k}\right) \hat{S}_{L,j}^{\bar{i}}\left (\boldsymbol{k}\right) e^{-i \boldsymbol{k} \cdot \left(\boldsymbol{r}_j-\boldsymbol{r}_i\right)}\right)\\
			& \times \left(\left[\bar{\sigma}^{\eta}\right]_{\alpha \beta}^\dagger \hat{S}_{L,s}^{\bar{j}}\left (\boldsymbol{k}\right) \hat{S}_{L,r}^{\bar{i} *}\left (\boldsymbol{k}\right) e^{-i \boldsymbol{k} \cdot \left(\boldsymbol{r}_{s}-\boldsymbol{r}_{r}\right)} \right.  
			 - \left. \left[\bar{\sigma}^{\eta}\right]_{\beta \alpha}^\dagger\hat{S}_{L,s}^{\bar{i}*}\left (\boldsymbol{k}\right) \hat{S}_{L,r}^{\bar{j}}\left (\boldsymbol{k}\right) e^{i \boldsymbol{k} \cdot \left(\boldsymbol{r}_{s}-\boldsymbol{r}_{r}\right)} \right)
		\end{split}
	\end{equation}
\begin{equation}\label{eq:SymPH}
	\begin{split}
		\mathcal{S}_{L,i j;L, sr }^{\bar{i},\bar{j},\eta,-}\left(\boldsymbol{k}\right) & = e^{-i\boldsymbol{k}\cdot\left(\boldsymbol{r}_j-\boldsymbol{r}_i\right) }   e^{i \boldsymbol{k} \cdot \left(\boldsymbol{r}_{s}-\boldsymbol{r}_{r}\right)} 
		\hat{S}_{L,i}^{\bar{i} *}\left (\boldsymbol{k}\right) \hat{S}_{L,j}^{\bar{j}}\left (\boldsymbol{k}\right)
		\hat{S}_{L,s}^{\bar{j}*}\left (\boldsymbol{k}\right) \hat{S}_{L,r}^{\bar{i}}\left (\boldsymbol{k}\right)
	\end{split}
\end{equation}
	The exponential factors in  $\mathcal{S}_{}^{\eta,-}$  differentiates between bond-independent direct (Hartree) particle-hole order parameter, ${m}_{i j}^{\eta}$,  and bond-dependent exchange (Fock) particle-hole order parameter, $\tilde{m}_{i j}^{\eta}$. We also note that  $\mathcal{S}^{\eta,-}$ is explicitly spin independent, implying the sign of the coupling differentiates between particle-hole orders.

	From Eq.~\eqref{eq:LGEapp} we see that each order parameter on the left-hand side of the equation is a linear combination of the order parameters on the right-hand side and each term in the linear combination is weighted by $\Gamma^{\eta,\pm} \mathcal{S}^{\eta,\pm}\chi^{\eta,\pm}$. Thus, the LGE in Eq.~\eqref{eq:LGEapp} is a system of linear equations and by collating the order parameters on both sides as column vectors, $\boldsymbol{D}^{\eta,\pm}$,  the LGE can be written in the compact form $\boldsymbol{D}_{}^{\eta,\pm} = \mathbb{M}_{}^{\eta,\pm} \boldsymbol{D}_{}^{\eta,\pm}$, which is presented in Eq.~\eqref{eq:StabMat} in the main text.
\twocolumngrid
\bibliography{BIB}

\begin{thebibliography}{124}%
\makeatletter
\providecommand \@ifxundefined [1]{%
 \@ifx{#1\undefined}
}%
\providecommand \@ifnum [1]{%
 \ifnum #1\expandafter \@firstoftwo
 \else \expandafter \@secondoftwo
 \fi
}%
\providecommand \@ifx [1]{%
 \ifx #1\expandafter \@firstoftwo
 \else \expandafter \@secondoftwo
 \fi
}%
\providecommand \natexlab [1]{#1}%
\providecommand \enquote  [1]{``#1''}%
\providecommand \bibnamefont  [1]{#1}%
\providecommand \bibfnamefont [1]{#1}%
\providecommand \citenamefont [1]{#1}%
\providecommand \href@noop [0]{\@secondoftwo}%
\providecommand \href [0]{\begingroup \@sanitize@url \@href}%
\providecommand \@href[1]{\@@startlink{#1}\@@href}%
\providecommand \@@href[1]{\endgroup#1\@@endlink}%
\providecommand \@sanitize@url [0]{\catcode `\\12\catcode `\$12\catcode
  `\&12\catcode `\#12\catcode `\^12\catcode `\_12\catcode `\%12\relax}%
\providecommand \@@startlink[1]{}%
\providecommand \@@endlink[0]{}%
\providecommand \url  [0]{\begingroup\@sanitize@url \@url }%
\providecommand \@url [1]{\endgroup\@href {#1}{\urlprefix }}%
\providecommand \urlprefix  [0]{URL }%
\providecommand \Eprint [0]{\href }%
\providecommand \doibase [0]{https://doi.org/}%
\providecommand \selectlanguage [0]{\@gobble}%
\providecommand \bibinfo  [0]{\@secondoftwo}%
\providecommand \bibfield  [0]{\@secondoftwo}%
\providecommand \translation [1]{[#1]}%
\providecommand \BibitemOpen [0]{}%
\providecommand \bibitemStop [0]{}%
\providecommand \bibitemNoStop [0]{.\EOS\space}%
\providecommand \EOS [0]{\spacefactor3000\relax}%
\providecommand \BibitemShut  [1]{\csname bibitem#1\endcsname}%
\let\auto@bib@innerbib\@empty
\bibitem [{\citenamefont {Novoselov}\ \emph {et~al.}(2004)\citenamefont
  {Novoselov}, \citenamefont {Geim}, \citenamefont {Morozov}, \citenamefont
  {Jiang}, \citenamefont {Zhang}, \citenamefont {Dubonos}, \citenamefont
  {Grigorieva},\ and\ \citenamefont {Firsov}}]{Novoselov2004Electric}%
  \BibitemOpen
  \bibfield  {author} {\bibinfo {author} {\bibfnamefont {K.~S.}\ \bibnamefont
  {Novoselov}}, \bibinfo {author} {\bibfnamefont {A.~K.}\ \bibnamefont {Geim}},
  \bibinfo {author} {\bibfnamefont {S.~V.}\ \bibnamefont {Morozov}}, \bibinfo
  {author} {\bibfnamefont {D.}~\bibnamefont {Jiang}}, \bibinfo {author}
  {\bibfnamefont {Y.}~\bibnamefont {Zhang}}, \bibinfo {author} {\bibfnamefont
  {S.~V.}\ \bibnamefont {Dubonos}}, \bibinfo {author} {\bibfnamefont {I.~V.}\
  \bibnamefont {Grigorieva}},\ and\ \bibinfo {author} {\bibfnamefont {A.~A.}\
  \bibnamefont {Firsov}},\ }\bibfield  {title} {\bibinfo {title} {Electric
  field effect in atomically thin carbon films},\ }\href
  {https://doi.org/10.1126/science.1102896} {\bibfield  {journal} {\bibinfo
  {journal} {Science}\ }\textbf {\bibinfo {volume} {306}},\ \bibinfo {pages}
  {666--669} (\bibinfo {year} {2004})}\BibitemShut {NoStop}%
\bibitem [{\citenamefont {Cao}\ \emph {et~al.}(2018{\natexlab{a}})\citenamefont
  {Cao}, \citenamefont {Fatemi}, \citenamefont {Demir}, \citenamefont {Fang},
  \citenamefont {Tomarken}, \citenamefont {Luo}, \citenamefont
  {Sanchez-Yamagishi}, \citenamefont {Watanabe}, \citenamefont {Taniguchi},
  \citenamefont {Kaxiras}, \citenamefont {Ashoori},\ and\ \citenamefont
  {Jarillo-Herrero}}]{Cao2018Correlated}%
  \BibitemOpen
  \bibfield  {author} {\bibinfo {author} {\bibfnamefont {Y.}~\bibnamefont
  {Cao}}, \bibinfo {author} {\bibfnamefont {V.}~\bibnamefont {Fatemi}},
  \bibinfo {author} {\bibfnamefont {A.}~\bibnamefont {Demir}}, \bibinfo
  {author} {\bibfnamefont {S.}~\bibnamefont {Fang}}, \bibinfo {author}
  {\bibfnamefont {S.~L.}\ \bibnamefont {Tomarken}}, \bibinfo {author}
  {\bibfnamefont {J.~Y.}\ \bibnamefont {Luo}}, \bibinfo {author} {\bibfnamefont
  {J.~D.}\ \bibnamefont {Sanchez-Yamagishi}}, \bibinfo {author} {\bibfnamefont
  {K.}~\bibnamefont {Watanabe}}, \bibinfo {author} {\bibfnamefont
  {T.}~\bibnamefont {Taniguchi}}, \bibinfo {author} {\bibfnamefont
  {E.}~\bibnamefont {Kaxiras}}, \bibinfo {author} {\bibfnamefont {R.~C.}\
  \bibnamefont {Ashoori}},\ and\ \bibinfo {author} {\bibfnamefont
  {P.}~\bibnamefont {Jarillo-Herrero}},\ }\bibfield  {title} {\bibinfo {title}
  {Correlated insulator behaviour at half-filling in magic-angle graphene
  superlattices},\ }\href {https://doi.org/10.1038/nature26154} {\bibfield
  {journal} {\bibinfo  {journal} {Nature}\ }\textbf {\bibinfo {volume} {556}},\
  \bibinfo {pages} {80--84} (\bibinfo {year} {2018}{\natexlab{a}})}\BibitemShut
  {NoStop}%
\bibitem [{\citenamefont {Cao}\ \emph {et~al.}(2018{\natexlab{b}})\citenamefont
  {Cao}, \citenamefont {Fatemi}, \citenamefont {Fang}, \citenamefont
  {Watanabe}, \citenamefont {Taniguchi}, \citenamefont {Kaxiras},\ and\
  \citenamefont {Jarillo-Herrero}}]{Cao2018Unconventional}%
  \BibitemOpen
  \bibfield  {author} {\bibinfo {author} {\bibfnamefont {Y.}~\bibnamefont
  {Cao}}, \bibinfo {author} {\bibfnamefont {V.}~\bibnamefont {Fatemi}},
  \bibinfo {author} {\bibfnamefont {S.}~\bibnamefont {Fang}}, \bibinfo {author}
  {\bibfnamefont {K.}~\bibnamefont {Watanabe}}, \bibinfo {author}
  {\bibfnamefont {T.}~\bibnamefont {Taniguchi}}, \bibinfo {author}
  {\bibfnamefont {E.}~\bibnamefont {Kaxiras}},\ and\ \bibinfo {author}
  {\bibfnamefont {P.}~\bibnamefont {Jarillo-Herrero}},\ }\bibfield  {title}
  {\bibinfo {title} {Unconventional superconductivity in magic-angle graphene
  superlattices},\ }\href {https://doi.org/10.1038/nature26160} {\bibfield
  {journal} {\bibinfo  {journal} {Nature}\ }\textbf {\bibinfo {volume} {556}},\
  \bibinfo {pages} {43--50} (\bibinfo {year} {2018}{\natexlab{b}})}\BibitemShut
  {NoStop}%
\bibitem [{\citenamefont {Hao}\ \emph {et~al.}(2021)\citenamefont {Hao},
  \citenamefont {Zimmerman}, \citenamefont {Ledwith}, \citenamefont {Khalaf},
  \citenamefont {Najafabadi}, \citenamefont {Watanabe}, \citenamefont
  {Taniguchi}, \citenamefont {Vishwanath},\ and\ \citenamefont
  {Kim}}]{Hao2021Electric}%
  \BibitemOpen
  \bibfield  {author} {\bibinfo {author} {\bibfnamefont {Z.}~\bibnamefont
  {Hao}}, \bibinfo {author} {\bibfnamefont {A.~M.}\ \bibnamefont {Zimmerman}},
  \bibinfo {author} {\bibfnamefont {P.}~\bibnamefont {Ledwith}}, \bibinfo
  {author} {\bibfnamefont {E.}~\bibnamefont {Khalaf}}, \bibinfo {author}
  {\bibfnamefont {D.~H.}\ \bibnamefont {Najafabadi}}, \bibinfo {author}
  {\bibfnamefont {K.}~\bibnamefont {Watanabe}}, \bibinfo {author}
  {\bibfnamefont {T.}~\bibnamefont {Taniguchi}}, \bibinfo {author}
  {\bibfnamefont {A.}~\bibnamefont {Vishwanath}},\ and\ \bibinfo {author}
  {\bibfnamefont {P.}~\bibnamefont {Kim}},\ }\bibfield  {title} {\bibinfo
  {title} {Electric field-tunable superconductivity in alternating-twist
  magic-angle trilayer graphene},\ }\href
  {https://doi.org/10.1126/science.abg0399} {\bibfield  {journal} {\bibinfo
  {journal} {Science}\ }\textbf {\bibinfo {volume} {371}},\ \bibinfo {pages}
  {1133--1138} (\bibinfo {year} {2021})}\BibitemShut {NoStop}%
\bibitem [{\citenamefont {Park}\ \emph {et~al.}(2021)\citenamefont {Park},
  \citenamefont {Cao}, \citenamefont {Watanabe}, \citenamefont {Taniguchi},\
  and\ \citenamefont {Jarillo-Herrero}}]{Park2021Tunable}%
  \BibitemOpen
  \bibfield  {author} {\bibinfo {author} {\bibfnamefont {J.~M.}\ \bibnamefont
  {Park}}, \bibinfo {author} {\bibfnamefont {Y.}~\bibnamefont {Cao}}, \bibinfo
  {author} {\bibfnamefont {K.}~\bibnamefont {Watanabe}}, \bibinfo {author}
  {\bibfnamefont {T.}~\bibnamefont {Taniguchi}},\ and\ \bibinfo {author}
  {\bibfnamefont {P.}~\bibnamefont {Jarillo-Herrero}},\ }\bibfield  {title}
  {\bibinfo {title} {Tunable strongly coupled superconductivity in magic-angle
  twisted trilayer graphene},\ }\href
  {https://doi.org/10.1038/s41586-021-03192-0} {\bibfield  {journal} {\bibinfo
  {journal} {Nature}\ }\textbf {\bibinfo {volume} {590}},\ \bibinfo {pages}
  {249--255} (\bibinfo {year} {2021})}\BibitemShut {NoStop}%
\bibitem [{\citenamefont {Kim}\ \emph {et~al.}(2022)\citenamefont {Kim},
  \citenamefont {Choi}, \citenamefont {Lewandowski}, \citenamefont {Thomson},
  \citenamefont {Zhang}, \citenamefont {Polski}, \citenamefont {Watanabe},
  \citenamefont {Taniguchi}, \citenamefont {Alicea},\ and\ \citenamefont
  {Nadj-Perge}}]{Kim2022Evidence}%
  \BibitemOpen
  \bibfield  {author} {\bibinfo {author} {\bibfnamefont {H.}~\bibnamefont
  {Kim}}, \bibinfo {author} {\bibfnamefont {Y.}~\bibnamefont {Choi}}, \bibinfo
  {author} {\bibfnamefont {C.}~\bibnamefont {Lewandowski}}, \bibinfo {author}
  {\bibfnamefont {A.}~\bibnamefont {Thomson}}, \bibinfo {author} {\bibfnamefont
  {Y.}~\bibnamefont {Zhang}}, \bibinfo {author} {\bibfnamefont
  {R.}~\bibnamefont {Polski}}, \bibinfo {author} {\bibfnamefont
  {K.}~\bibnamefont {Watanabe}}, \bibinfo {author} {\bibfnamefont
  {T.}~\bibnamefont {Taniguchi}}, \bibinfo {author} {\bibfnamefont
  {J.}~\bibnamefont {Alicea}},\ and\ \bibinfo {author} {\bibfnamefont
  {S.}~\bibnamefont {Nadj-Perge}},\ }\bibfield  {title} {\bibinfo {title}
  {Evidence for unconventional superconductivity in twisted trilayer
  graphene},\ }\href {https://doi.org/10.1038/s41586-022-04715-z} {\bibfield
  {journal} {\bibinfo  {journal} {Nature}\ }\textbf {\bibinfo {volume} {606}},\
  \bibinfo {pages} {494--500} (\bibinfo {year} {2022})}\BibitemShut {NoStop}%
\bibitem [{\citenamefont {Bistritzer}\ and\ \citenamefont
  {MacDonald}(2011)}]{Bistritzer2011Moire}%
  \BibitemOpen
  \bibfield  {author} {\bibinfo {author} {\bibfnamefont {R.}~\bibnamefont
  {Bistritzer}}\ and\ \bibinfo {author} {\bibfnamefont {A.~H.}\ \bibnamefont
  {MacDonald}},\ }\bibfield  {title} {\bibinfo {title} {moir\'e bands in
  twisted double-layer graphene},\ }\href
  {https://doi.org/10.1073/pnas.1108174108} {\bibfield  {journal} {\bibinfo
  {journal} {Proc. Nat. Acad. Sci.}\ }\textbf {\bibinfo {volume} {108}},\
  \bibinfo {pages} {12233--12237} (\bibinfo {year} {2011})}\BibitemShut
  {NoStop}%
\bibitem [{\citenamefont {Marchenko}\ \emph {et~al.}(2018)\citenamefont
  {Marchenko}, \citenamefont {Evtushinsky}, \citenamefont {Golias},
  \citenamefont {Varykhalov}, \citenamefont {Seyller},\ and\ \citenamefont
  {Rader}}]{Marchenko2018}%
  \BibitemOpen
  \bibfield  {author} {\bibinfo {author} {\bibfnamefont {D.}~\bibnamefont
  {Marchenko}}, \bibinfo {author} {\bibfnamefont {D.~V.}\ \bibnamefont
  {Evtushinsky}}, \bibinfo {author} {\bibfnamefont {E.}~\bibnamefont {Golias}},
  \bibinfo {author} {\bibfnamefont {A.}~\bibnamefont {Varykhalov}}, \bibinfo
  {author} {\bibfnamefont {T.}~\bibnamefont {Seyller}},\ and\ \bibinfo {author}
  {\bibfnamefont {O.}~\bibnamefont {Rader}},\ }\bibfield  {title} {\bibinfo
  {title} {Extremely flat band in bilayer graphene},\ }\bibfield  {journal}
  {\bibinfo  {journal} {Sci. Adv.}\ }\textbf {\bibinfo {volume} {4}},\ \href
  {https://doi.org/10.1126/sciadv.aau0059} {10.1126/sciadv.aau0059} (\bibinfo
  {year} {2018})\BibitemShut {NoStop}%
\bibitem [{\citenamefont {Zhou}\ \emph {et~al.}(2022)\citenamefont {Zhou},
  \citenamefont {Holleis}, \citenamefont {Saito}, \citenamefont {Cohen},
  \citenamefont {Huynh}, \citenamefont {Patterson}, \citenamefont {Yang},
  \citenamefont {Taniguchi}, \citenamefont {Watanabe},\ and\ \citenamefont
  {Young}}]{Zhou2022Isospin}%
  \BibitemOpen
  \bibfield  {author} {\bibinfo {author} {\bibfnamefont {H.}~\bibnamefont
  {Zhou}}, \bibinfo {author} {\bibfnamefont {L.}~\bibnamefont {Holleis}},
  \bibinfo {author} {\bibfnamefont {Y.}~\bibnamefont {Saito}}, \bibinfo
  {author} {\bibfnamefont {L.}~\bibnamefont {Cohen}}, \bibinfo {author}
  {\bibfnamefont {W.}~\bibnamefont {Huynh}}, \bibinfo {author} {\bibfnamefont
  {C.~L.}\ \bibnamefont {Patterson}}, \bibinfo {author} {\bibfnamefont
  {F.}~\bibnamefont {Yang}}, \bibinfo {author} {\bibfnamefont {T.}~\bibnamefont
  {Taniguchi}}, \bibinfo {author} {\bibfnamefont {K.}~\bibnamefont
  {Watanabe}},\ and\ \bibinfo {author} {\bibfnamefont {A.~F.}\ \bibnamefont
  {Young}},\ }\bibfield  {title} {\bibinfo {title} {Isospin magnetism and
  spin-polarized superconductivity in {B}ernal bilayer graphene},\ }\href
  {https://doi.org/10.1126/science.abm8386} {\bibfield  {journal} {\bibinfo
  {journal} {Science}\ }\textbf {\bibinfo {volume} {375}},\ \bibinfo {pages}
  {774--778} (\bibinfo {year} {2022})}\BibitemShut {NoStop}%
\bibitem [{\citenamefont {Zhang}\ \emph {et~al.}(2023)\citenamefont {Zhang},
  \citenamefont {Polski}, \citenamefont {Thomson}, \citenamefont
  {Lantagne-Hurtubise}, \citenamefont {Lewandowski}, \citenamefont {Zhou},
  \citenamefont {Watanabe}, \citenamefont {Taniguchi}, \citenamefont {Alicea},\
  and\ \citenamefont {Nadj-Perge}}]{Zhang2023Enhanced}%
  \BibitemOpen
  \bibfield  {author} {\bibinfo {author} {\bibfnamefont {Y.}~\bibnamefont
  {Zhang}}, \bibinfo {author} {\bibfnamefont {R.}~\bibnamefont {Polski}},
  \bibinfo {author} {\bibfnamefont {A.}~\bibnamefont {Thomson}}, \bibinfo
  {author} {\bibfnamefont {{\'{E}}.}~\bibnamefont {Lantagne-Hurtubise}},
  \bibinfo {author} {\bibfnamefont {C.}~\bibnamefont {Lewandowski}}, \bibinfo
  {author} {\bibfnamefont {H.}~\bibnamefont {Zhou}}, \bibinfo {author}
  {\bibfnamefont {K.}~\bibnamefont {Watanabe}}, \bibinfo {author}
  {\bibfnamefont {T.}~\bibnamefont {Taniguchi}}, \bibinfo {author}
  {\bibfnamefont {J.}~\bibnamefont {Alicea}},\ and\ \bibinfo {author}
  {\bibfnamefont {S.}~\bibnamefont {Nadj-Perge}},\ }\bibfield  {title}
  {\bibinfo {title} {Enhanced superconductivity in spin-orbit proximitized
  bilayer graphene},\ }\href {https://doi.org/10.1038/s41586-022-05446-x}
  {\bibfield  {journal} {\bibinfo  {journal} {Nature}\ }\textbf {\bibinfo
  {volume} {613}},\ \bibinfo {pages} {268--273} (\bibinfo {year}
  {2023})}\BibitemShut {NoStop}%
\bibitem [{\citenamefont {Zhou}\ \emph
  {et~al.}(2021{\natexlab{a}})\citenamefont {Zhou}, \citenamefont {Xie},
  \citenamefont {Taniguchi}, \citenamefont {Watanabe},\ and\ \citenamefont
  {Young}}]{Zhou2021Superconductivity}%
  \BibitemOpen
  \bibfield  {author} {\bibinfo {author} {\bibfnamefont {H.}~\bibnamefont
  {Zhou}}, \bibinfo {author} {\bibfnamefont {T.}~\bibnamefont {Xie}}, \bibinfo
  {author} {\bibfnamefont {T.}~\bibnamefont {Taniguchi}}, \bibinfo {author}
  {\bibfnamefont {K.}~\bibnamefont {Watanabe}},\ and\ \bibinfo {author}
  {\bibfnamefont {A.~F.}\ \bibnamefont {Young}},\ }\bibfield  {title} {\bibinfo
  {title} {Superconductivity in rhombohedral trilayer graphene.},\ }\href
  {https://doi.org/10.1038/s41586-021-03926-0} {\bibfield  {journal} {\bibinfo
  {journal} {Nature}\ }\textbf {\bibinfo {volume} {598}},\ \bibinfo {pages}
  {434--438} (\bibinfo {year} {2021}{\natexlab{a}})}\BibitemShut {NoStop}%
\bibitem [{\citenamefont {Zhou}\ \emph
  {et~al.}(2021{\natexlab{b}})\citenamefont {Zhou}, \citenamefont {Xie},
  \citenamefont {Ghazaryan}, \citenamefont {Holder}, \citenamefont {Ehrets},
  \citenamefont {Spanton}, \citenamefont {Taniguchi}, \citenamefont {Watanabe},
  \citenamefont {Berg}, \citenamefont {Serbyn},\ and\ \citenamefont
  {Young}}]{Zhou2021Half}%
  \BibitemOpen
  \bibfield  {author} {\bibinfo {author} {\bibfnamefont {H.}~\bibnamefont
  {Zhou}}, \bibinfo {author} {\bibfnamefont {T.}~\bibnamefont {Xie}}, \bibinfo
  {author} {\bibfnamefont {A.}~\bibnamefont {Ghazaryan}}, \bibinfo {author}
  {\bibfnamefont {T.}~\bibnamefont {Holder}}, \bibinfo {author} {\bibfnamefont
  {J.~R.}\ \bibnamefont {Ehrets}}, \bibinfo {author} {\bibfnamefont {E.~M.}\
  \bibnamefont {Spanton}}, \bibinfo {author} {\bibfnamefont {T.}~\bibnamefont
  {Taniguchi}}, \bibinfo {author} {\bibfnamefont {K.}~\bibnamefont {Watanabe}},
  \bibinfo {author} {\bibfnamefont {E.}~\bibnamefont {Berg}}, \bibinfo {author}
  {\bibfnamefont {M.}~\bibnamefont {Serbyn}},\ and\ \bibinfo {author}
  {\bibfnamefont {A.~F.}\ \bibnamefont {Young}},\ }\bibfield  {title} {\bibinfo
  {title} {Half- and quarter-metals in rhombohedral trilayer graphene},\ }\href
  {https://doi.org/10.1038/s41586-021-03938-w} {\bibfield  {journal} {\bibinfo
  {journal} {Nature}\ }\textbf {\bibinfo {volume} {598}},\ \bibinfo {pages}
  {429--433} (\bibinfo {year} {2021}{\natexlab{b}})}\BibitemShut {NoStop}%
\bibitem [{\citenamefont {Po}\ \emph {et~al.}(2018)\citenamefont {Po},
  \citenamefont {Zou}, \citenamefont {Vishwanath},\ and\ \citenamefont
  {Senthil}}]{Po2018Origin}%
  \BibitemOpen
  \bibfield  {author} {\bibinfo {author} {\bibfnamefont {H.~C.}\ \bibnamefont
  {Po}}, \bibinfo {author} {\bibfnamefont {L.}~\bibnamefont {Zou}}, \bibinfo
  {author} {\bibfnamefont {A.}~\bibnamefont {Vishwanath}},\ and\ \bibinfo
  {author} {\bibfnamefont {T.}~\bibnamefont {Senthil}},\ }\bibfield  {title}
  {\bibinfo {title} {Origin of mott insulating behavior and superconductivity
  in twisted bilayer graphene},\ }\href
  {https://doi.org/10.1103/PhysRevX.8.031089} {\bibfield  {journal} {\bibinfo
  {journal} {Phys. Rev. X}\ }\textbf {\bibinfo {volume} {8}},\ \bibinfo {pages}
  {031089} (\bibinfo {year} {2018})}\BibitemShut {NoStop}%
\bibitem [{\citenamefont {Guinea}\ and\ \citenamefont
  {Walet}(2018)}]{Guinea2018Electrostatic}%
  \BibitemOpen
  \bibfield  {author} {\bibinfo {author} {\bibfnamefont {F.}~\bibnamefont
  {Guinea}}\ and\ \bibinfo {author} {\bibfnamefont {N.~R.}\ \bibnamefont
  {Walet}},\ }\bibfield  {title} {\bibinfo {title} {Electrostatic effects, band
  distortions, and superconductivity in twisted graphene bilayers},\ }\href
  {https://doi.org/10.1073/pnas.1810947115} {\bibfield  {journal} {\bibinfo
  {journal} {Proc. Nat. Acad. Sci.}\ }\textbf {\bibinfo {volume} {115}},\
  \bibinfo {pages} {13174--13179} (\bibinfo {year} {2018})}\BibitemShut
  {NoStop}%
\bibitem [{\citenamefont {Wu}\ \emph {et~al.}(2018)\citenamefont {Wu},
  \citenamefont {MacDonald},\ and\ \citenamefont {Martin}}]{wu2018theory}%
  \BibitemOpen
  \bibfield  {author} {\bibinfo {author} {\bibfnamefont {F.}~\bibnamefont
  {Wu}}, \bibinfo {author} {\bibfnamefont {A.~H.}\ \bibnamefont {MacDonald}},\
  and\ \bibinfo {author} {\bibfnamefont {I.}~\bibnamefont {Martin}},\
  }\bibfield  {title} {\bibinfo {title} {Theory of phonon-mediated
  superconductivity in twisted bilayer graphene},\ }\href
  {https://doi.org/10.1103/PhysRevLett.121.257001} {\bibfield  {journal}
  {\bibinfo  {journal} {Phys. Rev. Lett.}\ }\textbf {\bibinfo {volume} {121}},\
  \bibinfo {pages} {257001} (\bibinfo {year} {2018})}\BibitemShut {NoStop}%
\bibitem [{\citenamefont {Peltonen}\ \emph {et~al.}(2018)\citenamefont
  {Peltonen}, \citenamefont {Ojaj\"arvi},\ and\ \citenamefont
  {Heikkil\"a}}]{Peltonen2018}%
  \BibitemOpen
  \bibfield  {author} {\bibinfo {author} {\bibfnamefont {T.~J.}\ \bibnamefont
  {Peltonen}}, \bibinfo {author} {\bibfnamefont {R.}~\bibnamefont
  {Ojaj\"arvi}},\ and\ \bibinfo {author} {\bibfnamefont {T.~T.}\ \bibnamefont
  {Heikkil\"a}},\ }\bibfield  {title} {\bibinfo {title} {Mean-field theory for
  superconductivity in twisted bilayer graphene},\ }\href
  {https://doi.org/10.1103/PhysRevB.98.220504} {\bibfield  {journal} {\bibinfo
  {journal} {Phys. Rev. B}\ }\textbf {\bibinfo {volume} {98}},\ \bibinfo
  {pages} {220504} (\bibinfo {year} {2018})}\BibitemShut {NoStop}%
\bibitem [{\citenamefont {Kennes}\ \emph {et~al.}(2018)\citenamefont {Kennes},
  \citenamefont {Lischner},\ and\ \citenamefont {Karrasch}}]{kennes2018strong}%
  \BibitemOpen
  \bibfield  {author} {\bibinfo {author} {\bibfnamefont {D.~M.}\ \bibnamefont
  {Kennes}}, \bibinfo {author} {\bibfnamefont {J.}~\bibnamefont {Lischner}},\
  and\ \bibinfo {author} {\bibfnamefont {C.}~\bibnamefont {Karrasch}},\
  }\bibfield  {title} {\bibinfo {title} {Strong correlations and $d+id$
  superconductivity in twisted bilayer graphene},\ }\href
  {https://doi.org/10.1103/PhysRevB.98.241407} {\bibfield  {journal} {\bibinfo
  {journal} {Phys. Rev. B}\ }\textbf {\bibinfo {volume} {98}},\ \bibinfo
  {pages} {241407} (\bibinfo {year} {2018})}\BibitemShut {NoStop}%
\bibitem [{\citenamefont {Lu}\ \emph {et~al.}(2019)\citenamefont {Lu},
  \citenamefont {Stepanov}, \citenamefont {Yang}, \citenamefont {Xie},
  \citenamefont {Aamir}, \citenamefont {Das}, \citenamefont {Urgell},
  \citenamefont {Watanabe}, \citenamefont {Taniguchi}, \citenamefont {Zhang},
  \citenamefont {Bachtold}, \citenamefont {MacDonald},\ and\ \citenamefont
  {Efetov}}]{Lu2019Superconductors}%
  \BibitemOpen
  \bibfield  {author} {\bibinfo {author} {\bibfnamefont {X.}~\bibnamefont
  {Lu}}, \bibinfo {author} {\bibfnamefont {P.}~\bibnamefont {Stepanov}},
  \bibinfo {author} {\bibfnamefont {W.}~\bibnamefont {Yang}}, \bibinfo {author}
  {\bibfnamefont {M.}~\bibnamefont {Xie}}, \bibinfo {author} {\bibfnamefont
  {M.~A.}\ \bibnamefont {Aamir}}, \bibinfo {author} {\bibfnamefont
  {I.}~\bibnamefont {Das}}, \bibinfo {author} {\bibfnamefont {C.}~\bibnamefont
  {Urgell}}, \bibinfo {author} {\bibfnamefont {K.}~\bibnamefont {Watanabe}},
  \bibinfo {author} {\bibfnamefont {T.}~\bibnamefont {Taniguchi}}, \bibinfo
  {author} {\bibfnamefont {G.}~\bibnamefont {Zhang}}, \bibinfo {author}
  {\bibfnamefont {A.}~\bibnamefont {Bachtold}}, \bibinfo {author}
  {\bibfnamefont {A.~H.}\ \bibnamefont {MacDonald}},\ and\ \bibinfo {author}
  {\bibfnamefont {D.~K.}\ \bibnamefont {Efetov}},\ }\bibfield  {title}
  {\bibinfo {title} {Superconductors, orbital magnets and correlated states in
  magic-angle bilayer graphene},\ }\href
  {https://doi.org/10.1038/s41586-019-1695-0} {\bibfield  {journal} {\bibinfo
  {journal} {Nature}\ }\textbf {\bibinfo {volume} {574}},\ \bibinfo {pages}
  {653--657} (\bibinfo {year} {2019})}\BibitemShut {NoStop}%
\bibitem [{\citenamefont {Yankowitz}\ \emph {et~al.}(2019)\citenamefont
  {Yankowitz}, \citenamefont {Chen}, \citenamefont {Polshyn}, \citenamefont
  {Zhang}, \citenamefont {Watanabe}, \citenamefont {Taniguchi}, \citenamefont
  {Graf}, \citenamefont {Young},\ and\ \citenamefont
  {Dean}}]{Yankowitz2019Tuning}%
  \BibitemOpen
  \bibfield  {author} {\bibinfo {author} {\bibfnamefont {M.}~\bibnamefont
  {Yankowitz}}, \bibinfo {author} {\bibfnamefont {S.}~\bibnamefont {Chen}},
  \bibinfo {author} {\bibfnamefont {H.}~\bibnamefont {Polshyn}}, \bibinfo
  {author} {\bibfnamefont {Y.}~\bibnamefont {Zhang}}, \bibinfo {author}
  {\bibfnamefont {K.}~\bibnamefont {Watanabe}}, \bibinfo {author}
  {\bibfnamefont {T.}~\bibnamefont {Taniguchi}}, \bibinfo {author}
  {\bibfnamefont {D.}~\bibnamefont {Graf}}, \bibinfo {author} {\bibfnamefont
  {A.~F.}\ \bibnamefont {Young}},\ and\ \bibinfo {author} {\bibfnamefont
  {C.~R.}\ \bibnamefont {Dean}},\ }\bibfield  {title} {\bibinfo {title} {Tuning
  superconductivity in twisted bilayer graphene},\ }\href
  {https://doi.org/10.1126/science.aav1910} {\bibfield  {journal} {\bibinfo
  {journal} {Science}\ }\textbf {\bibinfo {volume} {363}},\ \bibinfo {pages}
  {1059--1064} (\bibinfo {year} {2019})}\BibitemShut {NoStop}%
\bibitem [{\citenamefont {Chen}\ \emph {et~al.}(2019)\citenamefont {Chen},
  \citenamefont {Sharpe}, \citenamefont {Gallagher}, \citenamefont {Rosen},
  \citenamefont {Fox}, \citenamefont {Jiang}, \citenamefont {Lyu},
  \citenamefont {Li}, \citenamefont {Watanabe}, \citenamefont {Taniguchi},
  \citenamefont {Jung}, \citenamefont {Shi}, \citenamefont {Goldhaber-Gordon},
  \citenamefont {Zhang},\ and\ \citenamefont {Wang}}]{Chen2019Signatures}%
  \BibitemOpen
  \bibfield  {author} {\bibinfo {author} {\bibfnamefont {G.}~\bibnamefont
  {Chen}}, \bibinfo {author} {\bibfnamefont {A.~L.}\ \bibnamefont {Sharpe}},
  \bibinfo {author} {\bibfnamefont {P.}~\bibnamefont {Gallagher}}, \bibinfo
  {author} {\bibfnamefont {I.~T.}\ \bibnamefont {Rosen}}, \bibinfo {author}
  {\bibfnamefont {E.~J.}\ \bibnamefont {Fox}}, \bibinfo {author} {\bibfnamefont
  {L.}~\bibnamefont {Jiang}}, \bibinfo {author} {\bibfnamefont
  {B.}~\bibnamefont {Lyu}}, \bibinfo {author} {\bibfnamefont {H.}~\bibnamefont
  {Li}}, \bibinfo {author} {\bibfnamefont {K.}~\bibnamefont {Watanabe}},
  \bibinfo {author} {\bibfnamefont {T.}~\bibnamefont {Taniguchi}}, \bibinfo
  {author} {\bibfnamefont {J.}~\bibnamefont {Jung}}, \bibinfo {author}
  {\bibfnamefont {Z.}~\bibnamefont {Shi}}, \bibinfo {author} {\bibfnamefont
  {D.}~\bibnamefont {Goldhaber-Gordon}}, \bibinfo {author} {\bibfnamefont
  {Y.}~\bibnamefont {Zhang}},\ and\ \bibinfo {author} {\bibfnamefont
  {F.}~\bibnamefont {Wang}},\ }\bibfield  {title} {\bibinfo {title} {Signatures
  of tunable superconductivity in a trilayer graphene moir\'e superlattice},\
  }\href {https://doi.org/10.1038/s41586-019-1393-y} {\bibfield  {journal}
  {\bibinfo  {journal} {Nature}\ }\textbf {\bibinfo {volume} {572}},\ \bibinfo
  {pages} {215--219} (\bibinfo {year} {2019})}\BibitemShut {NoStop}%
\bibitem [{\citenamefont {Alidoust}\ \emph {et~al.}(2019)\citenamefont
  {Alidoust}, \citenamefont {Willatzen},\ and\ \citenamefont
  {Jauho}}]{Alidoust2019Symmetry}%
  \BibitemOpen
  \bibfield  {author} {\bibinfo {author} {\bibfnamefont {M.}~\bibnamefont
  {Alidoust}}, \bibinfo {author} {\bibfnamefont {M.}~\bibnamefont
  {Willatzen}},\ and\ \bibinfo {author} {\bibfnamefont {A.-P.}\ \bibnamefont
  {Jauho}},\ }\bibfield  {title} {\bibinfo {title} {Symmetry of superconducting
  correlations in displaced bilayers of graphene},\ }\href
  {https://doi.org/10.1103/PhysRevB.99.155413} {\bibfield  {journal} {\bibinfo
  {journal} {Phys. Rev. B}\ }\textbf {\bibinfo {volume} {99}},\ \bibinfo
  {pages} {155413} (\bibinfo {year} {2019})}\BibitemShut {NoStop}%
\bibitem [{\citenamefont {{Balents}}\ \emph {et~al.}(2020)\citenamefont
  {{Balents}}, \citenamefont {{Dean}}, \citenamefont {{Efetov}},\ and\
  \citenamefont {{Young}}}]{Balents2020Superconductivity}%
  \BibitemOpen
  \bibfield  {author} {\bibinfo {author} {\bibfnamefont {L.}~\bibnamefont
  {{Balents}}}, \bibinfo {author} {\bibfnamefont {C.~R.}\ \bibnamefont
  {{Dean}}}, \bibinfo {author} {\bibfnamefont {D.~K.}\ \bibnamefont
  {{Efetov}}},\ and\ \bibinfo {author} {\bibfnamefont {A.~F.}\ \bibnamefont
  {{Young}}},\ }\bibfield  {title} {\bibinfo {title} {Superconductivity and
  strong correlations in moir{\'e} flat bands},\ }\href
  {https://doi.org/10.1038/s41567-020-0906-9} {\bibfield  {journal} {\bibinfo
  {journal} {Nat. Phys.}\ }\textbf {\bibinfo {volume} {16}},\ \bibinfo {pages}
  {725--733} (\bibinfo {year} {2020})}\BibitemShut {NoStop}%
\bibitem [{\citenamefont {Andrei}\ and\ \citenamefont
  {MacDonald}(2020)}]{Andrei2020Graphene}%
  \BibitemOpen
  \bibfield  {author} {\bibinfo {author} {\bibfnamefont {E.~Y.}\ \bibnamefont
  {Andrei}}\ and\ \bibinfo {author} {\bibfnamefont {A.~H.}\ \bibnamefont
  {MacDonald}},\ }\bibfield  {title} {\bibinfo {title} {Graphene bilayers with
  a twist},\ }\href {https://doi.org/10.1038/s41563-020-00840-0} {\bibfield
  {journal} {\bibinfo  {journal} {Nat. Mater.}\ }\textbf {\bibinfo {volume}
  {19}},\ \bibinfo {pages} {1265--1275} (\bibinfo {year} {2020})}\BibitemShut
  {NoStop}%
\bibitem [{\citenamefont {Christos}\ \emph {et~al.}(2020)\citenamefont
  {Christos}, \citenamefont {Sachdev},\ and\ \citenamefont
  {Scheurer}}]{Christos2020Superconductivity}%
  \BibitemOpen
  \bibfield  {author} {\bibinfo {author} {\bibfnamefont {M.}~\bibnamefont
  {Christos}}, \bibinfo {author} {\bibfnamefont {S.}~\bibnamefont {Sachdev}},\
  and\ \bibinfo {author} {\bibfnamefont {M.~S.}\ \bibnamefont {Scheurer}},\
  }\bibfield  {title} {\bibinfo {title} {Superconductivity, correlated
  insulators, and wess{\textendash}zumino{\textendash}witten terms in twisted
  bilayer graphene},\ }\href {https://doi.org/10.1073/pnas.2014691117}
  {\bibfield  {journal} {\bibinfo  {journal} {Proc. Nat. Acad. Sci.}\ }\textbf
  {\bibinfo {volume} {117}},\ \bibinfo {pages} {29543--29554} (\bibinfo {year}
  {2020})}\BibitemShut {NoStop}%
\bibitem [{\citenamefont {Chichinadze}\ \emph {et~al.}(2020)\citenamefont
  {Chichinadze}, \citenamefont {Classen},\ and\ \citenamefont
  {Chubukov}}]{Chichinadze2020Nematic}%
  \BibitemOpen
  \bibfield  {author} {\bibinfo {author} {\bibfnamefont {D.~V.}\ \bibnamefont
  {Chichinadze}}, \bibinfo {author} {\bibfnamefont {L.}~\bibnamefont
  {Classen}},\ and\ \bibinfo {author} {\bibfnamefont {A.~V.}\ \bibnamefont
  {Chubukov}},\ }\bibfield  {title} {\bibinfo {title} {Nematic
  superconductivity in twisted bilayer graphene},\ }\href
  {https://doi.org/10.1103/PhysRevB.101.224513} {\bibfield  {journal} {\bibinfo
   {journal} {Phys. Rev. B}\ }\textbf {\bibinfo {volume} {101}},\ \bibinfo
  {pages} {224513} (\bibinfo {year} {2020})}\BibitemShut {NoStop}%
\bibitem [{\citenamefont {Wu}\ \emph {et~al.}(2020)\citenamefont {Wu},
  \citenamefont {Hanke}, \citenamefont {Fink}, \citenamefont {Klett},\ and\
  \citenamefont {Thomale}}]{Wu2020Harmonic}%
  \BibitemOpen
  \bibfield  {author} {\bibinfo {author} {\bibfnamefont {X.}~\bibnamefont
  {Wu}}, \bibinfo {author} {\bibfnamefont {W.}~\bibnamefont {Hanke}}, \bibinfo
  {author} {\bibfnamefont {M.}~\bibnamefont {Fink}}, \bibinfo {author}
  {\bibfnamefont {M.}~\bibnamefont {Klett}},\ and\ \bibinfo {author}
  {\bibfnamefont {R.}~\bibnamefont {Thomale}},\ }\bibfield  {title} {\bibinfo
  {title} {Harmonic fingerprint of unconventional superconductivity in twisted
  bilayer graphene},\ }\href {https://doi.org/10.1103/PhysRevB.101.134517}
  {\bibfield  {journal} {\bibinfo  {journal} {Phys. Rev. B}\ }\textbf {\bibinfo
  {volume} {101}},\ \bibinfo {pages} {134517} (\bibinfo {year}
  {2020})}\BibitemShut {NoStop}%
\bibitem [{\citenamefont {Fischer}\ \emph {et~al.}(2021)\citenamefont
  {Fischer}, \citenamefont {Klebl}, \citenamefont {Honerkamp},\ and\
  \citenamefont {Kennes}}]{Fischer2021}%
  \BibitemOpen
  \bibfield  {author} {\bibinfo {author} {\bibfnamefont {A.}~\bibnamefont
  {Fischer}}, \bibinfo {author} {\bibfnamefont {L.}~\bibnamefont {Klebl}},
  \bibinfo {author} {\bibfnamefont {C.}~\bibnamefont {Honerkamp}},\ and\
  \bibinfo {author} {\bibfnamefont {D.~M.}\ \bibnamefont {Kennes}},\ }\bibfield
   {title} {\bibinfo {title} {Spin-fluctuation-induced pairing in twisted
  bilayer graphene},\ }\href {https://doi.org/10.1103/physrevb.103.l041103}
  {\bibfield  {journal} {\bibinfo  {journal} {Phys. Rev. B}\ }\textbf {\bibinfo
  {volume} {103}},\ \bibinfo {pages} {l041103} (\bibinfo {year}
  {2021})}\BibitemShut {NoStop}%
\bibitem [{\citenamefont {Oh}\ \emph {et~al.}(2021)\citenamefont {Oh},
  \citenamefont {Nuckolls}, \citenamefont {Wong}, \citenamefont {Lee},
  \citenamefont {Liu}, \citenamefont {Watanabe}, \citenamefont {Taniguchi},\
  and\ \citenamefont {Yazdani}}]{Oh2021Evidence}%
  \BibitemOpen
  \bibfield  {author} {\bibinfo {author} {\bibfnamefont {M.}~\bibnamefont
  {Oh}}, \bibinfo {author} {\bibfnamefont {K.~P.}\ \bibnamefont {Nuckolls}},
  \bibinfo {author} {\bibfnamefont {D.}~\bibnamefont {Wong}}, \bibinfo {author}
  {\bibfnamefont {R.~L.}\ \bibnamefont {Lee}}, \bibinfo {author} {\bibfnamefont
  {X.}~\bibnamefont {Liu}}, \bibinfo {author} {\bibfnamefont {K.}~\bibnamefont
  {Watanabe}}, \bibinfo {author} {\bibfnamefont {T.}~\bibnamefont
  {Taniguchi}},\ and\ \bibinfo {author} {\bibfnamefont {A.}~\bibnamefont
  {Yazdani}},\ }\bibfield  {title} {\bibinfo {title} {Evidence for
  unconventional superconductivity in twisted bilayer graphene},\ }\href
  {https://doi.org/10.1038/s41586-021-04121-x} {\bibfield  {journal} {\bibinfo
  {journal} {Nature}\ }\textbf {\bibinfo {volume} {600}},\ \bibinfo {pages}
  {240--245} (\bibinfo {year} {2021})}\BibitemShut {NoStop}%
\bibitem [{\citenamefont {Cao}\ \emph {et~al.}(2021)\citenamefont {Cao},
  \citenamefont {Rodan-Legrain}, \citenamefont {Park}, \citenamefont {Yuan},
  \citenamefont {Watanabe}, \citenamefont {Taniguchi}, \citenamefont
  {Fernandes}, \citenamefont {Fu},\ and\ \citenamefont
  {Jarillo-Herrero}}]{Cao2021Nematicity}%
  \BibitemOpen
  \bibfield  {author} {\bibinfo {author} {\bibfnamefont {Y.}~\bibnamefont
  {Cao}}, \bibinfo {author} {\bibfnamefont {D.}~\bibnamefont {Rodan-Legrain}},
  \bibinfo {author} {\bibfnamefont {J.~M.}\ \bibnamefont {Park}}, \bibinfo
  {author} {\bibfnamefont {N.~F.~Q.}\ \bibnamefont {Yuan}}, \bibinfo {author}
  {\bibfnamefont {K.}~\bibnamefont {Watanabe}}, \bibinfo {author}
  {\bibfnamefont {T.}~\bibnamefont {Taniguchi}}, \bibinfo {author}
  {\bibfnamefont {R.~M.}\ \bibnamefont {Fernandes}}, \bibinfo {author}
  {\bibfnamefont {L.}~\bibnamefont {Fu}},\ and\ \bibinfo {author}
  {\bibfnamefont {P.}~\bibnamefont {Jarillo-Herrero}},\ }\bibfield  {title}
  {\bibinfo {title} {Nematicity and competing orders in superconducting
  magic-angle graphene},\ }\href {https://doi.org/10.1126/science.abc2836}
  {\bibfield  {journal} {\bibinfo  {journal} {Science}\ }\textbf {\bibinfo
  {volume} {372}},\ \bibinfo {pages} {264--271} (\bibinfo {year}
  {2021})}\BibitemShut {NoStop}%
\bibitem [{\citenamefont {Yu}\ \emph {et~al.}(2021)\citenamefont {Yu},
  \citenamefont {Kennes}, \citenamefont {Rubio},\ and\ \citenamefont
  {Sentef}}]{Yu2021Nematicity}%
  \BibitemOpen
  \bibfield  {author} {\bibinfo {author} {\bibfnamefont {T.}~\bibnamefont
  {Yu}}, \bibinfo {author} {\bibfnamefont {D.~M.}\ \bibnamefont {Kennes}},
  \bibinfo {author} {\bibfnamefont {A.}~\bibnamefont {Rubio}},\ and\ \bibinfo
  {author} {\bibfnamefont {M.~A.}\ \bibnamefont {Sentef}},\ }\bibfield  {title}
  {\bibinfo {title} {Nematicity arising from a chiral superconducting ground
  state in magic-angle twisted bilayer graphene under in-plane magnetic
  fields},\ }\href {https://doi.org/10.1103/PhysRevLett.127.127001} {\bibfield
  {journal} {\bibinfo  {journal} {Phys. Rev. Lett.}\ }\textbf {\bibinfo
  {volume} {127}},\ \bibinfo {pages} {127001} (\bibinfo {year}
  {2021})}\BibitemShut {NoStop}%
\bibitem [{\citenamefont {Khalaf}\ \emph {et~al.}(2021)\citenamefont {Khalaf},
  \citenamefont {Chatterjee}, \citenamefont {Bultinck}, \citenamefont
  {Zaletel},\ and\ \citenamefont {Vishwanath}}]{Khalaf2021Charged}%
  \BibitemOpen
  \bibfield  {author} {\bibinfo {author} {\bibfnamefont {E.}~\bibnamefont
  {Khalaf}}, \bibinfo {author} {\bibfnamefont {S.}~\bibnamefont {Chatterjee}},
  \bibinfo {author} {\bibfnamefont {N.}~\bibnamefont {Bultinck}}, \bibinfo
  {author} {\bibfnamefont {M.~P.}\ \bibnamefont {Zaletel}},\ and\ \bibinfo
  {author} {\bibfnamefont {A.}~\bibnamefont {Vishwanath}},\ }\bibfield  {title}
  {\bibinfo {title} {Charged skyrmions and topological origin of
  superconductivity in magic-angle graphene},\ }\bibfield  {journal} {\bibinfo
  {journal} {Sci. Adv.}\ }\textbf {\bibinfo {volume} {7}},\ \href
  {https://doi.org/10.1126/sciadv.abf5299} {10.1126/sciadv.abf5299} (\bibinfo
  {year} {2021})\BibitemShut {NoStop}%
\bibitem [{\citenamefont {Cea}\ and\ \citenamefont
  {Guinea}(2021)}]{cea2021coulomb}%
  \BibitemOpen
  \bibfield  {author} {\bibinfo {author} {\bibfnamefont {T.}~\bibnamefont
  {Cea}}\ and\ \bibinfo {author} {\bibfnamefont {F.}~\bibnamefont {Guinea}},\
  }\bibfield  {title} {\bibinfo {title} {Coulomb interaction, phonons, and
  superconductivity in twisted bilayer graphene},\ }\bibfield  {journal}
  {\bibinfo  {journal} {Proc. Nat. Acad. Sci.}\ }\textbf {\bibinfo {volume}
  {118}},\ \href {https://doi.org/10.1073/pnas.2107874118}
  {10.1073/pnas.2107874118} (\bibinfo {year} {2021})\BibitemShut {NoStop}%
\bibitem [{\citenamefont {Chou}\ \emph
  {et~al.}(2021{\natexlab{a}})\citenamefont {Chou}, \citenamefont {Wu},
  \citenamefont {Sau},\ and\ \citenamefont {Das~Sarma}}]{chou2021correlation}%
  \BibitemOpen
  \bibfield  {author} {\bibinfo {author} {\bibfnamefont {Y.-Z.}\ \bibnamefont
  {Chou}}, \bibinfo {author} {\bibfnamefont {F.}~\bibnamefont {Wu}}, \bibinfo
  {author} {\bibfnamefont {J.~D.}\ \bibnamefont {Sau}},\ and\ \bibinfo {author}
  {\bibfnamefont {S.}~\bibnamefont {Das~Sarma}},\ }\bibfield  {title} {\bibinfo
  {title} {Correlation-induced triplet pairing superconductivity in
  graphene-based moir\'e systems},\ }\href
  {https://doi.org/10.1103/PhysRevLett.127.217001} {\bibfield  {journal}
  {\bibinfo  {journal} {Phys. Rev. Lett.}\ }\textbf {\bibinfo {volume} {127}},\
  \bibinfo {pages} {217001} (\bibinfo {year} {2021}{\natexlab{a}})}\BibitemShut
  {NoStop}%
\bibitem [{\citenamefont {L{\"o}thman}\ \emph {et~al.}(2022)\citenamefont
  {L{\"o}thman}, \citenamefont {Schmidt}, \citenamefont {Parhizgar},\ and\
  \citenamefont {Black-Schaffer}}]{lothman2022nematic}%
  \BibitemOpen
  \bibfield  {author} {\bibinfo {author} {\bibfnamefont {T.}~\bibnamefont
  {L{\"o}thman}}, \bibinfo {author} {\bibfnamefont {J.}~\bibnamefont
  {Schmidt}}, \bibinfo {author} {\bibfnamefont {F.}~\bibnamefont {Parhizgar}},\
  and\ \bibinfo {author} {\bibfnamefont {A.~M.}\ \bibnamefont
  {Black-Schaffer}},\ }\bibfield  {title} {\bibinfo {title} {Nematic
  superconductivity in magic-angle twisted bilayer graphene from atomistic
  modeling},\ }\href {https://www.nature.com/articles/s42005-022-00860-z}
  {\bibfield  {journal} {\bibinfo  {journal} {Commun. Phys.}\ }\textbf
  {\bibinfo {volume} {5}},\ \bibinfo {pages} {1--11} (\bibinfo {year}
  {2022})}\BibitemShut {NoStop}%
\bibitem [{\citenamefont {Fischer}\ \emph {et~al.}(2022)\citenamefont
  {Fischer}, \citenamefont {Goodwin}, \citenamefont {Mostofi}, \citenamefont
  {Lischner}, \citenamefont {Kennes},\ and\ \citenamefont
  {Klebl}}]{Fischer2022Unconventional}%
  \BibitemOpen
  \bibfield  {author} {\bibinfo {author} {\bibfnamefont {A.}~\bibnamefont
  {Fischer}}, \bibinfo {author} {\bibfnamefont {Z.~A.~H.}\ \bibnamefont
  {Goodwin}}, \bibinfo {author} {\bibfnamefont {A.~A.}\ \bibnamefont
  {Mostofi}}, \bibinfo {author} {\bibfnamefont {J.}~\bibnamefont {Lischner}},
  \bibinfo {author} {\bibfnamefont {D.~M.}\ \bibnamefont {Kennes}},\ and\
  \bibinfo {author} {\bibfnamefont {L.}~\bibnamefont {Klebl}},\ }\bibfield
  {title} {\bibinfo {title} {Unconventional superconductivity in magic-angle
  twisted trilayer graphene},\ }\bibfield  {journal} {\bibinfo  {journal} {npj
  Quantum Mater.}\ }\textbf {\bibinfo {volume} {7}},\ \href
  {https://doi.org/10.1038/s41535-021-00410-w} {10.1038/s41535-021-00410-w}
  (\bibinfo {year} {2022})\BibitemShut {NoStop}%
\bibitem [{\citenamefont {Chatterjee}\ \emph {et~al.}(2022)\citenamefont
  {Chatterjee}, \citenamefont {Wang}, \citenamefont {Berg},\ and\ \citenamefont
  {Zaletel}}]{chatterjee2022intervalley}%
  \BibitemOpen
  \bibfield  {author} {\bibinfo {author} {\bibfnamefont {S.}~\bibnamefont
  {Chatterjee}}, \bibinfo {author} {\bibfnamefont {T.}~\bibnamefont {Wang}},
  \bibinfo {author} {\bibfnamefont {E.}~\bibnamefont {Berg}},\ and\ \bibinfo
  {author} {\bibfnamefont {M.~P.}\ \bibnamefont {Zaletel}},\ }\bibfield
  {title} {\bibinfo {title} {Inter-valley coherent order and isospin
  fluctuation mediated superconductivity in rhombohedral trilayer graphene},\
  }\bibfield  {journal} {\bibinfo  {journal} {Nat. Commun.}\ }\textbf {\bibinfo
  {volume} {13}},\ \href {https://doi.org/10.1038/s41467-022-33561-w}
  {10.1038/s41467-022-33561-w} (\bibinfo {year} {2022})\BibitemShut {NoStop}%
\bibitem [{\citenamefont {Chou}\ \emph
  {et~al.}(2021{\natexlab{b}})\citenamefont {Chou}, \citenamefont {Wu},
  \citenamefont {Sau},\ and\ \citenamefont {Das~Sarma}}]{chou2021Acoustic}%
  \BibitemOpen
  \bibfield  {author} {\bibinfo {author} {\bibfnamefont {Y.-Z.}\ \bibnamefont
  {Chou}}, \bibinfo {author} {\bibfnamefont {F.}~\bibnamefont {Wu}}, \bibinfo
  {author} {\bibfnamefont {J.~D.}\ \bibnamefont {Sau}},\ and\ \bibinfo {author}
  {\bibfnamefont {S.}~\bibnamefont {Das~Sarma}},\ }\bibfield  {title} {\bibinfo
  {title} {Acoustic-phonon-mediated superconductivity in rhombohedral trilayer
  graphene},\ }\href {https://doi.org/10.1103/PhysRevLett.127.187001}
  {\bibfield  {journal} {\bibinfo  {journal} {Phys. Rev. Lett.}\ }\textbf
  {\bibinfo {volume} {127}},\ \bibinfo {pages} {187001} (\bibinfo {year}
  {2021}{\natexlab{b}})}\BibitemShut {NoStop}%
\bibitem [{\citenamefont {You}\ and\ \citenamefont
  {Vishwanath}(2022)}]{you2021kohnluttinger}%
  \BibitemOpen
  \bibfield  {author} {\bibinfo {author} {\bibfnamefont {Y.-Z.}\ \bibnamefont
  {You}}\ and\ \bibinfo {author} {\bibfnamefont {A.}~\bibnamefont
  {Vishwanath}},\ }\bibfield  {title} {\bibinfo {title} {Kohn-luttinger
  superconductivity and intervalley coherence in rhombohedral trilayer
  graphene},\ }\href {https://doi.org/10.1103/PhysRevB.105.134524} {\bibfield
  {journal} {\bibinfo  {journal} {Phys. Rev. B}\ }\textbf {\bibinfo {volume}
  {105}},\ \bibinfo {pages} {134524} (\bibinfo {year} {2022})}\BibitemShut
  {NoStop}%
\bibitem [{\citenamefont {Ghazaryan}\ \emph {et~al.}(2021)\citenamefont
  {Ghazaryan}, \citenamefont {Holder}, \citenamefont {Serbyn},\ and\
  \citenamefont {Berg}}]{ghazaryan2021unconventional}%
  \BibitemOpen
  \bibfield  {author} {\bibinfo {author} {\bibfnamefont {A.}~\bibnamefont
  {Ghazaryan}}, \bibinfo {author} {\bibfnamefont {T.}~\bibnamefont {Holder}},
  \bibinfo {author} {\bibfnamefont {M.}~\bibnamefont {Serbyn}},\ and\ \bibinfo
  {author} {\bibfnamefont {E.}~\bibnamefont {Berg}},\ }\bibfield  {title}
  {\bibinfo {title} {Unconventional superconductivity in systems with annular
  {F}ermi surfaces: Application to rhombohedral trilayer graphene},\ }\href
  {https://doi.org/10.1103/PhysRevLett.127.247001} {\bibfield  {journal}
  {\bibinfo  {journal} {Phys. Rev. Lett.}\ }\textbf {\bibinfo {volume} {127}},\
  \bibinfo {pages} {247001} (\bibinfo {year} {2021})}\BibitemShut {NoStop}%
\bibitem [{\citenamefont {Szab\'o}\ and\ \citenamefont
  {Roy}(2022)}]{Szabo2022Metals}%
  \BibitemOpen
  \bibfield  {author} {\bibinfo {author} {\bibfnamefont {A.~L.}\ \bibnamefont
  {Szab\'o}}\ and\ \bibinfo {author} {\bibfnamefont {B.}~\bibnamefont {Roy}},\
  }\bibfield  {title} {\bibinfo {title} {Metals, fractional metals, and
  superconductivity in rhombohedral trilayer graphene},\ }\href
  {https://doi.org/10.1103/PhysRevB.105.L081407} {\bibfield  {journal}
  {\bibinfo  {journal} {Phys. Rev. B}\ }\textbf {\bibinfo {volume} {105}},\
  \bibinfo {pages} {L081407} (\bibinfo {year} {2022})}\BibitemShut {NoStop}%
\bibitem [{\citenamefont {Cea}\ \emph {et~al.}(2022)\citenamefont {Cea},
  \citenamefont {Pantale{\'{o}}n}, \citenamefont {Phong},\ and\ \citenamefont
  {Guinea}}]{Cea2022Superconductivity}%
  \BibitemOpen
  \bibfield  {author} {\bibinfo {author} {\bibfnamefont {T.}~\bibnamefont
  {Cea}}, \bibinfo {author} {\bibfnamefont {P.~A.}\ \bibnamefont
  {Pantale{\'{o}}n}}, \bibinfo {author} {\bibfnamefont {V.~T.}\ \bibnamefont
  {Phong}},\ and\ \bibinfo {author} {\bibfnamefont {F.}~\bibnamefont
  {Guinea}},\ }\bibfield  {title} {\bibinfo {title} {Superconductivity from
  repulsive interactions in rhombohedral trilayer graphene: A
  {K}ohn-{L}uttinger-like mechanism},\ }\href
  {https://doi.org/10.1103/physrevb.105.075432} {\bibfield  {journal} {\bibinfo
   {journal} {Phys. Rev. B}\ }\textbf {\bibinfo {volume} {105}},\ \bibinfo
  {pages} {075432} (\bibinfo {year} {2022})}\BibitemShut {NoStop}%
\bibitem [{\citenamefont {Lu}\ \emph {et~al.}(2022)\citenamefont {Lu},
  \citenamefont {Wang}, \citenamefont {Chatterjee},\ and\ \citenamefont
  {You}}]{Lu2022Correlated}%
  \BibitemOpen
  \bibfield  {author} {\bibinfo {author} {\bibfnamefont {D.-C.}\ \bibnamefont
  {Lu}}, \bibinfo {author} {\bibfnamefont {T.}~\bibnamefont {Wang}}, \bibinfo
  {author} {\bibfnamefont {S.}~\bibnamefont {Chatterjee}},\ and\ \bibinfo
  {author} {\bibfnamefont {Y.-Z.}\ \bibnamefont {You}},\ }\bibfield  {title}
  {\bibinfo {title} {Correlated metals and unconventional superconductivity in
  rhombohedral trilayer graphene: A renormalization group analysis},\ }\href
  {https://doi.org/10.1103/PhysRevB.106.155115} {\bibfield  {journal} {\bibinfo
   {journal} {Phys. Rev. B}\ }\textbf {\bibinfo {volume} {106}},\ \bibinfo
  {pages} {155115} (\bibinfo {year} {2022})}\BibitemShut {NoStop}%
\bibitem [{\citenamefont {Ghazaryan}\ \emph {et~al.}(2023)\citenamefont
  {Ghazaryan}, \citenamefont {Holder}, \citenamefont {Berg},\ and\
  \citenamefont {Serbyn}}]{Ghazaryan2023Multilayer}%
  \BibitemOpen
  \bibfield  {author} {\bibinfo {author} {\bibfnamefont {A.}~\bibnamefont
  {Ghazaryan}}, \bibinfo {author} {\bibfnamefont {T.}~\bibnamefont {Holder}},
  \bibinfo {author} {\bibfnamefont {E.}~\bibnamefont {Berg}},\ and\ \bibinfo
  {author} {\bibfnamefont {M.}~\bibnamefont {Serbyn}},\ }\bibfield  {title}
  {\bibinfo {title} {Multilayer graphenes as a platform for interaction-driven
  physics and topological superconductivity},\ }\href
  {https://doi.org/10.1103/PhysRevB.107.104502} {\bibfield  {journal} {\bibinfo
   {journal} {Phys. Rev. B}\ }\textbf {\bibinfo {volume} {107}},\ \bibinfo
  {pages} {104502} (\bibinfo {year} {2023})}\BibitemShut {NoStop}%
\bibitem [{\citenamefont {Xu}\ and\ \citenamefont
  {Balents}(2018)}]{Xu2018Topological}%
  \BibitemOpen
  \bibfield  {author} {\bibinfo {author} {\bibfnamefont {C.}~\bibnamefont
  {Xu}}\ and\ \bibinfo {author} {\bibfnamefont {L.}~\bibnamefont {Balents}},\
  }\bibfield  {title} {\bibinfo {title} {Topological superconductivity in
  twisted multilayer graphene},\ }\href
  {https://doi.org/10.1103/PhysRevLett.121.087001} {\bibfield  {journal}
  {\bibinfo  {journal} {Phys. Rev. Lett.}\ }\textbf {\bibinfo {volume} {121}},\
  \bibinfo {pages} {087001} (\bibinfo {year} {2018})}\BibitemShut {NoStop}%
\bibitem [{\citenamefont {Durajski}\ \emph {et~al.}(2019)\citenamefont
  {Durajski}, \citenamefont {Skoczylas},\ and\ \citenamefont
  {Szcz\c{e}\'sniak}}]{Durajski2019Superconductivity}%
  \BibitemOpen
  \bibfield  {author} {\bibinfo {author} {\bibfnamefont {A.~P.}\ \bibnamefont
  {Durajski}}, \bibinfo {author} {\bibfnamefont {K.~M.}\ \bibnamefont
  {Skoczylas}},\ and\ \bibinfo {author} {\bibfnamefont {R.~a.}\ \bibnamefont
  {Szcz\c{e}\'sniak}},\ }\bibfield  {title} {\bibinfo {title}
  {Superconductivity in bilayer graphene intercalated with alkali and alkaline
  earth metals},\ }\href {https://doi.org/10.1039/C9CP00176J} {\bibfield
  {journal} {\bibinfo  {journal} {Phys. Chem. Chem. Phys.}\ }\textbf {\bibinfo
  {volume} {21}},\ \bibinfo {pages} {5925--5931} (\bibinfo {year}
  {2019})}\BibitemShut {NoStop}%
\bibitem [{\citenamefont {Park}\ \emph {et~al.}(2022)\citenamefont {Park},
  \citenamefont {Cao}, \citenamefont {Xia}, \citenamefont {Sun}, \citenamefont
  {Watanabe}, \citenamefont {Taniguchi},\ and\ \citenamefont
  {Jarillo-Herrero}}]{Park2021Magic}%
  \BibitemOpen
  \bibfield  {author} {\bibinfo {author} {\bibfnamefont {J.~M.}\ \bibnamefont
  {Park}}, \bibinfo {author} {\bibfnamefont {Y.}~\bibnamefont {Cao}}, \bibinfo
  {author} {\bibfnamefont {L.}~\bibnamefont {Xia}}, \bibinfo {author}
  {\bibfnamefont {S.}~\bibnamefont {Sun}}, \bibinfo {author} {\bibfnamefont
  {K.}~\bibnamefont {Watanabe}}, \bibinfo {author} {\bibfnamefont
  {T.}~\bibnamefont {Taniguchi}},\ and\ \bibinfo {author} {\bibfnamefont
  {P.}~\bibnamefont {Jarillo-Herrero}},\ }\bibfield  {title} {\bibinfo {title}
  {Magic-angle multilayer graphene:{ A} robust family of moir\'e
  superconductors},\ }\href {https://doi.org/10.1038/s41563-022-01287-1}
  {\bibfield  {journal} {\bibinfo  {journal} {Nat. Mater.}\ }\textbf {\bibinfo
  {volume} {21}},\ \bibinfo {pages} {877–883} (\bibinfo {year}
  {2022})}\BibitemShut {NoStop}%
\bibitem [{\citenamefont {Chou}\ \emph
  {et~al.}(2022{\natexlab{a}})\citenamefont {Chou}, \citenamefont {Wu},
  \citenamefont {Sau},\ and\ \citenamefont {Das~Sarma}}]{Chou2022B}%
  \BibitemOpen
  \bibfield  {author} {\bibinfo {author} {\bibfnamefont {Y.-Z.}\ \bibnamefont
  {Chou}}, \bibinfo {author} {\bibfnamefont {F.}~\bibnamefont {Wu}}, \bibinfo
  {author} {\bibfnamefont {J.~D.}\ \bibnamefont {Sau}},\ and\ \bibinfo {author}
  {\bibfnamefont {S.}~\bibnamefont {Das~Sarma}},\ }\bibfield  {title} {\bibinfo
  {title} {Acoustic-phonon-mediated superconductivity in {B}ernal bilayer
  graphene},\ }\href {https://doi.org/10.1103/PhysRevB.105.L100503} {\bibfield
  {journal} {\bibinfo  {journal} {Phys. Rev. B}\ }\textbf {\bibinfo {volume}
  {105}},\ \bibinfo {pages} {L100503} (\bibinfo {year}
  {2022}{\natexlab{a}})}\BibitemShut {NoStop}%
\bibitem [{\citenamefont {Chou}\ \emph
  {et~al.}(2022{\natexlab{b}})\citenamefont {Chou}, \citenamefont {Wu},
  \citenamefont {Sau},\ and\ \citenamefont {Das~Sarma}}]{Chou2022C}%
  \BibitemOpen
  \bibfield  {author} {\bibinfo {author} {\bibfnamefont {Y.-Z.}\ \bibnamefont
  {Chou}}, \bibinfo {author} {\bibfnamefont {F.}~\bibnamefont {Wu}}, \bibinfo
  {author} {\bibfnamefont {J.~D.}\ \bibnamefont {Sau}},\ and\ \bibinfo {author}
  {\bibfnamefont {S.}~\bibnamefont {Das~Sarma}},\ }\bibfield  {title} {\bibinfo
  {title} {Acoustic-phonon-mediated superconductivity in moir\'eless graphene
  multilayers},\ }\href {https://doi.org/10.1103/PhysRevB.106.024507}
  {\bibfield  {journal} {\bibinfo  {journal} {Phys. Rev. B}\ }\textbf {\bibinfo
  {volume} {106}},\ \bibinfo {pages} {024507} (\bibinfo {year}
  {2022}{\natexlab{b}})}\BibitemShut {NoStop}%
\bibitem [{\citenamefont {Su}\ \emph {et~al.}(2023)\citenamefont {Su},
  \citenamefont {Kuiri}, \citenamefont {Watanabe}, \citenamefont {Taniguchi},\
  and\ \citenamefont {Folk}}]{Su2022Superconductivity}%
  \BibitemOpen
  \bibfield  {author} {\bibinfo {author} {\bibfnamefont {R.}~\bibnamefont
  {Su}}, \bibinfo {author} {\bibfnamefont {M.}~\bibnamefont {Kuiri}}, \bibinfo
  {author} {\bibfnamefont {K.}~\bibnamefont {Watanabe}}, \bibinfo {author}
  {\bibfnamefont {T.}~\bibnamefont {Taniguchi}},\ and\ \bibinfo {author}
  {\bibfnamefont {J.}~\bibnamefont {Folk}},\ }\bibfield  {title} {\bibinfo
  {title} {Superconductivity in twisted double bilayer graphene stabilized by
  {WSe$_2$}},\ }\href {https://doi.org/10.1038/s41563-023-01653-7} {\bibfield
  {journal} {\bibinfo  {journal} {Nat. Mater.}\ } (\bibinfo {year}
  {2023})}\BibitemShut {NoStop}%
\bibitem [{\citenamefont {Cea}(2023)}]{Cea2023Superconductivity}%
  \BibitemOpen
  \bibfield  {author} {\bibinfo {author} {\bibfnamefont {T.}~\bibnamefont
  {Cea}},\ }\bibfield  {title} {\bibinfo {title} {Superconductivity induced by
  the intervalley {C}oulomb scattering in a few layers of graphene},\ }\href
  {https://doi.org/10.1103/PhysRevB.107.L041111} {\bibfield  {journal}
  {\bibinfo  {journal} {Phys. Rev. B}\ }\textbf {\bibinfo {volume} {107}},\
  \bibinfo {pages} {L041111} (\bibinfo {year} {2023})}\BibitemShut {NoStop}%
\bibitem [{\citenamefont {Pangburn}\ \emph
  {et~al.}(2022{\natexlab{a}})\citenamefont {Pangburn}, \citenamefont {Haurie},
  \citenamefont {Cr\'epieux}, \citenamefont {Awoga}, \citenamefont
  {Black-Schaffer}, \citenamefont {P\'epin},\ and\ \citenamefont
  {Bena}}]{pangburn2022superconductivity}%
  \BibitemOpen
  \bibfield  {author} {\bibinfo {author} {\bibfnamefont {E.}~\bibnamefont
  {Pangburn}}, \bibinfo {author} {\bibfnamefont {L.}~\bibnamefont {Haurie}},
  \bibinfo {author} {\bibfnamefont {A.}~\bibnamefont {Cr\'epieux}}, \bibinfo
  {author} {\bibfnamefont {O.~A.}\ \bibnamefont {Awoga}}, \bibinfo {author}
  {\bibfnamefont {A.~M.}\ \bibnamefont {Black-Schaffer}}, \bibinfo {author}
  {\bibfnamefont {C.}~\bibnamefont {P\'epin}},\ and\ \bibinfo {author}
  {\bibfnamefont {C.}~\bibnamefont {Bena}},\ }\bibfield  {title} {\bibinfo
  {title} {Superconductivity in monolayer and few-layer graphene: {I}. review
  of possible pairing symmetries and basic electronic properties},\ }\href@noop
  {} {\bibfield  {journal} {\bibinfo  {journal} {{arXiv}}\ } (\bibinfo {year}
  {2022}{\natexlab{a}})},\ \Eprint {https://arxiv.org/abs/2211.05146}
  {arXiv:2211.05146} \BibitemShut {NoStop}%
\bibitem [{\citenamefont {Cr\'epieux}\ \emph {et~al.}(2022)\citenamefont
  {Cr\'epieux}, \citenamefont {Pangburn}, \citenamefont {Haurie}, \citenamefont
  {Awoga}, \citenamefont {Black-Schaffer}, \citenamefont {Sedlmayr},
  \citenamefont {P\'epin},\ and\ \citenamefont
  {Bena}}]{crepieux2022superconductivity}%
  \BibitemOpen
  \bibfield  {author} {\bibinfo {author} {\bibfnamefont {A.}~\bibnamefont
  {Cr\'epieux}}, \bibinfo {author} {\bibfnamefont {E.}~\bibnamefont
  {Pangburn}}, \bibinfo {author} {\bibfnamefont {L.}~\bibnamefont {Haurie}},
  \bibinfo {author} {\bibfnamefont {O.~A.}\ \bibnamefont {Awoga}}, \bibinfo
  {author} {\bibfnamefont {A.~M.}\ \bibnamefont {Black-Schaffer}}, \bibinfo
  {author} {\bibfnamefont {N.}~\bibnamefont {Sedlmayr}}, \bibinfo {author}
  {\bibfnamefont {C.}~\bibnamefont {P\'epin}},\ and\ \bibinfo {author}
  {\bibfnamefont {C.}~\bibnamefont {Bena}},\ }\bibfield  {title} {\bibinfo
  {title} {Superconductivity in monolayer and few-layer graphene: {II}.
  topological edge states and chern numbers},\ }\href@noop {} {\bibfield
  {journal} {\bibinfo  {journal} {{arXiv}}\ } (\bibinfo {year} {2022})},\
  \Eprint {https://arxiv.org/abs/2211.11778} {arXiv:2211.11778} \BibitemShut
  {NoStop}%
\bibitem [{\citenamefont {Pangburn}\ \emph
  {et~al.}(2022{\natexlab{b}})\citenamefont {Pangburn}, \citenamefont {Haurie},
  \citenamefont {Cr\'epieux}, \citenamefont {Awoga}, \citenamefont {Sedlmayr},
  \citenamefont {Black-Schaffer}, \citenamefont {P\'epin},\ and\ \citenamefont
  {Bena}}]{pangburn2022superconductivityIII}%
  \BibitemOpen
  \bibfield  {author} {\bibinfo {author} {\bibfnamefont {E.}~\bibnamefont
  {Pangburn}}, \bibinfo {author} {\bibfnamefont {L.}~\bibnamefont {Haurie}},
  \bibinfo {author} {\bibfnamefont {A.}~\bibnamefont {Cr\'epieux}}, \bibinfo
  {author} {\bibfnamefont {O.~A.}\ \bibnamefont {Awoga}}, \bibinfo {author}
  {\bibfnamefont {N.}~\bibnamefont {Sedlmayr}}, \bibinfo {author}
  {\bibfnamefont {A.~M.}\ \bibnamefont {Black-Schaffer}}, \bibinfo {author}
  {\bibfnamefont {C.}~\bibnamefont {P\'epin}},\ and\ \bibinfo {author}
  {\bibfnamefont {C.}~\bibnamefont {Bena}},\ }\bibfield  {title} {\bibinfo
  {title} {Superconductivity in monolayer and few-layer graphene: {III}.
  impurity-induced subgap states and quasi-particle interference patterns},\
  }\href@noop {} {\bibfield  {journal} {\bibinfo  {journal} {{arXiv}}\ }
  (\bibinfo {year} {2022}{\natexlab{b}})},\ \Eprint
  {https://arxiv.org/abs/2212.07445} {arXiv:2212.07445} \BibitemShut {NoStop}%
\bibitem [{\citenamefont {Lipson}\ and\ \citenamefont
  {Stokes}(1942)}]{Lipson1942Structure}%
  \BibitemOpen
  \bibfield  {author} {\bibinfo {author} {\bibfnamefont {H.~S.}\ \bibnamefont
  {Lipson}}\ and\ \bibinfo {author} {\bibfnamefont {A.~R.}\ \bibnamefont
  {Stokes}},\ }\bibfield  {title} {\bibinfo {title} {The structure of
  graphite},\ }\href {https://doi.org/10.1098/rspa.1942.0063} {\bibfield
  {journal} {\bibinfo  {journal} {Proc. R. Soc. Lond. A}\ }\textbf {\bibinfo
  {volume} {181}},\ \bibinfo {pages} {101--105} (\bibinfo {year}
  {1942})}\BibitemShut {NoStop}%
\bibitem [{\citenamefont {Heikkilä}\ and\ \citenamefont
  {Volovik}(2011)}]{Heikkila2011Dimensional}%
  \BibitemOpen
  \bibfield  {author} {\bibinfo {author} {\bibfnamefont {T.~T.}\ \bibnamefont
  {Heikkilä}}\ and\ \bibinfo {author} {\bibfnamefont {G.~E.}\ \bibnamefont
  {Volovik}},\ }\bibfield  {title} {\bibinfo {title} {Dimensional crossover in
  topological matter: Evolution of the multiple {D}irac point in the layered
  system to the flat band on the surface},\ }\href
  {https://doi.org/10.1134/s002136401102007x} {\bibfield  {journal} {\bibinfo
  {journal} {JETP Lett.}\ }\textbf {\bibinfo {volume} {93}},\ \bibinfo {pages}
  {59--65} (\bibinfo {year} {2011})}\BibitemShut {NoStop}%
\bibitem [{\citenamefont {Heikkilä}\ \emph {et~al.}(2011)\citenamefont
  {Heikkilä}, \citenamefont {Kopnin},\ and\ \citenamefont
  {Volovik}}]{Heikkila2011Flat}%
  \BibitemOpen
  \bibfield  {author} {\bibinfo {author} {\bibfnamefont {T.~T.}\ \bibnamefont
  {Heikkilä}}, \bibinfo {author} {\bibfnamefont {N.~B.}\ \bibnamefont
  {Kopnin}},\ and\ \bibinfo {author} {\bibfnamefont {G.~E.}\ \bibnamefont
  {Volovik}},\ }\bibfield  {title} {\bibinfo {title} {Flat bands in topological
  media},\ }\href {https://doi.org/10.1134/s0021364011150045} {\bibfield
  {journal} {\bibinfo  {journal} {JETP Lett.}\ }\textbf {\bibinfo {volume}
  {94}},\ \bibinfo {pages} {233--239} (\bibinfo {year} {2011})}\BibitemShut
  {NoStop}%
\bibitem [{\citenamefont {Pierucci}\ \emph {et~al.}(2015)\citenamefont
  {Pierucci}, \citenamefont {Sediri}, \citenamefont {Hajlaoui}, \citenamefont
  {Girard}, \citenamefont {Brumme}, \citenamefont {Calandra}, \citenamefont
  {Velez-Fort}, \citenamefont {Patriarche}, \citenamefont {Silly},
  \citenamefont {Ferro}, \citenamefont {Soulière}, \citenamefont {Marangolo},
  \citenamefont {Sirotti}, \citenamefont {Mauri},\ and\ \citenamefont
  {Ouerghi}}]{Pierucci2015Evidence}%
  \BibitemOpen
  \bibfield  {author} {\bibinfo {author} {\bibfnamefont {D.}~\bibnamefont
  {Pierucci}}, \bibinfo {author} {\bibfnamefont {H.}~\bibnamefont {Sediri}},
  \bibinfo {author} {\bibfnamefont {M.}~\bibnamefont {Hajlaoui}}, \bibinfo
  {author} {\bibfnamefont {J.-C.}\ \bibnamefont {Girard}}, \bibinfo {author}
  {\bibfnamefont {T.}~\bibnamefont {Brumme}}, \bibinfo {author} {\bibfnamefont
  {M.}~\bibnamefont {Calandra}}, \bibinfo {author} {\bibfnamefont
  {E.}~\bibnamefont {Velez-Fort}}, \bibinfo {author} {\bibfnamefont
  {G.}~\bibnamefont {Patriarche}}, \bibinfo {author} {\bibfnamefont {M.~G.}\
  \bibnamefont {Silly}}, \bibinfo {author} {\bibfnamefont {G.}~\bibnamefont
  {Ferro}}, \bibinfo {author} {\bibfnamefont {V.}~\bibnamefont {Soulière}},
  \bibinfo {author} {\bibfnamefont {M.}~\bibnamefont {Marangolo}}, \bibinfo
  {author} {\bibfnamefont {F.}~\bibnamefont {Sirotti}}, \bibinfo {author}
  {\bibfnamefont {F.}~\bibnamefont {Mauri}},\ and\ \bibinfo {author}
  {\bibfnamefont {A.}~\bibnamefont {Ouerghi}},\ }\bibfield  {title} {\bibinfo
  {title} {Evidence for flat bands near the {F}ermi level in epitaxial
  rhombohedral multilayer graphene},\ }\href
  {https://doi.org/10.1021/acsnano.5b01239} {\bibfield  {journal} {\bibinfo
  {journal} {ACS Nano}\ }\textbf {\bibinfo {volume} {9}},\ \bibinfo {pages}
  {5432--5439} (\bibinfo {year} {2015})}\BibitemShut {NoStop}%
\bibitem [{\citenamefont {Henck}\ \emph {et~al.}(2018)\citenamefont {Henck},
  \citenamefont {Avila}, \citenamefont {Ben~Aziza}, \citenamefont {Pierucci},
  \citenamefont {Baima}, \citenamefont {Pamuk}, \citenamefont {Chaste},
  \citenamefont {Utt}, \citenamefont {Bartos}, \citenamefont {Nogajewski},
  \citenamefont {Piot}, \citenamefont {Orlita}, \citenamefont {Potemski},
  \citenamefont {Calandra}, \citenamefont {Asensio}, \citenamefont {Mauri},
  \citenamefont {Faugeras},\ and\ \citenamefont {Ouerghi}}]{Henck2018Flat}%
  \BibitemOpen
  \bibfield  {author} {\bibinfo {author} {\bibfnamefont {H.}~\bibnamefont
  {Henck}}, \bibinfo {author} {\bibfnamefont {J.}~\bibnamefont {Avila}},
  \bibinfo {author} {\bibfnamefont {Z.}~\bibnamefont {Ben~Aziza}}, \bibinfo
  {author} {\bibfnamefont {D.}~\bibnamefont {Pierucci}}, \bibinfo {author}
  {\bibfnamefont {J.}~\bibnamefont {Baima}}, \bibinfo {author} {\bibfnamefont
  {B.}~\bibnamefont {Pamuk}}, \bibinfo {author} {\bibfnamefont
  {J.}~\bibnamefont {Chaste}}, \bibinfo {author} {\bibfnamefont
  {D.}~\bibnamefont {Utt}}, \bibinfo {author} {\bibfnamefont {M.}~\bibnamefont
  {Bartos}}, \bibinfo {author} {\bibfnamefont {K.}~\bibnamefont {Nogajewski}},
  \bibinfo {author} {\bibfnamefont {B.~A.}\ \bibnamefont {Piot}}, \bibinfo
  {author} {\bibfnamefont {M.}~\bibnamefont {Orlita}}, \bibinfo {author}
  {\bibfnamefont {M.}~\bibnamefont {Potemski}}, \bibinfo {author}
  {\bibfnamefont {M.}~\bibnamefont {Calandra}}, \bibinfo {author}
  {\bibfnamefont {M.~C.}\ \bibnamefont {Asensio}}, \bibinfo {author}
  {\bibfnamefont {F.}~\bibnamefont {Mauri}}, \bibinfo {author} {\bibfnamefont
  {C.}~\bibnamefont {Faugeras}},\ and\ \bibinfo {author} {\bibfnamefont
  {A.}~\bibnamefont {Ouerghi}},\ }\bibfield  {title} {\bibinfo {title} {Flat
  electronic bands in long sequences of rhombohedral-stacked graphene},\ }\href
  {https://doi.org/10.1103/PhysRevB.97.245421} {\bibfield  {journal} {\bibinfo
  {journal} {Phys. Rev. B}\ }\textbf {\bibinfo {volume} {97}},\ \bibinfo
  {pages} {245421} (\bibinfo {year} {2018})}\BibitemShut {NoStop}%
\bibitem [{\citenamefont {Norimatsu}\ and\ \citenamefont
  {Kusunoki}(2010)}]{Norimatsu2010}%
  \BibitemOpen
  \bibfield  {author} {\bibinfo {author} {\bibfnamefont {W.}~\bibnamefont
  {Norimatsu}}\ and\ \bibinfo {author} {\bibfnamefont {M.}~\bibnamefont
  {Kusunoki}},\ }\bibfield  {title} {\bibinfo {title} {Selective formation of
  {ABC}-stacked graphene layers on {SiC}(0001)},\ }\href
  {https://doi.org/10.1103/physrevb.81.161410} {\bibfield  {journal} {\bibinfo
  {journal} {Phys. Rev. B}\ }\textbf {\bibinfo {volume} {81}},\ \bibinfo
  {pages} {161410} (\bibinfo {year} {2010})}\BibitemShut {NoStop}%
\bibitem [{\citenamefont {Henni}\ \emph {et~al.}(2016)\citenamefont {Henni},
  \citenamefont {Collado}, \citenamefont {Nogajewski}, \citenamefont {Molas},
  \citenamefont {Usaj}, \citenamefont {Balseiro}, \citenamefont {Orlita},
  \citenamefont {Potemski},\ and\ \citenamefont {Faugeras}}]{Henni2016}%
  \BibitemOpen
  \bibfield  {author} {\bibinfo {author} {\bibfnamefont {Y.}~\bibnamefont
  {Henni}}, \bibinfo {author} {\bibfnamefont {H.~P.~O.}\ \bibnamefont
  {Collado}}, \bibinfo {author} {\bibfnamefont {K.}~\bibnamefont {Nogajewski}},
  \bibinfo {author} {\bibfnamefont {M.~R.}\ \bibnamefont {Molas}}, \bibinfo
  {author} {\bibfnamefont {G.}~\bibnamefont {Usaj}}, \bibinfo {author}
  {\bibfnamefont {C.~A.}\ \bibnamefont {Balseiro}}, \bibinfo {author}
  {\bibfnamefont {M.}~\bibnamefont {Orlita}}, \bibinfo {author} {\bibfnamefont
  {M.}~\bibnamefont {Potemski}},\ and\ \bibinfo {author} {\bibfnamefont
  {C.}~\bibnamefont {Faugeras}},\ }\bibfield  {title} {\bibinfo {title}
  {Rhombohedral multilayer graphene: A magneto-raman scattering study},\ }\href
  {https://doi.org/10.1021/acs.nanolett.6b01041} {\bibfield  {journal}
  {\bibinfo  {journal} {Nano Lett.}\ }\textbf {\bibinfo {volume} {16}},\
  \bibinfo {pages} {3710--3716} (\bibinfo {year} {2016})}\BibitemShut {NoStop}%
\bibitem [{\citenamefont {Yang}\ \emph {et~al.}(2019)\citenamefont {Yang},
  \citenamefont {Zou}, \citenamefont {Woods}, \citenamefont {Shi},
  \citenamefont {Yin}, \citenamefont {Xu}, \citenamefont {Ozdemir},
  \citenamefont {Taniguchi}, \citenamefont {Watanabe}, \citenamefont {Geim},
  \citenamefont {Novoselov}, \citenamefont {Haigh},\ and\ \citenamefont
  {Mishchenko}}]{Yang2019}%
  \BibitemOpen
  \bibfield  {author} {\bibinfo {author} {\bibfnamefont {Y.}~\bibnamefont
  {Yang}}, \bibinfo {author} {\bibfnamefont {Y.-C.}\ \bibnamefont {Zou}},
  \bibinfo {author} {\bibfnamefont {C.~R.}\ \bibnamefont {Woods}}, \bibinfo
  {author} {\bibfnamefont {Y.}~\bibnamefont {Shi}}, \bibinfo {author}
  {\bibfnamefont {J.}~\bibnamefont {Yin}}, \bibinfo {author} {\bibfnamefont
  {S.}~\bibnamefont {Xu}}, \bibinfo {author} {\bibfnamefont {S.}~\bibnamefont
  {Ozdemir}}, \bibinfo {author} {\bibfnamefont {T.}~\bibnamefont {Taniguchi}},
  \bibinfo {author} {\bibfnamefont {K.}~\bibnamefont {Watanabe}}, \bibinfo
  {author} {\bibfnamefont {A.~K.}\ \bibnamefont {Geim}}, \bibinfo {author}
  {\bibfnamefont {K.~S.}\ \bibnamefont {Novoselov}}, \bibinfo {author}
  {\bibfnamefont {S.~J.}\ \bibnamefont {Haigh}},\ and\ \bibinfo {author}
  {\bibfnamefont {A.}~\bibnamefont {Mishchenko}},\ }\bibfield  {title}
  {\bibinfo {title} {Stacking order in graphite films controlled by van der
  waals technology},\ }\href {https://doi.org/10.1021/acs.nanolett.9b03014}
  {\bibfield  {journal} {\bibinfo  {journal} {Nano Lett.}\ }\textbf {\bibinfo
  {volume} {19}},\ \bibinfo {pages} {8526--8532} (\bibinfo {year}
  {2019})}\BibitemShut {NoStop}%
\bibitem [{\citenamefont {{Shi}}\ \emph {et~al.}(2020)\citenamefont {{Shi}},
  \citenamefont {{Xu}}, \citenamefont {{Yang}}, \citenamefont {{Slizovskiy}},
  \citenamefont {{Morozov}}, \citenamefont {{Son}}, \citenamefont {{Ozdemir}},
  \citenamefont {{Mullan}}, \citenamefont {{Barrier}}, \citenamefont {{Yin}},
  \citenamefont {{Berdyugin}}, \citenamefont {{Piot}}, \citenamefont
  {{Taniguchi}}, \citenamefont {{Watanabe}}, \citenamefont {{Fal'ko}},
  \citenamefont {{Novoselov}}, \citenamefont {{Geim}},\ and\ \citenamefont
  {{Mishchenko}}}]{Yanmeng2020Electronic}%
  \BibitemOpen
  \bibfield  {author} {\bibinfo {author} {\bibfnamefont {Y.}~\bibnamefont
  {{Shi}}}, \bibinfo {author} {\bibfnamefont {S.}~\bibnamefont {{Xu}}},
  \bibinfo {author} {\bibfnamefont {Y.}~\bibnamefont {{Yang}}}, \bibinfo
  {author} {\bibfnamefont {S.}~\bibnamefont {{Slizovskiy}}}, \bibinfo {author}
  {\bibfnamefont {S.~V.}\ \bibnamefont {{Morozov}}}, \bibinfo {author}
  {\bibfnamefont {S.-K.}\ \bibnamefont {{Son}}}, \bibinfo {author}
  {\bibfnamefont {S.}~\bibnamefont {{Ozdemir}}}, \bibinfo {author}
  {\bibfnamefont {C.}~\bibnamefont {{Mullan}}}, \bibinfo {author}
  {\bibfnamefont {J.}~\bibnamefont {{Barrier}}}, \bibinfo {author}
  {\bibfnamefont {J.}~\bibnamefont {{Yin}}}, \bibinfo {author} {\bibfnamefont
  {A.~I.}\ \bibnamefont {{Berdyugin}}}, \bibinfo {author} {\bibfnamefont
  {B.~A.}\ \bibnamefont {{Piot}}}, \bibinfo {author} {\bibfnamefont
  {T.}~\bibnamefont {{Taniguchi}}}, \bibinfo {author} {\bibfnamefont
  {K.}~\bibnamefont {{Watanabe}}}, \bibinfo {author} {\bibfnamefont {V.~I.}\
  \bibnamefont {{Fal'ko}}}, \bibinfo {author} {\bibfnamefont {K.~S.}\
  \bibnamefont {{Novoselov}}}, \bibinfo {author} {\bibfnamefont {A.~K.}\
  \bibnamefont {{Geim}}},\ and\ \bibinfo {author} {\bibfnamefont
  {A.}~\bibnamefont {{Mishchenko}}},\ }\bibfield  {title} {\bibinfo {title}
  {Electronic phase separation in multilayer rhombohedral graphite},\ }\href
  {https://doi.org/10.1038/s41586-020-2568-2} {\bibfield  {journal} {\bibinfo
  {journal} {Nature}\ }\textbf {\bibinfo {volume} {584}},\ \bibinfo {pages}
  {210--214} (\bibinfo {year} {2020})}\BibitemShut {NoStop}%
\bibitem [{\citenamefont {Bouhafs}\ \emph {et~al.}(2021)\citenamefont
  {Bouhafs}, \citenamefont {Pezzini}, \citenamefont {Geisenhof}, \citenamefont
  {Mishra}, \citenamefont {Mi{\v{s}}eikis}, \citenamefont {Niu}, \citenamefont
  {Struzzi}, \citenamefont {Weitz}, \citenamefont {Zakharov}, \citenamefont
  {Forti},\ and\ \citenamefont {Coletti}}]{Bouhafs2021}%
  \BibitemOpen
  \bibfield  {author} {\bibinfo {author} {\bibfnamefont {C.}~\bibnamefont
  {Bouhafs}}, \bibinfo {author} {\bibfnamefont {S.}~\bibnamefont {Pezzini}},
  \bibinfo {author} {\bibfnamefont {F.~R.}\ \bibnamefont {Geisenhof}}, \bibinfo
  {author} {\bibfnamefont {N.}~\bibnamefont {Mishra}}, \bibinfo {author}
  {\bibfnamefont {V.}~\bibnamefont {Mi{\v{s}}eikis}}, \bibinfo {author}
  {\bibfnamefont {Y.}~\bibnamefont {Niu}}, \bibinfo {author} {\bibfnamefont
  {C.}~\bibnamefont {Struzzi}}, \bibinfo {author} {\bibfnamefont {R.~T.}\
  \bibnamefont {Weitz}}, \bibinfo {author} {\bibfnamefont {A.~A.}\ \bibnamefont
  {Zakharov}}, \bibinfo {author} {\bibfnamefont {S.}~\bibnamefont {Forti}},\
  and\ \bibinfo {author} {\bibfnamefont {C.}~\bibnamefont {Coletti}},\
  }\bibfield  {title} {\bibinfo {title} {Synthesis of large-area rhombohedral
  few-layer graphene by chemical vapor deposition on copper},\ }\href
  {https://doi.org/10.1016/j.carbon.2021.02.082} {\bibfield  {journal}
  {\bibinfo  {journal} {Carbon}\ }\textbf {\bibinfo {volume} {177}},\ \bibinfo
  {pages} {282--290} (\bibinfo {year} {2021})}\BibitemShut {NoStop}%
\bibitem [{\citenamefont {Hagymási}\ \emph {et~al.}(2022)\citenamefont
  {Hagymási}, \citenamefont {Isa}, \citenamefont {Tajkov}, \citenamefont
  {Márity}, \citenamefont {László}, \citenamefont {Koltai}, \citenamefont
  {Alassaf}, \citenamefont {Kun}, \citenamefont {Kandrai}, \citenamefont
  {Pálinkás}, \citenamefont {Vancsó}, \citenamefont {Tapasztó},\ and\
  \citenamefont {Nemes-Incze}}]{hagymasi2022signature}%
  \BibitemOpen
  \bibfield  {author} {\bibinfo {author} {\bibfnamefont {I.}~\bibnamefont
  {Hagymási}}, \bibinfo {author} {\bibfnamefont {M.~S.~M.}\ \bibnamefont
  {Isa}}, \bibinfo {author} {\bibfnamefont {Z.}~\bibnamefont {Tajkov}},
  \bibinfo {author} {\bibfnamefont {K.}~\bibnamefont {Márity}}, \bibinfo
  {author} {\bibfnamefont {O.}~\bibnamefont {László}}, \bibinfo {author}
  {\bibfnamefont {J.}~\bibnamefont {Koltai}}, \bibinfo {author} {\bibfnamefont
  {A.}~\bibnamefont {Alassaf}}, \bibinfo {author} {\bibfnamefont
  {P.}~\bibnamefont {Kun}}, \bibinfo {author} {\bibfnamefont {K.}~\bibnamefont
  {Kandrai}}, \bibinfo {author} {\bibfnamefont {A.}~\bibnamefont {Pálinkás}},
  \bibinfo {author} {\bibfnamefont {P.}~\bibnamefont {Vancsó}}, \bibinfo
  {author} {\bibfnamefont {L.}~\bibnamefont {Tapasztó}},\ and\ \bibinfo
  {author} {\bibfnamefont {P.}~\bibnamefont {Nemes-Incze}},\ }\bibfield
  {title} {\bibinfo {title} {Signature of quantum magnetism in thick
  rhombohedral graphite},\ }\href@noop {} {\bibfield  {journal} {\bibinfo
  {journal} {{arXiv}}\ } (\bibinfo {year} {2022})},\ \Eprint
  {https://arxiv.org/abs/2201.10844} {arXiv:2201.10844} \BibitemShut {NoStop}%
\bibitem [{\citenamefont {Blundell}(1967)}]{Blundell1967Mag}%
  \BibitemOpen
  \bibfield  {author} {\bibinfo {author} {\bibfnamefont {S.}~\bibnamefont
  {Blundell}},\ }\href@noop {} {\emph {\bibinfo {title} {Magnetism in Condensed
  Matter}}}\ (\bibinfo  {publisher} {Oxford University Press, Oxford},\
  \bibinfo {year} {1967})\BibitemShut {NoStop}%
\bibitem [{\citenamefont {Otani}\ \emph {et~al.}(2010)\citenamefont {Otani},
  \citenamefont {Koshino}, \citenamefont {Takagi},\ and\ \citenamefont
  {Okada}}]{Otani2010Intrinsic}%
  \BibitemOpen
  \bibfield  {author} {\bibinfo {author} {\bibfnamefont {M.}~\bibnamefont
  {Otani}}, \bibinfo {author} {\bibfnamefont {M.}~\bibnamefont {Koshino}},
  \bibinfo {author} {\bibfnamefont {Y.}~\bibnamefont {Takagi}},\ and\ \bibinfo
  {author} {\bibfnamefont {S.}~\bibnamefont {Okada}},\ }\bibfield  {title}
  {\bibinfo {title} {Intrinsic magnetic moment on (0001) surfaces of
  rhombohedral graphite},\ }\href {https://doi.org/10.1103/PhysRevB.81.161403}
  {\bibfield  {journal} {\bibinfo  {journal} {Phys. Rev. B (R)}\ }\textbf
  {\bibinfo {volume} {81}},\ \bibinfo {pages} {161403} (\bibinfo {year}
  {2010})}\BibitemShut {NoStop}%
\bibitem [{\citenamefont {Xiao}\ \emph {et~al.}(2011)\citenamefont {Xiao},
  \citenamefont {Tasn\'adi}, \citenamefont {Koepernik}, \citenamefont
  {Venderbos}, \citenamefont {Richter},\ and\ \citenamefont
  {Taut}}]{Xiao2011Density}%
  \BibitemOpen
  \bibfield  {author} {\bibinfo {author} {\bibfnamefont {R.}~\bibnamefont
  {Xiao}}, \bibinfo {author} {\bibfnamefont {F.}~\bibnamefont {Tasn\'adi}},
  \bibinfo {author} {\bibfnamefont {K.}~\bibnamefont {Koepernik}}, \bibinfo
  {author} {\bibfnamefont {J.~W.~F.}\ \bibnamefont {Venderbos}}, \bibinfo
  {author} {\bibfnamefont {M.}~\bibnamefont {Richter}},\ and\ \bibinfo {author}
  {\bibfnamefont {M.}~\bibnamefont {Taut}},\ }\bibfield  {title} {\bibinfo
  {title} {Density functional investigation of rhombohedral stacks of graphene:
  Topological surface states, nonlinear dielectric response, and bulk limit},\
  }\href {https://doi.org/10.1103/PhysRevB.84.165404} {\bibfield  {journal}
  {\bibinfo  {journal} {Phys. Rev. B}\ }\textbf {\bibinfo {volume} {84}},\
  \bibinfo {pages} {165404} (\bibinfo {year} {2011})}\BibitemShut {NoStop}%
\bibitem [{\citenamefont {Cuong}\ \emph {et~al.}(2012)\citenamefont {Cuong},
  \citenamefont {Otani},\ and\ \citenamefont {Okada}}]{Cuong2012Magnetic}%
  \BibitemOpen
  \bibfield  {author} {\bibinfo {author} {\bibfnamefont {N.~T.}\ \bibnamefont
  {Cuong}}, \bibinfo {author} {\bibfnamefont {M.}~\bibnamefont {Otani}},\ and\
  \bibinfo {author} {\bibfnamefont {S.}~\bibnamefont {Okada}},\ }\bibfield
  {title} {\bibinfo {title} {Magnetic-state tuning of the rhombohedral graphite
  film by interlayer spacing and thickness},\ }\href
  {https://doi.org/10.1016/j.susc.2011.10.001} {\bibfield  {journal} {\bibinfo
  {journal} {Surf. Sci.}\ }\textbf {\bibinfo {volume} {606}},\ \bibinfo {pages}
  {253--257} (\bibinfo {year} {2012})}\BibitemShut {NoStop}%
\bibitem [{\citenamefont {Pamuk}\ \emph {et~al.}(2017)\citenamefont {Pamuk},
  \citenamefont {Baima}, \citenamefont {Mauri},\ and\ \citenamefont
  {Calandra}}]{Pamuk2017}%
  \BibitemOpen
  \bibfield  {author} {\bibinfo {author} {\bibfnamefont {B.}~\bibnamefont
  {Pamuk}}, \bibinfo {author} {\bibfnamefont {J.}~\bibnamefont {Baima}},
  \bibinfo {author} {\bibfnamefont {F.}~\bibnamefont {Mauri}},\ and\ \bibinfo
  {author} {\bibfnamefont {M.}~\bibnamefont {Calandra}},\ }\bibfield  {title}
  {\bibinfo {title} {Magnetic gap opening in rhombohedral-stacked multilayer
  graphene from first principles},\ }\href
  {https://doi.org/10.1103/PhysRevB.95.075422} {\bibfield  {journal} {\bibinfo
  {journal} {Phys. Rev. B}\ }\textbf {\bibinfo {volume} {95}},\ \bibinfo
  {pages} {075422} (\bibinfo {year} {2017})}\BibitemShut {NoStop}%
\bibitem [{\citenamefont {Lee}\ \emph {et~al.}(2014)\citenamefont {Lee},
  \citenamefont {Tran}, \citenamefont {Myhro}, \citenamefont {Velasco},
  \citenamefont {Gillgren}, \citenamefont {Lau}, \citenamefont {Barlas},
  \citenamefont {Poumirol}, \citenamefont {Smirnov},\ and\ \citenamefont
  {Guinea}}]{Lee2014competition}%
  \BibitemOpen
  \bibfield  {author} {\bibinfo {author} {\bibfnamefont {Y.}~\bibnamefont
  {Lee}}, \bibinfo {author} {\bibfnamefont {D.}~\bibnamefont {Tran}}, \bibinfo
  {author} {\bibfnamefont {K.}~\bibnamefont {Myhro}}, \bibinfo {author}
  {\bibfnamefont {J.}~\bibnamefont {Velasco}}, \bibinfo {author} {\bibfnamefont
  {N.}~\bibnamefont {Gillgren}}, \bibinfo {author} {\bibfnamefont
  {C.}~\bibnamefont {Lau}}, \bibinfo {author} {\bibfnamefont {Y.}~\bibnamefont
  {Barlas}}, \bibinfo {author} {\bibfnamefont {J.}~\bibnamefont {Poumirol}},
  \bibinfo {author} {\bibfnamefont {D.}~\bibnamefont {Smirnov}},\ and\ \bibinfo
  {author} {\bibfnamefont {F.}~\bibnamefont {Guinea}},\ }\bibfield  {title}
  {\bibinfo {title} {Competition between spontaneous symmetry breaking and
  single-particle gaps in trilayer graphene},\ }\href
  {https://www.nature.com/articles/ncomms6656} {\bibfield  {journal} {\bibinfo
  {journal} {Nat. Commun.}\ }\textbf {\bibinfo {volume} {5}},\ \bibinfo {pages}
  {1--5} (\bibinfo {year} {2014})}\BibitemShut {NoStop}%
\bibitem [{\citenamefont {Lee}\ \emph {et~al.}(2019)\citenamefont {Lee},
  \citenamefont {Che}, \citenamefont {Jr.}, \citenamefont {Tran}, \citenamefont
  {Baima}, \citenamefont {Mauri}, \citenamefont {Calandra}, \citenamefont
  {Bockrath},\ and\ \citenamefont {Lau}}]{lee2019gate}%
  \BibitemOpen
  \bibfield  {author} {\bibinfo {author} {\bibfnamefont {Y.}~\bibnamefont
  {Lee}}, \bibinfo {author} {\bibfnamefont {S.}~\bibnamefont {Che}}, \bibinfo
  {author} {\bibfnamefont {J.~V.}\ \bibnamefont {Jr.}}, \bibinfo {author}
  {\bibfnamefont {D.}~\bibnamefont {Tran}}, \bibinfo {author} {\bibfnamefont
  {J.}~\bibnamefont {Baima}}, \bibinfo {author} {\bibfnamefont
  {F.}~\bibnamefont {Mauri}}, \bibinfo {author} {\bibfnamefont
  {M.}~\bibnamefont {Calandra}}, \bibinfo {author} {\bibfnamefont
  {M.}~\bibnamefont {Bockrath}},\ and\ \bibinfo {author} {\bibfnamefont
  {C.~N.}\ \bibnamefont {Lau}},\ }\bibfield  {title} {\bibinfo {title} {Gate
  tunable magnetism and giant magnetoresistance in {ABC}-stacked few-layer
  graphene},\ }\href@noop {} {\bibfield  {journal} {\bibinfo  {journal}
  {{arXiv}}\ } (\bibinfo {year} {2019})},\ \Eprint
  {https://arxiv.org/abs/1911.04450} {arXiv:1911.04450} \BibitemShut {NoStop}%
\bibitem [{\citenamefont {Myhro}\ \emph {et~al.}(2018)\citenamefont {Myhro},
  \citenamefont {Che}, \citenamefont {Shi}, \citenamefont {Lee}, \citenamefont
  {Thilahar}, \citenamefont {Bleich}, \citenamefont {Smirnov},\ and\
  \citenamefont {Lau}}]{Myhro2018Large}%
  \BibitemOpen
  \bibfield  {author} {\bibinfo {author} {\bibfnamefont {K.}~\bibnamefont
  {Myhro}}, \bibinfo {author} {\bibfnamefont {S.}~\bibnamefont {Che}}, \bibinfo
  {author} {\bibfnamefont {Y.}~\bibnamefont {Shi}}, \bibinfo {author}
  {\bibfnamefont {Y.}~\bibnamefont {Lee}}, \bibinfo {author} {\bibfnamefont
  {K.}~\bibnamefont {Thilahar}}, \bibinfo {author} {\bibfnamefont
  {K.}~\bibnamefont {Bleich}}, \bibinfo {author} {\bibfnamefont
  {D.}~\bibnamefont {Smirnov}},\ and\ \bibinfo {author} {\bibfnamefont {C.~N.}\
  \bibnamefont {Lau}},\ }\bibfield  {title} {\bibinfo {title} {Large tunable
  intrinsic gap in rhombohedral-stacked tetralayer graphene at half filling},\
  }\href {https://doi.org/10.1088/2053-1583/aad2f2} {\bibfield  {journal}
  {\bibinfo  {journal} {2D Mater.}\ }\textbf {\bibinfo {volume} {5}},\ \bibinfo
  {pages} {045013} (\bibinfo {year} {2018})}\BibitemShut {NoStop}%
\bibitem [{\citenamefont {L\"othman}\ and\ \citenamefont
  {Black-Schaffer}(2017)}]{Lothman2017Universal}%
  \BibitemOpen
  \bibfield  {author} {\bibinfo {author} {\bibfnamefont {T.}~\bibnamefont
  {L\"othman}}\ and\ \bibinfo {author} {\bibfnamefont {A.~M.}\ \bibnamefont
  {Black-Schaffer}},\ }\bibfield  {title} {\bibinfo {title} {Universal phase
  diagrams with superconducting domes for electronic flat bands},\ }\href
  {https://doi.org/10.1103/PhysRevB.96.064505} {\bibfield  {journal} {\bibinfo
  {journal} {Phys. Rev. B}\ }\textbf {\bibinfo {volume} {96}},\ \bibinfo
  {pages} {064505} (\bibinfo {year} {2017})}\BibitemShut {NoStop}%
\bibitem [{\citenamefont {Kopnin}\ \emph {et~al.}(2011)\citenamefont {Kopnin},
  \citenamefont {Heikkil\"a},\ and\ \citenamefont {Volovik}}]{Kopnin2011High}%
  \BibitemOpen
  \bibfield  {author} {\bibinfo {author} {\bibfnamefont {N.~B.}\ \bibnamefont
  {Kopnin}}, \bibinfo {author} {\bibfnamefont {T.~T.}\ \bibnamefont
  {Heikkil\"a}},\ and\ \bibinfo {author} {\bibfnamefont {G.~E.}\ \bibnamefont
  {Volovik}},\ }\bibfield  {title} {\bibinfo {title} {High-temperature surface
  superconductivity in topological flat-band systems},\ }\href
  {https://doi.org/10.1103/PhysRevB.83.220503} {\bibfield  {journal} {\bibinfo
  {journal} {Phys. Rev. B}\ }\textbf {\bibinfo {volume} {83}},\ \bibinfo
  {pages} {220503} (\bibinfo {year} {2011})}\BibitemShut {NoStop}%
\bibitem [{\citenamefont {Kopnin}\ \emph {et~al.}(2013)\citenamefont {Kopnin},
  \citenamefont {Ij\"as}, \citenamefont {Harju},\ and\ \citenamefont
  {Heikkil\"a}}]{Kopnin2013High}%
  \BibitemOpen
  \bibfield  {author} {\bibinfo {author} {\bibfnamefont {N.~B.}\ \bibnamefont
  {Kopnin}}, \bibinfo {author} {\bibfnamefont {M.}~\bibnamefont {Ij\"as}},
  \bibinfo {author} {\bibfnamefont {A.}~\bibnamefont {Harju}},\ and\ \bibinfo
  {author} {\bibfnamefont {T.~T.}\ \bibnamefont {Heikkil\"a}},\ }\bibfield
  {title} {\bibinfo {title} {High-temperature surface superconductivity in
  rhombohedral graphite},\ }\href {https://doi.org/10.1103/PhysRevB.87.140503}
  {\bibfield  {journal} {\bibinfo  {journal} {Phys. Rev. B}\ }\textbf {\bibinfo
  {volume} {87}},\ \bibinfo {pages} {140503} (\bibinfo {year}
  {2013})}\BibitemShut {NoStop}%
\bibitem [{\citenamefont {Mu\~noz}\ \emph {et~al.}(2013)\citenamefont
  {Mu\~noz}, \citenamefont {Covaci},\ and\ \citenamefont
  {Peeters}}]{Munoz2013Tight}%
  \BibitemOpen
  \bibfield  {author} {\bibinfo {author} {\bibfnamefont {W.~A.}\ \bibnamefont
  {Mu\~noz}}, \bibinfo {author} {\bibfnamefont {L.}~\bibnamefont {Covaci}},\
  and\ \bibinfo {author} {\bibfnamefont {F.~M.}\ \bibnamefont {Peeters}},\
  }\bibfield  {title} {\bibinfo {title} {Tight-binding description of intrinsic
  superconducting correlations in multilayer graphene},\ }\href
  {https://doi.org/10.1103/PhysRevB.87.134509} {\bibfield  {journal} {\bibinfo
  {journal} {Phys. Rev. B}\ }\textbf {\bibinfo {volume} {87}},\ \bibinfo
  {pages} {134509} (\bibinfo {year} {2013})}\BibitemShut {NoStop}%
\bibitem [{\citenamefont {Scalapino}\ \emph {et~al.}(1992)\citenamefont
  {Scalapino}, \citenamefont {White},\ and\ \citenamefont
  {Zhang}}]{Scalapino1992Superfluid}%
  \BibitemOpen
  \bibfield  {author} {\bibinfo {author} {\bibfnamefont {D.~J.}\ \bibnamefont
  {Scalapino}}, \bibinfo {author} {\bibfnamefont {S.~R.}\ \bibnamefont
  {White}},\ and\ \bibinfo {author} {\bibfnamefont {S.~C.}\ \bibnamefont
  {Zhang}},\ }\bibfield  {title} {\bibinfo {title} {Superfluid density and the
  {D}rude weight of the {H}ubbard model},\ }\href
  {https://doi.org/10.1103/PhysRevLett.68.2830} {\bibfield  {journal} {\bibinfo
   {journal} {Phys. Rev. Lett.}\ }\textbf {\bibinfo {volume} {68}},\ \bibinfo
  {pages} {2830--2833} (\bibinfo {year} {1992})}\BibitemShut {NoStop}%
\bibitem [{\citenamefont {Peotta}\ and\ \citenamefont
  {Törmä}(2015)}]{Peotta2015Superfluidity}%
  \BibitemOpen
  \bibfield  {author} {\bibinfo {author} {\bibfnamefont {S.}~\bibnamefont
  {Peotta}}\ and\ \bibinfo {author} {\bibfnamefont {P.}~\bibnamefont
  {Törmä}},\ }\bibfield  {title} {\bibinfo {title} {Superfluidity in
  topologically nontrivial flat bands},\ }\bibfield  {journal} {\bibinfo
  {journal} {Nat. Commun.}\ }\textbf {\bibinfo {volume} {6}},\ \href
  {https://doi.org/10.1038/ncomms9944} {10.1038/ncomms9944} (\bibinfo {year}
  {2015})\BibitemShut {NoStop}%
\bibitem [{\citenamefont {Liang}\ \emph {et~al.}(2017)\citenamefont {Liang},
  \citenamefont {Vanhala}, \citenamefont {Peotta}, \citenamefont {Siro},
  \citenamefont {Harju},\ and\ \citenamefont {T\"orm\"a}}]{Liang2017Band}%
  \BibitemOpen
  \bibfield  {author} {\bibinfo {author} {\bibfnamefont {L.}~\bibnamefont
  {Liang}}, \bibinfo {author} {\bibfnamefont {T.~I.}\ \bibnamefont {Vanhala}},
  \bibinfo {author} {\bibfnamefont {S.}~\bibnamefont {Peotta}}, \bibinfo
  {author} {\bibfnamefont {T.}~\bibnamefont {Siro}}, \bibinfo {author}
  {\bibfnamefont {A.}~\bibnamefont {Harju}},\ and\ \bibinfo {author}
  {\bibfnamefont {P.}~\bibnamefont {T\"orm\"a}},\ }\bibfield  {title} {\bibinfo
  {title} {Band geometry, {B}erry curvature, and superfluid weight},\ }\href
  {https://doi.org/10.1103/PhysRevB.95.024515} {\bibfield  {journal} {\bibinfo
  {journal} {Phys. Rev. B}\ }\textbf {\bibinfo {volume} {95}},\ \bibinfo
  {pages} {024515} (\bibinfo {year} {2017})}\BibitemShut {NoStop}%
\bibitem [{\citenamefont {Hu}\ \emph {et~al.}(2019)\citenamefont {Hu},
  \citenamefont {Hyart}, \citenamefont {Pikulin},\ and\ \citenamefont
  {Rossi}}]{Xiang2019Geometrical}%
  \BibitemOpen
  \bibfield  {author} {\bibinfo {author} {\bibfnamefont {X.}~\bibnamefont
  {Hu}}, \bibinfo {author} {\bibfnamefont {T.}~\bibnamefont {Hyart}}, \bibinfo
  {author} {\bibfnamefont {D.~I.}\ \bibnamefont {Pikulin}},\ and\ \bibinfo
  {author} {\bibfnamefont {E.}~\bibnamefont {Rossi}},\ }\bibfield  {title}
  {\bibinfo {title} {Geometric and conventional contribution to the superfluid
  weight in twisted bilayer graphene},\ }\href
  {https://doi.org/10.1103/PhysRevLett.123.237002} {\bibfield  {journal}
  {\bibinfo  {journal} {Phys. Rev. Lett.}\ }\textbf {\bibinfo {volume} {123}},\
  \bibinfo {pages} {237002} (\bibinfo {year} {2019})}\BibitemShut {NoStop}%
\bibitem [{\citenamefont {Verma}\ \emph {et~al.}(2021)\citenamefont {Verma},
  \citenamefont {Hazra},\ and\ \citenamefont {Randeria}}]{Verma2021}%
  \BibitemOpen
  \bibfield  {author} {\bibinfo {author} {\bibfnamefont {N.}~\bibnamefont
  {Verma}}, \bibinfo {author} {\bibfnamefont {T.}~\bibnamefont {Hazra}},\ and\
  \bibinfo {author} {\bibfnamefont {M.}~\bibnamefont {Randeria}},\ }\bibfield
  {title} {\bibinfo {title} {Optical spectral weight, phase stiffness, and {Tc}
  bounds for trivial and topological flat band superconductors},\ }\bibfield
  {journal} {\bibinfo  {journal} {Proc. Nat. Acad. Sci.}\ }\textbf {\bibinfo
  {volume} {118}},\ \href {https://doi.org/10.1073/pnas.2106744118}
  {10.1073/pnas.2106744118} (\bibinfo {year} {2021})\BibitemShut {NoStop}%
\bibitem [{\citenamefont {Kopnin}(2011)}]{Kopnin2011}%
  \BibitemOpen
  \bibfield  {author} {\bibinfo {author} {\bibfnamefont {N.~B.}\ \bibnamefont
  {Kopnin}},\ }\bibfield  {title} {\bibinfo {title} {Surface superconductivity
  in multilayered rhombohedral graphene: Supercurrent},\ }\href
  {https://doi.org/10.1134/s002136401113011x} {\bibfield  {journal} {\bibinfo
  {journal} {{JETP} Lett.}\ }\textbf {\bibinfo {volume} {94}},\ \bibinfo
  {pages} {81--85} (\bibinfo {year} {2011})}\BibitemShut {NoStop}%
\bibitem [{\citenamefont {Black-Schaffer}\ and\ \citenamefont
  {Honerkamp}(2014)}]{BlackSchafferandHonerkamp2014}%
  \BibitemOpen
  \bibfield  {author} {\bibinfo {author} {\bibfnamefont {A.~M.}\ \bibnamefont
  {Black-Schaffer}}\ and\ \bibinfo {author} {\bibfnamefont {C.}~\bibnamefont
  {Honerkamp}},\ }\bibfield  {title} {\bibinfo {title} {Chiral $d$-wave
  superconductivity in doped graphene},\ }\href
  {http://stacks.iop.org/0953-8984/26/i=42/a=423201} {\bibfield  {journal}
  {\bibinfo  {journal} {J. Phys. Condens. Matter}\ }\textbf {\bibinfo {volume}
  {26}},\ \bibinfo {pages} {423201} (\bibinfo {year} {2014})}\BibitemShut
  {NoStop}%
\bibitem [{\citenamefont {Dai}\ \emph {et~al.}(2021)\citenamefont {Dai},
  \citenamefont {Hou}, \citenamefont {Zhang}, \citenamefont {Liang},\ and\
  \citenamefont {Ma}}]{Dai2021Mott}%
  \BibitemOpen
  \bibfield  {author} {\bibinfo {author} {\bibfnamefont {H.}~\bibnamefont
  {Dai}}, \bibinfo {author} {\bibfnamefont {J.}~\bibnamefont {Hou}}, \bibinfo
  {author} {\bibfnamefont {X.}~\bibnamefont {Zhang}}, \bibinfo {author}
  {\bibfnamefont {Y.}~\bibnamefont {Liang}},\ and\ \bibinfo {author}
  {\bibfnamefont {T.}~\bibnamefont {Ma}},\ }\bibfield  {title} {\bibinfo
  {title} {Mott insulating state and $d+id$ superconductivity in an {ABC}
  graphene trilayer},\ }\href {https://doi.org/10.1103/PhysRevB.104.035104}
  {\bibfield  {journal} {\bibinfo  {journal} {Phys. Rev. B}\ }\textbf {\bibinfo
  {volume} {104}},\ \bibinfo {pages} {035104} (\bibinfo {year}
  {2021})}\BibitemShut {NoStop}%
\bibitem [{\citenamefont {Ojajärvi}\ \emph {et~al.}(2018)\citenamefont
  {Ojajärvi}, \citenamefont {Hyart}, \citenamefont {Silaev},\ and\
  \citenamefont {Heikkilä}}]{Ojajaervi2018}%
  \BibitemOpen
  \bibfield  {author} {\bibinfo {author} {\bibfnamefont {R.}~\bibnamefont
  {Ojajärvi}}, \bibinfo {author} {\bibfnamefont {T.}~\bibnamefont {Hyart}},
  \bibinfo {author} {\bibfnamefont {M.~A.}\ \bibnamefont {Silaev}},\ and\
  \bibinfo {author} {\bibfnamefont {T.~T.}\ \bibnamefont {Heikkilä}},\
  }\bibfield  {title} {\bibinfo {title} {Competition of electron-phonon
  mediated superconductivity and stoner magnetism on a flat band},\ }\href
  {https://doi.org/10.1103/physrevb.98.054515} {\bibfield  {journal} {\bibinfo
  {journal} {Phys Rev B}\ }\textbf {\bibinfo {volume} {98}},\ \bibinfo {pages}
  {054515} (\bibinfo {year} {2018})}\BibitemShut {NoStop}%
\bibitem [{\citenamefont {Wehling}\ \emph {et~al.}(2011)\citenamefont
  {Wehling}, \citenamefont {\ifmmode \mbox{\c{S}}\else \c{S}\fi{}a\ifmmode
  \mbox{\c{s}}\else \c{s}\fi{}\ifmmode \imath \else \i
  \fi{}o\ifmmode~\breve{g}\else \u{g}\fi{}lu}, \citenamefont {Friedrich},
  \citenamefont {Lichtenstein}, \citenamefont {Katsnelson},\ and\ \citenamefont
  {Bl\"ugel}}]{Wehling2011strength}%
  \BibitemOpen
  \bibfield  {author} {\bibinfo {author} {\bibfnamefont {T.~O.}\ \bibnamefont
  {Wehling}}, \bibinfo {author} {\bibfnamefont {E.}~\bibnamefont {\ifmmode
  \mbox{\c{S}}\else \c{S}\fi{}a\ifmmode \mbox{\c{s}}\else \c{s}\fi{}\ifmmode
  \imath \else \i \fi{}o\ifmmode~\breve{g}\else \u{g}\fi{}lu}}, \bibinfo
  {author} {\bibfnamefont {C.}~\bibnamefont {Friedrich}}, \bibinfo {author}
  {\bibfnamefont {A.~I.}\ \bibnamefont {Lichtenstein}}, \bibinfo {author}
  {\bibfnamefont {M.~I.}\ \bibnamefont {Katsnelson}},\ and\ \bibinfo {author}
  {\bibfnamefont {S.}~\bibnamefont {Bl\"ugel}},\ }\bibfield  {title} {\bibinfo
  {title} {Strength of effective {C}oulomb interactions in graphene and
  graphite},\ }\href {https://doi.org/10.1103/PhysRevLett.106.236805}
  {\bibfield  {journal} {\bibinfo  {journal} {Phys. Rev. Lett.}\ }\textbf
  {\bibinfo {volume} {106}},\ \bibinfo {pages} {236805} (\bibinfo {year}
  {2011})}\BibitemShut {NoStop}%
\bibitem [{\citenamefont {Xu}\ \emph {et~al.}(2012)\citenamefont {Xu},
  \citenamefont {Yuan}, \citenamefont {Yao}, \citenamefont {Zhou},
  \citenamefont {Gao},\ and\ \citenamefont {Zhang}}]{Xu2012}%
  \BibitemOpen
  \bibfield  {author} {\bibinfo {author} {\bibfnamefont {D.-H.}\ \bibnamefont
  {Xu}}, \bibinfo {author} {\bibfnamefont {J.}~\bibnamefont {Yuan}}, \bibinfo
  {author} {\bibfnamefont {Z.-J.}\ \bibnamefont {Yao}}, \bibinfo {author}
  {\bibfnamefont {Y.}~\bibnamefont {Zhou}}, \bibinfo {author} {\bibfnamefont
  {J.-H.}\ \bibnamefont {Gao}},\ and\ \bibinfo {author} {\bibfnamefont {F.-C.}\
  \bibnamefont {Zhang}},\ }\bibfield  {title} {\bibinfo {title} {Stacking
  order, interaction, and weak surface magnetism in layered graphene sheets},\
  }\href {https://doi.org/10.1103/physrevb.86.201404} {\bibfield  {journal}
  {\bibinfo  {journal} {Phys. Rev. B}\ }\textbf {\bibinfo {volume} {86}},\
  \bibinfo {pages} {201404} (\bibinfo {year} {2012})}\BibitemShut {NoStop}%
\bibitem [{\citenamefont {McClure}(1969)}]{MCCLURE1969Electron}%
  \BibitemOpen
  \bibfield  {author} {\bibinfo {author} {\bibfnamefont {J.}~\bibnamefont
  {McClure}},\ }\bibfield  {title} {\bibinfo {title} {Electron energy band
  structure and electronic properties of rhombohedral graphite},\ }\href
  {https://doi.org/10.1016/0008-6223(69)90073-6} {\bibfield  {journal}
  {\bibinfo  {journal} {Carbon}\ }\textbf {\bibinfo {volume} {7}},\ \bibinfo
  {pages} {425--432} (\bibinfo {year} {1969})}\BibitemShut {NoStop}%
\bibitem [{\citenamefont {Su}\ \emph {et~al.}(1979)\citenamefont {Su},
  \citenamefont {Schrieffer},\ and\ \citenamefont {Heeger}}]{Su1979}%
  \BibitemOpen
  \bibfield  {author} {\bibinfo {author} {\bibfnamefont {W.~P.}\ \bibnamefont
  {Su}}, \bibinfo {author} {\bibfnamefont {J.~R.}\ \bibnamefont {Schrieffer}},\
  and\ \bibinfo {author} {\bibfnamefont {A.~J.}\ \bibnamefont {Heeger}},\
  }\bibfield  {title} {\bibinfo {title} {Solitons in polyacetylene},\ }\href
  {https://doi.org/10.1103/physrevlett.42.1698} {\bibfield  {journal} {\bibinfo
   {journal} {Phys. Rev. Lett.}\ }\textbf {\bibinfo {volume} {42}},\ \bibinfo
  {pages} {1698--1701} (\bibinfo {year} {1979})}\BibitemShut {NoStop}%
\bibitem [{\citenamefont {Tancogne-Dejean}\ and\ \citenamefont
  {Rubio}(2020)}]{TancogneDejean2020}%
  \BibitemOpen
  \bibfield  {author} {\bibinfo {author} {\bibfnamefont {N.}~\bibnamefont
  {Tancogne-Dejean}}\ and\ \bibinfo {author} {\bibfnamefont {A.}~\bibnamefont
  {Rubio}},\ }\bibfield  {title} {\bibinfo {title} {Parameter-free hybridlike
  functional based on an extended {H}ubbard model: {DFT+U+V}},\ }\href
  {https://doi.org/10.1103/physrevb.102.155117} {\bibfield  {journal} {\bibinfo
   {journal} {Phys. Rev. B}\ }\textbf {\bibinfo {volume} {102}},\ \bibinfo
  {pages} {155117} (\bibinfo {year} {2020})}\BibitemShut {NoStop}%
\bibitem [{\citenamefont {Zhang}\ \emph {et~al.}(2010)\citenamefont {Zhang},
  \citenamefont {Sahu}, \citenamefont {Min},\ and\ \citenamefont
  {MacDonald}}]{Zhang2010Band}%
  \BibitemOpen
  \bibfield  {author} {\bibinfo {author} {\bibfnamefont {F.}~\bibnamefont
  {Zhang}}, \bibinfo {author} {\bibfnamefont {B.}~\bibnamefont {Sahu}},
  \bibinfo {author} {\bibfnamefont {H.}~\bibnamefont {Min}},\ and\ \bibinfo
  {author} {\bibfnamefont {A.~H.}\ \bibnamefont {MacDonald}},\ }\bibfield
  {title} {\bibinfo {title} {Band structure of {ABC}-stacked graphene
  trilayers},\ }\href {https://doi.org/10.1103/PhysRevB.82.035409} {\bibfield
  {journal} {\bibinfo  {journal} {Phys. Rev. B}\ }\textbf {\bibinfo {volume}
  {82}},\ \bibinfo {pages} {035409} (\bibinfo {year} {2010})}\BibitemShut
  {NoStop}%
\bibitem [{\citenamefont {Wang}\ \emph {et~al.}(2013)\citenamefont {Wang},
  \citenamefont {Gao},\ and\ \citenamefont {Zhang}}]{Wang2013Flat}%
  \BibitemOpen
  \bibfield  {author} {\bibinfo {author} {\bibfnamefont {H.}~\bibnamefont
  {Wang}}, \bibinfo {author} {\bibfnamefont {J.-H.}\ \bibnamefont {Gao}},\ and\
  \bibinfo {author} {\bibfnamefont {F.-C.}\ \bibnamefont {Zhang}},\ }\bibfield
  {title} {\bibinfo {title} {Flat band electrons and interactions in
  rhombohedral trilayer graphene},\ }\href
  {https://doi.org/10.1103/PhysRevB.87.155116} {\bibfield  {journal} {\bibinfo
  {journal} {Phys. Rev. B}\ }\textbf {\bibinfo {volume} {87}},\ \bibinfo
  {pages} {155116} (\bibinfo {year} {2013})}\BibitemShut {NoStop}%
\bibitem [{\citenamefont {Black-Schaffer}\ and\ \citenamefont
  {Doniach}(2007)}]{Black-Schaffer07}%
  \BibitemOpen
  \bibfield  {author} {\bibinfo {author} {\bibfnamefont {A.~M.}\ \bibnamefont
  {Black-Schaffer}}\ and\ \bibinfo {author} {\bibfnamefont {S.}~\bibnamefont
  {Doniach}},\ }\bibfield  {title} {\bibinfo {title} {Resonating valence bonds
  and mean-field $d$-wave superconductivity in graphite},\ }\href
  {https://doi.org/10.1103/PhysRevB.75.134512} {\bibfield  {journal} {\bibinfo
  {journal} {Phys. Rev. B}\ }\textbf {\bibinfo {volume} {75}},\ \bibinfo
  {pages} {134512} (\bibinfo {year} {2007})}\BibitemShut {NoStop}%
\bibitem [{\citenamefont {L\"othman}\ and\ \citenamefont
  {Black-Schaffer}(2014)}]{Lothman14}%
  \BibitemOpen
  \bibfield  {author} {\bibinfo {author} {\bibfnamefont {T.}~\bibnamefont
  {L\"othman}}\ and\ \bibinfo {author} {\bibfnamefont {A.~M.}\ \bibnamefont
  {Black-Schaffer}},\ }\bibfield  {title} {\bibinfo {title} {Defects in the
  $d+id$-wave superconducting state in heavily doped graphene},\ }\href
  {https://doi.org/10.1103/PhysRevB.90.224504} {\bibfield  {journal} {\bibinfo
  {journal} {Phys. Rev. B}\ }\textbf {\bibinfo {volume} {90}},\ \bibinfo
  {pages} {224504} (\bibinfo {year} {2014})}\BibitemShut {NoStop}%
\bibitem [{\citenamefont {Awoga}\ \emph {et~al.}(2017)\citenamefont {Awoga},
  \citenamefont {Bouhon},\ and\ \citenamefont
  {Black-Schaffer}}]{Awoga2017Domain}%
  \BibitemOpen
  \bibfield  {author} {\bibinfo {author} {\bibfnamefont {O.~A.}\ \bibnamefont
  {Awoga}}, \bibinfo {author} {\bibfnamefont {A.}~\bibnamefont {Bouhon}},\ and\
  \bibinfo {author} {\bibfnamefont {A.~M.}\ \bibnamefont {Black-Schaffer}},\
  }\bibfield  {title} {\bibinfo {title} {Domain walls in a chiral $d$-wave
  superconductor on the honeycomb lattice},\ }\href
  {https://doi.org/10.1103/PhysRevB.96.014521} {\bibfield  {journal} {\bibinfo
  {journal} {Phys. Rev. B}\ }\textbf {\bibinfo {volume} {96}},\ \bibinfo
  {pages} {014521} (\bibinfo {year} {2017})}\BibitemShut {NoStop}%
\bibitem [{\citenamefont {Awoga}\ and\ \citenamefont
  {Black-Schaffer}(2018)}]{Awoga2018Probing}%
  \BibitemOpen
  \bibfield  {author} {\bibinfo {author} {\bibfnamefont {O.~A.}\ \bibnamefont
  {Awoga}}\ and\ \bibinfo {author} {\bibfnamefont {A.~M.}\ \bibnamefont
  {Black-Schaffer}},\ }\bibfield  {title} {\bibinfo {title} {Probing
  unconventional superconductivity in proximitized graphene by impurity
  scattering},\ }\href {https://doi.org/10.1103/PhysRevB.97.214515} {\bibfield
  {journal} {\bibinfo  {journal} {Phys. Rev. B}\ }\textbf {\bibinfo {volume}
  {97}},\ \bibinfo {pages} {214515} (\bibinfo {year} {2018})}\BibitemShut
  {NoStop}%
\bibitem [{\citenamefont {Sigrist}\ and\ \citenamefont
  {Ueda}(1991)}]{SigristUeda1991}%
  \BibitemOpen
  \bibfield  {author} {\bibinfo {author} {\bibfnamefont {M.}~\bibnamefont
  {Sigrist}}\ and\ \bibinfo {author} {\bibfnamefont {K.}~\bibnamefont {Ueda}},\
  }\bibfield  {title} {\bibinfo {title} {Phenomenological theory of
  unconventional superconductivity},\ }\href
  {https://doi.org/10.1103/RevModPhys.63.239} {\bibfield  {journal} {\bibinfo
  {journal} {Rev. Mod. Phys.}\ }\textbf {\bibinfo {volume} {63}},\ \bibinfo
  {pages} {239--311} (\bibinfo {year} {1991})}\BibitemShut {NoStop}%
\bibitem [{\citenamefont {de~Gennes}(1999)}]{DeGennes}%
  \BibitemOpen
  \bibfield  {author} {\bibinfo {author} {\bibfnamefont {P.-G.}\ \bibnamefont
  {de~Gennes}},\ }\href@noop {} {\emph {\bibinfo {title} {Superconductivity of
  metals and alloys}}}\ (\bibinfo  {publisher} {Westview Press, Florida},\
  \bibinfo {year} {1999})\BibitemShut {NoStop}%
\bibitem [{\citenamefont {Bursill}\ \emph {et~al.}(1998)\citenamefont
  {Bursill}, \citenamefont {Castleton},\ and\ \citenamefont
  {Barford}}]{Bursill1998}%
  \BibitemOpen
  \bibfield  {author} {\bibinfo {author} {\bibfnamefont {R.~J.}\ \bibnamefont
  {Bursill}}, \bibinfo {author} {\bibfnamefont {C.}~\bibnamefont {Castleton}},\
  and\ \bibinfo {author} {\bibfnamefont {W.}~\bibnamefont {Barford}},\
  }\bibfield  {title} {\bibinfo {title} {Optimal parametrisation of the
  {P}ariser{\textendash}{P}arr{\textendash}{P}ople model for benzene and
  biphenyl},\ }\href {https://doi.org/10.1016/s0009-2614(98)00903-8} {\bibfield
   {journal} {\bibinfo  {journal} {Chem. Phys. Lett.}\ }\textbf {\bibinfo
  {volume} {294}},\ \bibinfo {pages} {305--313} (\bibinfo {year}
  {1998})}\BibitemShut {NoStop}%
\bibitem [{\citenamefont {Verg{\'{e}}s}\ \emph {et~al.}(2010)\citenamefont
  {Verg{\'{e}}s}, \citenamefont {SanFabi{\'{a}}n}, \citenamefont {Chiappe},\
  and\ \citenamefont {Louis}}]{Verges2010}%
  \BibitemOpen
  \bibfield  {author} {\bibinfo {author} {\bibfnamefont {J.~A.}\ \bibnamefont
  {Verg{\'{e}}s}}, \bibinfo {author} {\bibfnamefont {E.}~\bibnamefont
  {SanFabi{\'{a}}n}}, \bibinfo {author} {\bibfnamefont {G.}~\bibnamefont
  {Chiappe}},\ and\ \bibinfo {author} {\bibfnamefont {E.}~\bibnamefont
  {Louis}},\ }\bibfield  {title} {\bibinfo {title} {Fit of
  {P}ariser-{P}arr-{P}ople and {H}ubbard model {H}amiltonians to charge and
  spin states of polycyclic aromatic hydrocarbons},\ }\href
  {https://doi.org/10.1103/physrevb.81.085120} {\bibfield  {journal} {\bibinfo
  {journal} {Phys. Rev. B}\ }\textbf {\bibinfo {volume} {81}},\ \bibinfo
  {pages} {085120} (\bibinfo {year} {2010})}\BibitemShut {NoStop}%
\bibitem [{\citenamefont {Pietronero}\ \emph {et~al.}(1980)\citenamefont
  {Pietronero}, \citenamefont {Strässler}, \citenamefont {Zeller},\ and\
  \citenamefont {Rice}}]{Pietronero1980}%
  \BibitemOpen
  \bibfield  {author} {\bibinfo {author} {\bibfnamefont {L.}~\bibnamefont
  {Pietronero}}, \bibinfo {author} {\bibfnamefont {S.}~\bibnamefont
  {Strässler}}, \bibinfo {author} {\bibfnamefont {H.~R.}\ \bibnamefont
  {Zeller}},\ and\ \bibinfo {author} {\bibfnamefont {M.~J.}\ \bibnamefont
  {Rice}},\ }\bibfield  {title} {\bibinfo {title} {Electrical conductivity of a
  graphite layer},\ }\href {https://doi.org/10.1103/physrevb.22.904} {\bibfield
   {journal} {\bibinfo  {journal} {Phys. Rev. B}\ }\textbf {\bibinfo {volume}
  {22}},\ \bibinfo {pages} {904--910} (\bibinfo {year} {1980})}\BibitemShut
  {NoStop}%
\bibitem [{\citenamefont {Thingstad}\ \emph {et~al.}(2020)\citenamefont
  {Thingstad}, \citenamefont {Kamra}, \citenamefont {Wells},\ and\
  \citenamefont {Sudb{\o}}}]{Thingstad2020}%
  \BibitemOpen
  \bibfield  {author} {\bibinfo {author} {\bibfnamefont {E.}~\bibnamefont
  {Thingstad}}, \bibinfo {author} {\bibfnamefont {A.}~\bibnamefont {Kamra}},
  \bibinfo {author} {\bibfnamefont {J.~W.}\ \bibnamefont {Wells}},\ and\
  \bibinfo {author} {\bibfnamefont {A.}~\bibnamefont {Sudb{\o}}},\ }\bibfield
  {title} {\bibinfo {title} {Phonon-mediated superconductivity in doped
  monolayer materials},\ }\href {https://doi.org/10.1103/physrevb.101.214513}
  {\bibfield  {journal} {\bibinfo  {journal} {Phys. Rev. B}\ }\textbf {\bibinfo
  {volume} {101}},\ \bibinfo {pages} {214513} (\bibinfo {year}
  {2020})}\BibitemShut {NoStop}%
\bibitem [{\citenamefont {Falkovsky}(2008)}]{Falkovsky2008}%
  \BibitemOpen
  \bibfield  {author} {\bibinfo {author} {\bibfnamefont {L.~A.}\ \bibnamefont
  {Falkovsky}},\ }\bibfield  {title} {\bibinfo {title} {Phonon dispersion in
  graphene},\ }\href {https://doi.org/10.1121/1.2934282} {\bibfield  {journal}
  {\bibinfo  {journal} {J. Acoust. Soc. Am.}\ }\textbf {\bibinfo {volume}
  {123}},\ \bibinfo {pages} {3453--3453} (\bibinfo {year} {2008})}\BibitemShut
  {NoStop}%
\bibitem [{\citenamefont {O}\ \emph {et~al.}(2021)\citenamefont {O},
  \citenamefont {Kim}, \citenamefont {Pak}, \citenamefont {Jong}, \citenamefont
  {Ri},\ and\ \citenamefont {Pak}}]{SongJin2021}%
  \BibitemOpen
  \bibfield  {author} {\bibinfo {author} {\bibfnamefont {S.-J.}\ \bibnamefont
  {O}}, \bibinfo {author} {\bibfnamefont {Y.-H.}\ \bibnamefont {Kim}}, \bibinfo
  {author} {\bibfnamefont {O.-G.}\ \bibnamefont {Pak}}, \bibinfo {author}
  {\bibfnamefont {K.-H.}\ \bibnamefont {Jong}}, \bibinfo {author}
  {\bibfnamefont {C.-W.}\ \bibnamefont {Ri}},\ and\ \bibinfo {author}
  {\bibfnamefont {H.-C.}\ \bibnamefont {Pak}},\ }\bibfield  {title} {\bibinfo
  {title} {Competing electronic orders on a heavily doped honeycomb lattice
  with enhanced exchange coupling},\ }\href
  {https://doi.org/10.1103/physrevb.103.235150} {\bibfield  {journal} {\bibinfo
   {journal} {Phys. Rev. B}\ }\textbf {\bibinfo {volume} {103}},\ \bibinfo
  {pages} {235150} (\bibinfo {year} {2021})}\BibitemShut {NoStop}%
\bibitem [{\citenamefont {Lee}\ \emph {et~al.}(2006)\citenamefont {Lee},
  \citenamefont {Nagaosa},\ and\ \citenamefont {Wen}}]{Lee2006}%
  \BibitemOpen
  \bibfield  {author} {\bibinfo {author} {\bibfnamefont {P.~A.}\ \bibnamefont
  {Lee}}, \bibinfo {author} {\bibfnamefont {N.}~\bibnamefont {Nagaosa}},\ and\
  \bibinfo {author} {\bibfnamefont {X.-G.}\ \bibnamefont {Wen}},\ }\bibfield
  {title} {\bibinfo {title} {Doping a mott insulator: Physics of
  high-temperature superconductivity},\ }\href
  {https://doi.org/10.1103/revmodphys.78.17} {\bibfield  {journal} {\bibinfo
  {journal} {Rev. Mod. Phys.}\ }\textbf {\bibinfo {volume} {78}},\ \bibinfo
  {pages} {17--85} (\bibinfo {year} {2006})}\BibitemShut {NoStop}%
\bibitem [{\citenamefont {Hadipour}\ and\ \citenamefont
  {Jafari}(2015)}]{Hadipour2015}%
  \BibitemOpen
  \bibfield  {author} {\bibinfo {author} {\bibfnamefont {H.}~\bibnamefont
  {Hadipour}}\ and\ \bibinfo {author} {\bibfnamefont {S.~A.}\ \bibnamefont
  {Jafari}},\ }\bibfield  {title} {\bibinfo {title} {The importance of electron
  correlation in graphene and hydrogenated graphene},\ }\bibfield  {journal}
  {\bibinfo  {journal} {Eur. Phys. J. B}\ }\textbf {\bibinfo {volume} {88}},\
  \href {https://doi.org/10.1140/epjb/e2015-60454-1}
  {10.1140/epjb/e2015-60454-1} (\bibinfo {year} {2015})\BibitemShut {NoStop}%
\bibitem [{\citenamefont {Reich}\ \emph {et~al.}(2002)\citenamefont {Reich},
  \citenamefont {Maultzsch}, \citenamefont {Thomsen},\ and\ \citenamefont
  {Ordej{\'{o}}n}}]{Reich2002}%
  \BibitemOpen
  \bibfield  {author} {\bibinfo {author} {\bibfnamefont {S.}~\bibnamefont
  {Reich}}, \bibinfo {author} {\bibfnamefont {J.}~\bibnamefont {Maultzsch}},
  \bibinfo {author} {\bibfnamefont {C.}~\bibnamefont {Thomsen}},\ and\ \bibinfo
  {author} {\bibfnamefont {P.}~\bibnamefont {Ordej{\'{o}}n}},\ }\bibfield
  {title} {\bibinfo {title} {Tight-binding description of graphene},\ }\href
  {https://doi.org/10.1103/physrevb.66.035412} {\bibfield  {journal} {\bibinfo
  {journal} {Phys. Rev. B}\ }\textbf {\bibinfo {volume} {66}},\ \bibinfo
  {pages} {035412} (\bibinfo {year} {2002})}\BibitemShut {NoStop}%
\bibitem [{\citenamefont {Neto}\ \emph {et~al.}(2009)\citenamefont {Neto},
  \citenamefont {Guinea}, \citenamefont {Peres}, \citenamefont {Novoselov},\
  and\ \citenamefont {Geim}}]{Neto2009}%
  \BibitemOpen
  \bibfield  {author} {\bibinfo {author} {\bibfnamefont {A.~H.~C.}\
  \bibnamefont {Neto}}, \bibinfo {author} {\bibfnamefont {F.}~\bibnamefont
  {Guinea}}, \bibinfo {author} {\bibfnamefont {N.~M.~R.}\ \bibnamefont
  {Peres}}, \bibinfo {author} {\bibfnamefont {K.~S.}\ \bibnamefont
  {Novoselov}},\ and\ \bibinfo {author} {\bibfnamefont {A.~K.}\ \bibnamefont
  {Geim}},\ }\bibfield  {title} {\bibinfo {title} {The electronic properties of
  graphene},\ }\href {https://doi.org/10.1103/revmodphys.81.109} {\bibfield
  {journal} {\bibinfo  {journal} {Rev. Mod. Phys.}\ }\textbf {\bibinfo {volume}
  {81}},\ \bibinfo {pages} {109--162} (\bibinfo {year} {2009})}\BibitemShut
  {NoStop}%
\bibitem [{\citenamefont {Saremi}(2007)}]{Saremi2007}%
  \BibitemOpen
  \bibfield  {author} {\bibinfo {author} {\bibfnamefont {S.}~\bibnamefont
  {Saremi}},\ }\bibfield  {title} {\bibinfo {title} {{RKKY} in half-filled
  bipartite lattices: Graphene as an example},\ }\href
  {https://doi.org/10.1103/physrevb.76.184430} {\bibfield  {journal} {\bibinfo
  {journal} {Phys. Rev. B}\ }\textbf {\bibinfo {volume} {76}},\ \bibinfo
  {pages} {184430} (\bibinfo {year} {2007})}\BibitemShut {NoStop}%
\bibitem [{\citenamefont {Scalapino}(1995)}]{Scalapino1995}%
  \BibitemOpen
  \bibfield  {author} {\bibinfo {author} {\bibfnamefont {D.}~\bibnamefont
  {Scalapino}},\ }\bibfield  {title} {\bibinfo {title} {The case for
  $\mathrm{d_{x^2 - y^2}}$ pairing in the cuprate superconductors},\ }\href
  {https://doi.org/10.1016/0370-1573(94)00086-i} {\bibfield  {journal}
  {\bibinfo  {journal} {Phys. Rep.}\ }\textbf {\bibinfo {volume} {250}},\
  \bibinfo {pages} {329--365} (\bibinfo {year} {1995})}\BibitemShut {NoStop}%
\bibitem [{\citenamefont {Bickers}\ \emph {et~al.}(1989)\citenamefont
  {Bickers}, \citenamefont {Scalapino},\ and\ \citenamefont
  {White}}]{Bickers1989}%
  \BibitemOpen
  \bibfield  {author} {\bibinfo {author} {\bibfnamefont {N.~E.}\ \bibnamefont
  {Bickers}}, \bibinfo {author} {\bibfnamefont {D.~J.}\ \bibnamefont
  {Scalapino}},\ and\ \bibinfo {author} {\bibfnamefont {S.~R.}\ \bibnamefont
  {White}},\ }\bibfield  {title} {\bibinfo {title} {Conserving approximations
  for strongly correlated electron systems: {B}ethe-{S}alpeter equation and
  dynamics for the two-dimensional {H}ubbard model},\ }\href
  {https://doi.org/10.1103/physrevlett.62.961} {\bibfield  {journal} {\bibinfo
  {journal} {Phys. Rev. Lett.}\ }\textbf {\bibinfo {volume} {62}},\ \bibinfo
  {pages} {961--964} (\bibinfo {year} {1989})}\BibitemShut {NoStop}%
\bibitem [{\citenamefont {R{\o}mer}\ \emph {et~al.}(2020)\citenamefont
  {R{\o}mer}, \citenamefont {Maier}, \citenamefont {Kreisel}, \citenamefont
  {Eremin}, \citenamefont {Hirschfeld},\ and\ \citenamefont
  {Andersen}}]{Roemer2020}%
  \BibitemOpen
  \bibfield  {author} {\bibinfo {author} {\bibfnamefont {A.~T.}\ \bibnamefont
  {R{\o}mer}}, \bibinfo {author} {\bibfnamefont {T.~A.}\ \bibnamefont {Maier}},
  \bibinfo {author} {\bibfnamefont {A.}~\bibnamefont {Kreisel}}, \bibinfo
  {author} {\bibfnamefont {I.}~\bibnamefont {Eremin}}, \bibinfo {author}
  {\bibfnamefont {P.~J.}\ \bibnamefont {Hirschfeld}},\ and\ \bibinfo {author}
  {\bibfnamefont {B.~M.}\ \bibnamefont {Andersen}},\ }\bibfield  {title}
  {\bibinfo {title} {Pairing in the two-dimensional {H}ubbard model from weak
  to strong coupling},\ }\href
  {https://doi.org/10.1103/physrevresearch.2.013108} {\bibfield  {journal}
  {\bibinfo  {journal} {Phys. Rev. Res.}\ }\textbf {\bibinfo {volume} {2}},\
  \bibinfo {pages} {013108} (\bibinfo {year} {2020})}\BibitemShut {NoStop}%
\bibitem [{\citenamefont {Honerkamp}(2008)}]{Honerkamp2008}%
  \BibitemOpen
  \bibfield  {author} {\bibinfo {author} {\bibfnamefont {C.}~\bibnamefont
  {Honerkamp}},\ }\bibfield  {title} {\bibinfo {title} {Density waves and
  cooper pairing on the honeycomb lattice},\ }\href
  {https://doi.org/10.1103/physrevlett.100.146404} {\bibfield  {journal}
  {\bibinfo  {journal} {Phys. Rev. Lett.}\ }\textbf {\bibinfo {volume} {100}},\
  \bibinfo {pages} {146404} (\bibinfo {year} {2008})}\BibitemShut {NoStop}%
\bibitem [{\citenamefont {Scherer}\ \emph
  {et~al.}(2012{\natexlab{a}})\citenamefont {Scherer}, \citenamefont
  {Uebelacker},\ and\ \citenamefont {Honerkamp}}]{Scherer2012}%
  \BibitemOpen
  \bibfield  {author} {\bibinfo {author} {\bibfnamefont {M.~M.}\ \bibnamefont
  {Scherer}}, \bibinfo {author} {\bibfnamefont {S.}~\bibnamefont
  {Uebelacker}},\ and\ \bibinfo {author} {\bibfnamefont {C.}~\bibnamefont
  {Honerkamp}},\ }\bibfield  {title} {\bibinfo {title} {Instabilities of
  interacting electrons on the honeycomb bilayer},\ }\href
  {https://doi.org/10.1103/physrevb.85.235408} {\bibfield  {journal} {\bibinfo
  {journal} {Phys. Rev. B}\ }\textbf {\bibinfo {volume} {85}},\ \bibinfo
  {pages} {235408} (\bibinfo {year} {2012}{\natexlab{a}})}\BibitemShut
  {NoStop}%
\bibitem [{\citenamefont {Scherer}\ \emph
  {et~al.}(2012{\natexlab{b}})\citenamefont {Scherer}, \citenamefont
  {Uebelacker}, \citenamefont {Scherer},\ and\ \citenamefont
  {Honerkamp}}]{Scherer2012a}%
  \BibitemOpen
  \bibfield  {author} {\bibinfo {author} {\bibfnamefont {M.~M.}\ \bibnamefont
  {Scherer}}, \bibinfo {author} {\bibfnamefont {S.}~\bibnamefont {Uebelacker}},
  \bibinfo {author} {\bibfnamefont {D.~D.}\ \bibnamefont {Scherer}},\ and\
  \bibinfo {author} {\bibfnamefont {C.}~\bibnamefont {Honerkamp}},\ }\bibfield
  {title} {\bibinfo {title} {Interacting electrons on trilayer honeycomb
  lattices},\ }\href {https://doi.org/10.1103/physrevb.86.155415} {\bibfield
  {journal} {\bibinfo  {journal} {Phys. Rev. B}\ }\textbf {\bibinfo {volume}
  {86}},\ \bibinfo {pages} {155415} (\bibinfo {year}
  {2012}{\natexlab{b}})}\BibitemShut {NoStop}%
\bibitem [{\citenamefont {Oostinga}\ \emph {et~al.}(2008)\citenamefont
  {Oostinga}, \citenamefont {Heersche}, \citenamefont {Liu}, \citenamefont
  {Morpurgo},\ and\ \citenamefont {Vandersypen}}]{Oostinga2008}%
  \BibitemOpen
  \bibfield  {author} {\bibinfo {author} {\bibfnamefont {J.~B.}\ \bibnamefont
  {Oostinga}}, \bibinfo {author} {\bibfnamefont {H.~B.}\ \bibnamefont
  {Heersche}}, \bibinfo {author} {\bibfnamefont {X.}~\bibnamefont {Liu}},
  \bibinfo {author} {\bibfnamefont {A.~F.}\ \bibnamefont {Morpurgo}},\ and\
  \bibinfo {author} {\bibfnamefont {L.~M.}\ \bibnamefont {Vandersypen}},\
  }\bibfield  {title} {\bibinfo {title} {Gate-induced insulating state in
  bilayer graphene devices},\ }\href {https://doi.org/10.1038/nmat2082}
  {\bibfield  {journal} {\bibinfo  {journal} {Nat. Mater.}\ }\textbf {\bibinfo
  {volume} {7}},\ \bibinfo {pages} {151} (\bibinfo {year} {2008})}\BibitemShut
  {NoStop}%
\bibitem [{\citenamefont {Koshino}(2010)}]{Koshino2010}%
  \BibitemOpen
  \bibfield  {author} {\bibinfo {author} {\bibfnamefont {M.}~\bibnamefont
  {Koshino}},\ }\bibfield  {title} {\bibinfo {title} {Interlayer screening
  effect in graphene multilayers with {ABA} and {ABC} stacking},\ }\href
  {https://doi.org/10.1103/physrevb.81.125304} {\bibfield  {journal} {\bibinfo
  {journal} {Phys. Rev. B}\ }\textbf {\bibinfo {volume} {81}},\ \bibinfo
  {pages} {125304} (\bibinfo {year} {2010})}\BibitemShut {NoStop}%
\bibitem [{\citenamefont {Aoki}\ and\ \citenamefont
  {Amawashi}(2007)}]{Aoki2007}%
  \BibitemOpen
  \bibfield  {author} {\bibinfo {author} {\bibfnamefont {M.}~\bibnamefont
  {Aoki}}\ and\ \bibinfo {author} {\bibfnamefont {H.}~\bibnamefont
  {Amawashi}},\ }\bibfield  {title} {\bibinfo {title} {Dependence of band
  structures on stacking and field in layered graphene},\ }\href
  {https://doi.org/10.1016/j.ssc.2007.02.013} {\bibfield  {journal} {\bibinfo
  {journal} {Solid State Commun.}\ }\textbf {\bibinfo {volume} {142}},\
  \bibinfo {pages} {123--127} (\bibinfo {year} {2007})}\BibitemShut {NoStop}%
\bibitem [{\citenamefont {Giannozzi}\ \emph {et~al.}(2009)\citenamefont
  {Giannozzi}, \citenamefont {Baroni}, \citenamefont {Bonini}, \citenamefont
  {Calandra}, \citenamefont {Car}, \citenamefont {Cavazzoni}, \citenamefont
  {Ceresoli}, \citenamefont {Chiarotti}, \citenamefont {Cococcioni},
  \citenamefont {Dabo}, \citenamefont {Corso}, \citenamefont {de~Gironcoli},
  \citenamefont {Fabris}, \citenamefont {Fratesi}, \citenamefont {Gebauer},
  \citenamefont {Gerstmann}, \citenamefont {Gougoussis}, \citenamefont
  {Kokalj}, \citenamefont {Lazzeri}, \citenamefont {Martin-Samos},
  \citenamefont {Marzari}, \citenamefont {Mauri}, \citenamefont {Mazzarello},
  \citenamefont {Paolini}, \citenamefont {Pasquarello}, \citenamefont
  {Paulatto}, \citenamefont {Sbraccia}, \citenamefont {Scandolo}, \citenamefont
  {Sclauzero}, \citenamefont {Seitsonen}, \citenamefont {Smogunov},
  \citenamefont {Umari},\ and\ \citenamefont {Wentzcovitch}}]{Giannozzi2009}%
  \BibitemOpen
  \bibfield  {author} {\bibinfo {author} {\bibfnamefont {P.}~\bibnamefont
  {Giannozzi}}, \bibinfo {author} {\bibfnamefont {S.}~\bibnamefont {Baroni}},
  \bibinfo {author} {\bibfnamefont {N.}~\bibnamefont {Bonini}}, \bibinfo
  {author} {\bibfnamefont {M.}~\bibnamefont {Calandra}}, \bibinfo {author}
  {\bibfnamefont {R.}~\bibnamefont {Car}}, \bibinfo {author} {\bibfnamefont
  {C.}~\bibnamefont {Cavazzoni}}, \bibinfo {author} {\bibfnamefont
  {D.}~\bibnamefont {Ceresoli}}, \bibinfo {author} {\bibfnamefont {G.~L.}\
  \bibnamefont {Chiarotti}}, \bibinfo {author} {\bibfnamefont {M.}~\bibnamefont
  {Cococcioni}}, \bibinfo {author} {\bibfnamefont {I.}~\bibnamefont {Dabo}},
  \bibinfo {author} {\bibfnamefont {A.~D.}\ \bibnamefont {Corso}}, \bibinfo
  {author} {\bibfnamefont {S.}~\bibnamefont {de~Gironcoli}}, \bibinfo {author}
  {\bibfnamefont {S.}~\bibnamefont {Fabris}}, \bibinfo {author} {\bibfnamefont
  {G.}~\bibnamefont {Fratesi}}, \bibinfo {author} {\bibfnamefont
  {R.}~\bibnamefont {Gebauer}}, \bibinfo {author} {\bibfnamefont
  {U.}~\bibnamefont {Gerstmann}}, \bibinfo {author} {\bibfnamefont
  {C.}~\bibnamefont {Gougoussis}}, \bibinfo {author} {\bibfnamefont
  {A.}~\bibnamefont {Kokalj}}, \bibinfo {author} {\bibfnamefont
  {M.}~\bibnamefont {Lazzeri}}, \bibinfo {author} {\bibfnamefont
  {L.}~\bibnamefont {Martin-Samos}}, \bibinfo {author} {\bibfnamefont
  {N.}~\bibnamefont {Marzari}}, \bibinfo {author} {\bibfnamefont
  {F.}~\bibnamefont {Mauri}}, \bibinfo {author} {\bibfnamefont
  {R.}~\bibnamefont {Mazzarello}}, \bibinfo {author} {\bibfnamefont
  {S.}~\bibnamefont {Paolini}}, \bibinfo {author} {\bibfnamefont
  {A.}~\bibnamefont {Pasquarello}}, \bibinfo {author} {\bibfnamefont
  {L.}~\bibnamefont {Paulatto}}, \bibinfo {author} {\bibfnamefont
  {C.}~\bibnamefont {Sbraccia}}, \bibinfo {author} {\bibfnamefont
  {S.}~\bibnamefont {Scandolo}}, \bibinfo {author} {\bibfnamefont
  {G.}~\bibnamefont {Sclauzero}}, \bibinfo {author} {\bibfnamefont {A.~P.}\
  \bibnamefont {Seitsonen}}, \bibinfo {author} {\bibfnamefont {A.}~\bibnamefont
  {Smogunov}}, \bibinfo {author} {\bibfnamefont {P.}~\bibnamefont {Umari}},\
  and\ \bibinfo {author} {\bibfnamefont {R.~M.}\ \bibnamefont {Wentzcovitch}},\
  }\bibfield  {title} {\bibinfo {title} {{QUANTUM} {ESPRESSO}: a modular and
  open-source software project for quantum simulations of materials},\ }\href
  {https://doi.org/10.1088/0953-8984/21/39/395502} {\bibfield  {journal}
  {\bibinfo  {journal} {J. Phys: Condens Matter}\ }\textbf {\bibinfo {volume}
  {21}},\ \bibinfo {pages} {395502} (\bibinfo {year} {2009})}\BibitemShut
  {NoStop}%
\bibitem [{\citenamefont {Giannozzi}\ \emph {et~al.}(2017)\citenamefont
  {Giannozzi}, \citenamefont {Andreussi}, \citenamefont {Brumme}, \citenamefont
  {Bunau}, \citenamefont {Nardelli}, \citenamefont {Calandra}, \citenamefont
  {Car}, \citenamefont {Cavazzoni}, \citenamefont {Ceresoli}, \citenamefont
  {Cococcioni}, \citenamefont {Colonna}, \citenamefont {Carnimeo},
  \citenamefont {Corso}, \citenamefont {de~Gironcoli}, \citenamefont {Delugas},
  \citenamefont {DiStasio}, \citenamefont {Ferretti}, \citenamefont {Floris},
  \citenamefont {Fratesi}, \citenamefont {Fugallo}, \citenamefont {Gebauer},
  \citenamefont {Gerstmann}, \citenamefont {Giustino}, \citenamefont {Gorni},
  \citenamefont {Jia}, \citenamefont {Kawamura}, \citenamefont {Ko},
  \citenamefont {Kokalj}, \citenamefont {Kü{\c{c}}ükbenli}, \citenamefont
  {Lazzeri}, \citenamefont {Marsili}, \citenamefont {Marzari}, \citenamefont
  {Mauri}, \citenamefont {Nguyen}, \citenamefont {Nguyen}, \citenamefont {de-la
  Roza}, \citenamefont {Paulatto}, \citenamefont {Ponc{\'{e}}}, \citenamefont
  {Rocca}, \citenamefont {Sabatini}, \citenamefont {Santra}, \citenamefont
  {Schlipf}, \citenamefont {Seitsonen}, \citenamefont {Smogunov}, \citenamefont
  {Timrov}, \citenamefont {Thonhauser}, \citenamefont {Umari}, \citenamefont
  {Vast}, \citenamefont {Wu},\ and\ \citenamefont {Baroni}}]{Giannozzi2017}%
  \BibitemOpen
  \bibfield  {author} {\bibinfo {author} {\bibfnamefont {P.}~\bibnamefont
  {Giannozzi}}, \bibinfo {author} {\bibfnamefont {O.}~\bibnamefont
  {Andreussi}}, \bibinfo {author} {\bibfnamefont {T.}~\bibnamefont {Brumme}},
  \bibinfo {author} {\bibfnamefont {O.}~\bibnamefont {Bunau}}, \bibinfo
  {author} {\bibfnamefont {M.~B.}\ \bibnamefont {Nardelli}}, \bibinfo {author}
  {\bibfnamefont {M.}~\bibnamefont {Calandra}}, \bibinfo {author}
  {\bibfnamefont {R.}~\bibnamefont {Car}}, \bibinfo {author} {\bibfnamefont
  {C.}~\bibnamefont {Cavazzoni}}, \bibinfo {author} {\bibfnamefont
  {D.}~\bibnamefont {Ceresoli}}, \bibinfo {author} {\bibfnamefont
  {M.}~\bibnamefont {Cococcioni}}, \bibinfo {author} {\bibfnamefont
  {N.}~\bibnamefont {Colonna}}, \bibinfo {author} {\bibfnamefont
  {I.}~\bibnamefont {Carnimeo}}, \bibinfo {author} {\bibfnamefont {A.~D.}\
  \bibnamefont {Corso}}, \bibinfo {author} {\bibfnamefont {S.}~\bibnamefont
  {de~Gironcoli}}, \bibinfo {author} {\bibfnamefont {P.}~\bibnamefont
  {Delugas}}, \bibinfo {author} {\bibfnamefont {R.~A.}\ \bibnamefont
  {DiStasio}}, \bibinfo {author} {\bibfnamefont {A.}~\bibnamefont {Ferretti}},
  \bibinfo {author} {\bibfnamefont {A.}~\bibnamefont {Floris}}, \bibinfo
  {author} {\bibfnamefont {G.}~\bibnamefont {Fratesi}}, \bibinfo {author}
  {\bibfnamefont {G.}~\bibnamefont {Fugallo}}, \bibinfo {author} {\bibfnamefont
  {R.}~\bibnamefont {Gebauer}}, \bibinfo {author} {\bibfnamefont
  {U.}~\bibnamefont {Gerstmann}}, \bibinfo {author} {\bibfnamefont
  {F.}~\bibnamefont {Giustino}}, \bibinfo {author} {\bibfnamefont
  {T.}~\bibnamefont {Gorni}}, \bibinfo {author} {\bibfnamefont
  {J.}~\bibnamefont {Jia}}, \bibinfo {author} {\bibfnamefont {M.}~\bibnamefont
  {Kawamura}}, \bibinfo {author} {\bibfnamefont {H.-Y.}\ \bibnamefont {Ko}},
  \bibinfo {author} {\bibfnamefont {A.}~\bibnamefont {Kokalj}}, \bibinfo
  {author} {\bibfnamefont {E.}~\bibnamefont {Kü{\c{c}}ükbenli}}, \bibinfo
  {author} {\bibfnamefont {M.}~\bibnamefont {Lazzeri}}, \bibinfo {author}
  {\bibfnamefont {M.}~\bibnamefont {Marsili}}, \bibinfo {author} {\bibfnamefont
  {N.}~\bibnamefont {Marzari}}, \bibinfo {author} {\bibfnamefont
  {F.}~\bibnamefont {Mauri}}, \bibinfo {author} {\bibfnamefont {N.~L.}\
  \bibnamefont {Nguyen}}, \bibinfo {author} {\bibfnamefont {H.-V.}\
  \bibnamefont {Nguyen}}, \bibinfo {author} {\bibfnamefont {A.~O.}\
  \bibnamefont {de-la Roza}}, \bibinfo {author} {\bibfnamefont
  {L.}~\bibnamefont {Paulatto}}, \bibinfo {author} {\bibfnamefont
  {S.}~\bibnamefont {Ponc{\'{e}}}}, \bibinfo {author} {\bibfnamefont
  {D.}~\bibnamefont {Rocca}}, \bibinfo {author} {\bibfnamefont
  {R.}~\bibnamefont {Sabatini}}, \bibinfo {author} {\bibfnamefont
  {B.}~\bibnamefont {Santra}}, \bibinfo {author} {\bibfnamefont
  {M.}~\bibnamefont {Schlipf}}, \bibinfo {author} {\bibfnamefont {A.~P.}\
  \bibnamefont {Seitsonen}}, \bibinfo {author} {\bibfnamefont {A.}~\bibnamefont
  {Smogunov}}, \bibinfo {author} {\bibfnamefont {I.}~\bibnamefont {Timrov}},
  \bibinfo {author} {\bibfnamefont {T.}~\bibnamefont {Thonhauser}}, \bibinfo
  {author} {\bibfnamefont {P.}~\bibnamefont {Umari}}, \bibinfo {author}
  {\bibfnamefont {N.}~\bibnamefont {Vast}}, \bibinfo {author} {\bibfnamefont
  {X.}~\bibnamefont {Wu}},\ and\ \bibinfo {author} {\bibfnamefont
  {S.}~\bibnamefont {Baroni}},\ }\bibfield  {title} {\bibinfo {title} {Advanced
  capabilities for materials modelling with quantum {ESPRESSO}},\ }\href
  {https://doi.org/10.1088/1361-648x/aa8f79} {\bibfield  {journal} {\bibinfo
  {journal} {J. Phys: Condens Matter}\ }\textbf {\bibinfo {volume} {29}},\
  \bibinfo {pages} {465901} (\bibinfo {year} {2017})}\BibitemShut {NoStop}%
\bibitem [{\citenamefont {Grimme}(2006)}]{Grimme2006Semi}%
  \BibitemOpen
  \bibfield  {author} {\bibinfo {author} {\bibfnamefont {S.}~\bibnamefont
  {Grimme}},\ }\bibfield  {title} {\bibinfo {title} {Semiempirical {GGA}-type
  density functional constructed with a long-range dispersion correction},\
  }\href {https://doi.org/10.1002/jcc.20495} {\bibfield  {journal} {\bibinfo
  {journal} {J. Comput. Chem.}\ }\textbf {\bibinfo {volume} {27}},\ \bibinfo
  {pages} {1787--1799} (\bibinfo {year} {2006})}\BibitemShut {NoStop}%
\bibitem [{\citenamefont {Prandini}\ \emph {et~al.}(2018)\citenamefont
  {Prandini}, \citenamefont {Marrazzo}, \citenamefont {Castelli}, \citenamefont
  {Mounet},\ and\ \citenamefont {Marzari}}]{Prandini2018Precision}%
  \BibitemOpen
  \bibfield  {author} {\bibinfo {author} {\bibfnamefont {G.}~\bibnamefont
  {Prandini}}, \bibinfo {author} {\bibfnamefont {A.}~\bibnamefont {Marrazzo}},
  \bibinfo {author} {\bibfnamefont {I.~E.}\ \bibnamefont {Castelli}}, \bibinfo
  {author} {\bibfnamefont {N.}~\bibnamefont {Mounet}},\ and\ \bibinfo {author}
  {\bibfnamefont {N.}~\bibnamefont {Marzari}},\ }\bibfield  {title} {\bibinfo
  {title} {Precision and efficiency in solid-state pseudopotential
  calculations},\ }\bibfield  {journal} {\bibinfo  {journal} {npj Comput.
  Mater.}\ }\textbf {\bibinfo {volume} {4}},\ \href
  {https://doi.org/10.1038/s41524-018-0127-2} {10.1038/s41524-018-0127-2}
  (\bibinfo {year} {2018})\BibitemShut {NoStop}%
\bibitem [{\citenamefont {Prandini}\ \emph {et~al.}(2023)\citenamefont
  {Prandini}, \citenamefont {Marrazzo}, \citenamefont {Castelli}, \citenamefont
  {Mounet}, \citenamefont {Passaro}, \citenamefont {Yu},\ and\ \citenamefont
  {Marzari}}]{Prandini2018A}%
  \BibitemOpen
  \bibfield  {author} {\bibinfo {author} {\bibfnamefont {G.}~\bibnamefont
  {Prandini}}, \bibinfo {author} {\bibfnamefont {A.}~\bibnamefont {Marrazzo}},
  \bibinfo {author} {\bibfnamefont {I.~E.}\ \bibnamefont {Castelli}}, \bibinfo
  {author} {\bibfnamefont {N.}~\bibnamefont {Mounet}}, \bibinfo {author}
  {\bibfnamefont {E.}~\bibnamefont {Passaro}}, \bibinfo {author} {\bibfnamefont
  {J.}~\bibnamefont {Yu}},\ and\ \bibinfo {author} {\bibfnamefont
  {N.}~\bibnamefont {Marzari}},\ }\bibfield  {title} {\bibinfo {title}
  {Standard solid state pseudopotentials ({SSSP}) library optimized for
  precision and efficiency},\ }\bibfield  {journal} {\bibinfo  {journal}
  {Materials Cloud Archive}\ }\textbf {\bibinfo {volume} {38}},\ \href
  {https://doi.org/10.24435/materialscloud:gv-gp}
  {10.24435/materialscloud:gv-gp} (\bibinfo {year} {2023})\BibitemShut
  {NoStop}%
\bibitem [{\citenamefont {Feynman}(1998)}]{Feynman1998Stat}%
  \BibitemOpen
  \bibfield  {author} {\bibinfo {author} {\bibfnamefont {R.~P.}\ \bibnamefont
  {Feynman}},\ }\href@noop {} {\emph {\bibinfo {title} {Statistical Mechanics,
  revised ed.}}}\ (\bibinfo  {publisher} {Advanced Books Classics (Westview
  Press)},\ \bibinfo {year} {1998})\BibitemShut {NoStop}%
\end{thebibliography}%
\end{document}